\documentclass[aps,prb,twocolumn,amsmath,amssymb,eqsecnum,groupedaddress,showpacs,floats,floatfix]{revtex4}

\usepackage{graphics}

\usepackage{epsf}

\newcommand{\tr}{\mbox{tr}\, }
\newcommand{\sign}{\mbox{sign}}
\newcommand*{\cN}{{\cal N}}
\newcommand*{\cS}{{\cal S}}
\newcommand{\rc}{r_{\rm c}^{\vphantom{\dagger}}}
\newcommand{\rcd}{r_{\rm c}^{{\dagger}}}

\begin{document}

\title{Nonequilibrium theory of Coulomb blockade in open quantum dots}

\author{Piet W.\ Brouwer}
\affiliation{Laboratory of Atomic and Solid State Physics,
Cornell University, Ithaca, New York 14853-2501}
\email[]{brouwer@ccmr.cornell.edu}

\author{Austen Lamacraft}
\affiliation{Department of Physics, Princeton University, Princeton, NJ
  08544, USA}
\email[]{alamacra@princeton.edu}

\author{Karsten Flensberg}
\affiliation{Nano-Science Center, Niels Bohr Institute,
Universitetsparken 5, 2100 Copenhagen, Denmark}

\begin{abstract}

We develop a non-equilibrium theory to describe weak Coulomb
blockade effects in open quantum dots. Working within the
bosonized description of electrons in the point contacts, we
expose deficiencies in earlier applications of this method, and
address them using a $1/N$ expansion in the inverse number of
channels. At leading order this yields the self-consistent
potential for the charging interaction. Coulomb blockade effects
arise as quantum corrections to transport at the next order. Our
approach unifies the phase functional and bosonization approaches
to the problem, as well as providing a simple picture for the
conductance corrections in terms of renormalization of the dot's
elastic scattering matrix, which is obtained also by elementary
perturbation theory. For the case of ideal contacts, a symmetry
argument immediately allows us to conclude that interactions give
no signature in the averaged conductance. Non-equilibrium
applications to the pumped current in a quantum pump are worked
out in detail. 
\end{abstract}

\maketitle

\section{Introduction}

Coulomb blockade is the phenomenon that transport through an
almost isolated system is prohibited by the energy cost $E_{\rm
c}$ to add or remove an electron. The Coulomb blockade can be
lifted by fine tuning a gate voltage to a point of charge
degeneracy, where the energy cost of adding one electron is zero,
or by raising the bias voltage or the temperature above $E_{\rm
c}$.  A fundamental question is to what extent the quantization of
charge on the system, and hence the Coulomb blockade, is affected
by the inclusion of that system in an electrical circuit. This
question has been addressed both for metal particles coupled to
electrodes via tunnel barriers, as a function of the dimensionless
conductance of the tunnel barrier, and for semiconductor quantum
dots coupled to electrodes via ballistic point contacts, as a
function of the number of propagating channels in the point
contacts. In each case, the Coulomb blockade is lifted when the
conductance of the contact to the electrodes is larger than the
conductance quantum $e^2/h$. Such quantum dots (or metal
particles) are referred to as `open', in contrast to `closed'
quantum dots, which are connected to the electron reservoirs
through tunneling contacts with a conductance smaller than the
conductance quantum. The present paper deals with Coulomb blockade
in open semiconductor quantum dots, which we shall refer to as
`weak Coulomb blockade'.

In the literature, two approaches have been taken to this problem.
One theory of `weak Coulomb blockade' was proposed by
Flensberg\cite{kn:flensberg1993}, Matveev\cite{kn:matveev1995} and
Furusaki and Matveev,\cite{kn:furusaki1995a,kn:furusaki1995b} who
realized that a tractable description of the strong quantum charge
fluctuations can be formulated using the one-dimensional nature of
the point contacts. Whereas these theories accounted for the
electron-electron interactions non-perturbatively, they neglected
the effect of electrons coherently traversing the dot or being
backscattered from inside. Extensions of the theories of Refs.\
\onlinecite{kn:flensberg1993,kn:matveev1995,kn:furusaki1995a,kn:furusaki1995b}
to the case of open quantum dots with coherent scattering of
electrons were given by Yi and Kane\cite{kn:yi1996} for quantum
dots in the quantum Hall regime and by Aleiner and
Glazman\cite{kn:aleiner1998a} and Brouwer and
Aleiner\cite{kn:brouwer1999c} for the general case (see Ref.\
\onlinecite{kn:aleiner2002} for a review).

The other theoretical approach starts from an effective theory in
which the primary dynamic variables are the potential differences
between the quantum dot and the electrodes.\cite{kn:schoen1990}
This approach has mainly been used to study metal grains coupled
to the reservoirs via tunnel barriers with many
channels,\cite{kn:zaikin1993,kn:golubev1996,kn:wang1997b,
kn:wang1997c,kn:golubev1997,kn:panyukov1991,kn:falci1995,kn:wang1996,
kn:wang1997d,kn:koenig1998,kn:beloborodov2003}, but recently it
has been applied to semiconductor quantum dots, both
without\cite{kn:nazarov1999,kn:levyyeyati2001,kn:golubev2001} and
with\cite{kn:golubev2004} coherent scattering of electrons from
inside the quantum dot. The general conclusions of both approaches
are the same: For the incoherent case and ideal point contacts (no
backscattering in the contact), it was found that all charge
quantization effects completely
vanishes.\cite{kn:flensberg1993,kn:matveev1995,kn:furusaki1995a,kn:furusaki1995b,kn:nazarov1999}
In all other cases --- contacts with a small amplitude $r_{\rm c}$
for backscattering in the contact or coherent scattering from
inside the dot ---, charge quantization gives a correction to the
dot's capacitance or conductance. This correction is small, but it
becomes more important as the temperature is lowered or the number
of channels in the point contacts is reduced.

The previous works on `weak Coulomb blockade' all dealt with
thermodynamic properties (capacitance) or time-independent
transport. It is the goal of the present work to develop the general
non-equilibrium theory for Coulomb blockade in open quantum dots,
accounting for coherent scattering of electrons inside the dot.
Our main result is an expansion in powers of $1/N$ for the current
through the dot, where $N$ is the total number of channels in the
point contacts.

Two examples of non-equilibrium problems
involving quantum interference in open and interacting mesoscopic
systems have been a particular motivation for this work. First, the
possibility of applying time dependent potentials via local gating
of quantum dots led to the theoretical and experimental
investigation of `quantum pumps', in which a periodic perturbation
of potentials inside the quantum dot, combined with quantum
interference, leads to a dc current through the
dot.\cite{kn:switkes1999} For quantum dots with time dependent
potentials, effects of coherent scattering inside the quantum dot,
charge quantization, and electron-electron interactions were
considered separately theoretically (see, {\em e.g.}, Refs.\
\onlinecite{kn:buettiker1993a,kn:buettiker1993b,
kn:buettiker1994,kn:aleiner1998b,kn:brouwer1998,kn:andreev2001}),
but not together. The second example that motivated this work is
the experiment of Ref.~\onlinecite{kn:pothier1997}, where the
nonequilibrium electron distribution function of a current
carrying diffusive wire was measured by tunneling spectroscopy.
This allowed an experimental determination of the
electron-electron collision integral, and led to the discovery of
a new mechanism of energy
relaxation.\cite{kn:kaminski2001,kn:anthore2003}

Following previous studies of Coulomb blockade in open quantum dots,
the only electron-electron interaction term we consider is the
capacitive charging energy of the quantum dot. This requires
relatively large quantum dots: For open quantum dots Coulomb blockade
effects are small in the inverse of the number of channels  $N$ in the
contacts connecting the dot to the electron reservoirs, whereas
the residual parts of the Coulomb interaction are small in the inverse
dimensionless conductance of the `closed'
dot.\cite{kn:aleiner1998a,kn:kurland2000,kn:aleiner2002} The ratio of
these two conductances equals the ratio of the dot's dwell time and
ergodic time, which is large for large quantum dots. This parameter justifies the use of the random matrix description of scattering by the dot, and means that the problem is effectively `zero dimensional'. The dimensionality of the system appears only in the corrections to this picture.

Unlike for the case of linear time-independent transport, where
the capacitive interaction gives corrections to transport properties
beyond the Hartree level only, interaction corrections to nonlinear
or time-dependent transport exist already when interactions
are described by a self-consistent (Hartree) potential $V_{\rm d}$.
As pointed out by B\"uttiker and coworkers, the reason is that, in
non-equilibrium or time-dependent problems, a change in bias voltage
or a change in a gate voltage
potentials may change the number of electrons on the dot,
which, in turn, changes $V_{\rm d}$,\cite{kn:buettiker1993a,kn:buettiker1993b,kn:buettiker1994}
\begin{equation} \label{sc_pot}
   V_{\mathrm{d}}(t)=-\frac{e}{C}\left(
  \langle\hat N_{\mathrm{dot}}(t)\rangle-\cN \right).
\end{equation}
Here $C$ is the dot's capacitance, $-e\langle\hat
N_{\mathrm{dot}}\rangle$ is the dot charge, and $-e{\cal N}$ the
offset charge induced by the gates. The effect of a time-dependent
or bias-dependent Hartree potential $V_{\rm d}$ on mesoscopic
fluctuations of various transport properties --- but neglecting the
`weak Coulomb blockade' --- was investigated in Refs.\
\onlinecite{kn:brouwer1997b,kn:brouwer1997d,kn:pedersen1998,kn:brouwer1998,kn:pilgram2002}.
To leading order in $1/N$, our theory recovers the
self-consistent theory with the Hartree potential (\ref{sc_pot}).
In this sense, our result can be viewed as a formal confirmation
of the self-consistent theory. Further, it gives a controlled
method to find interaction corrections that cannot be described by
means of a self-consistent potential.

A second motivation of this work is to compare and unify the results from
two different approaches to these corrections --- the
`one-dimensional' and `environmental' formalisms. Whereas both
approaches have given equal results for the case of quantum dots
without coherent scattering from inside the dot, results for the dc
conductance of coherent dots, as reported by Brouwer and
Aleiner\cite{kn:brouwer1999c} and Golubev and
Zaikin\cite{kn:golubev2004}, are in disagreement. Using a formalism
closely related to that of Ref.\ \onlinecite{kn:brouwer1999c}, we are
able to present detailed calculations in two limits: A formal
expansion in the scattering matrix of the dot for an arbitrary number
of channels $N$ in the contacts, and as an expansion in powers of
$1/N$.
The former limit reproduces the result of Ref.\
\onlinecite{kn:brouwer1999c}, whereas the latter limit agrees with
Ref.\ \onlinecite{kn:golubev2004}. The formal expansion in the
scattering matrix $S$, however, is only allowed if the probability of
coherent (back)scattering from either the quantum dot or the contact
is small, and does not describe interaction corrections to the
conductance of a fully coherent quantum dot ({\em i.e.}, a dot with
a unitary scattering matrix). With the same arguments,
the result for the interaction correction to the dot's capacitance,
which was calculated in Ref.\ \onlinecite{kn:aleiner2002} using a
formal expansion in $S$, is not applicable to a quantum dot with a
unitary scattering matrix. The correct theory, in the limit of large
$N$, will be given here. We will not compare non-perturbative results in the
two formalisms, that were obtained
using refermionization\cite{kn:furusaki1995b} or
instanton solutions.\cite{kn:nazarov1999,kn:golubev2001}

With regard to the interaction corrections to conductance, our
result is particularly simple: for a fully coherent dot there
is no correction to the {\em ensemble averaged} conductance, to 
leading and sub-leading order in $1/N$. In particular, this applies
to weak localization, the
quantum correction to the ensemble averaged conductance that
is suppressed by the application of a magnetic field. The reason for 
the absence of an interaction correction to the ensemble averaged
conductance is the following. In the Landauer description of transport, 
conductance is related to a sum of squares of moduli of elements 
$S_{ij}(\varepsilon)$ of the scattering matrix $S(\varepsilon)$, 
appropriately integrated over energy if the temperature is nonzero 
(see, {\em e.g.}, Ref.~\onlinecite{kn:beenakker1997}). The ensemble 
average of the conductance depends on the ensemble averages $\langle
|S(\varepsilon)_{ij}|^2 \rangle$ only. According to the random matrix
theory of quantum transport, such an average is determined by 
symmetries only (presence or absence of time-reversal and
spin-rotation symmetry). At leading and sub-leading order in $1/N$, 
we find that the interaction corrections can be expressed through a
renormalized scattering matrix: $S(\varepsilon)\to S'(\varepsilon)
\equiv S(\varepsilon)+\delta S(\varepsilon)$ [see Eq.\ (\ref{deltaS})
below]. Although this interaction correction changes the scattering 
matrix, and hence conductance, of a specific dot, it does not change
the {\em symmetry} of the scattering matrix ensemble, leaving the 
ensemble averaged conductance (including the weak-localization
correction) unchanged. We will revisit this argument in more detail 
in Sec.\ \ref{sec:conclusion}. 

The outline of our paper is as follows. In Sec.\
\ref{sec:previous} we present a detailed discussion of the
relevant existing works in the literature. This discussion will
serve to introduce the necessary concepts and notations, and to
explain the disagreement between the two existing theories of
`weak Coulomb blockade' in coherent quantum dots. In Sec.\
\ref{sec:formalism} we proceed with a precise definition of the
problem and an exposition of the formalism. In the diagrammatic
language that we will introduce later, the leading interaction
correction to the current arises from a Fock-type diagram. The
idea that interaction corrections to conductance in the presence
of impurities should be thought of as due to scattering from an
Hartree-Fock potential was introduced in the work of
Matveev, Yue, and Glazman.\cite{kn:matveev1993,kn:yue1994} In
order to make contact to that work, we have added to Sec.\
\ref{sec:formalism} a calculation similar in spirit to that of
Refs.\ \onlinecite{kn:matveev1993,kn:yue1994} that already
contains the basic structure of the interaction correction to the
current. The full calculation of the current through the quantum
dot, for the general non-equilibrium and time-dependent situation
is then described in Sec.\ \ref{sec:current}. Sections
\ref{sec:steadystate} and \ref{sec:adiabatic} contain a detailed
analysis of these results for nonlinear steady-state transport and
adiabatic time-dependent transport, respectively. We conclude in
Sec.\ \ref{sec:conclusion}. Finally, appendices \ref{app:boson},
\ref{app:a}, and \ref{app:c} contain materials on the correlation
functions in the `one-dimensional' formalism, the ensemble
average, and the case of a quantum dot with partially coherent
scattering, respectively. A summary of our results, focusing on
the relation between our work and previous approaches to `weak
Coulomb blockade' appeared in Ref.\ \onlinecite{kn:brouwer2005}.

\section{Previous work}
\label{sec:previous}

In this section, we present a more detailed overview of the
results published in the literature. This will also serve to
introduce some of the needed concepts and notations.

\begin{figure}
\epsfxsize=0.7\hsize
\epsffile{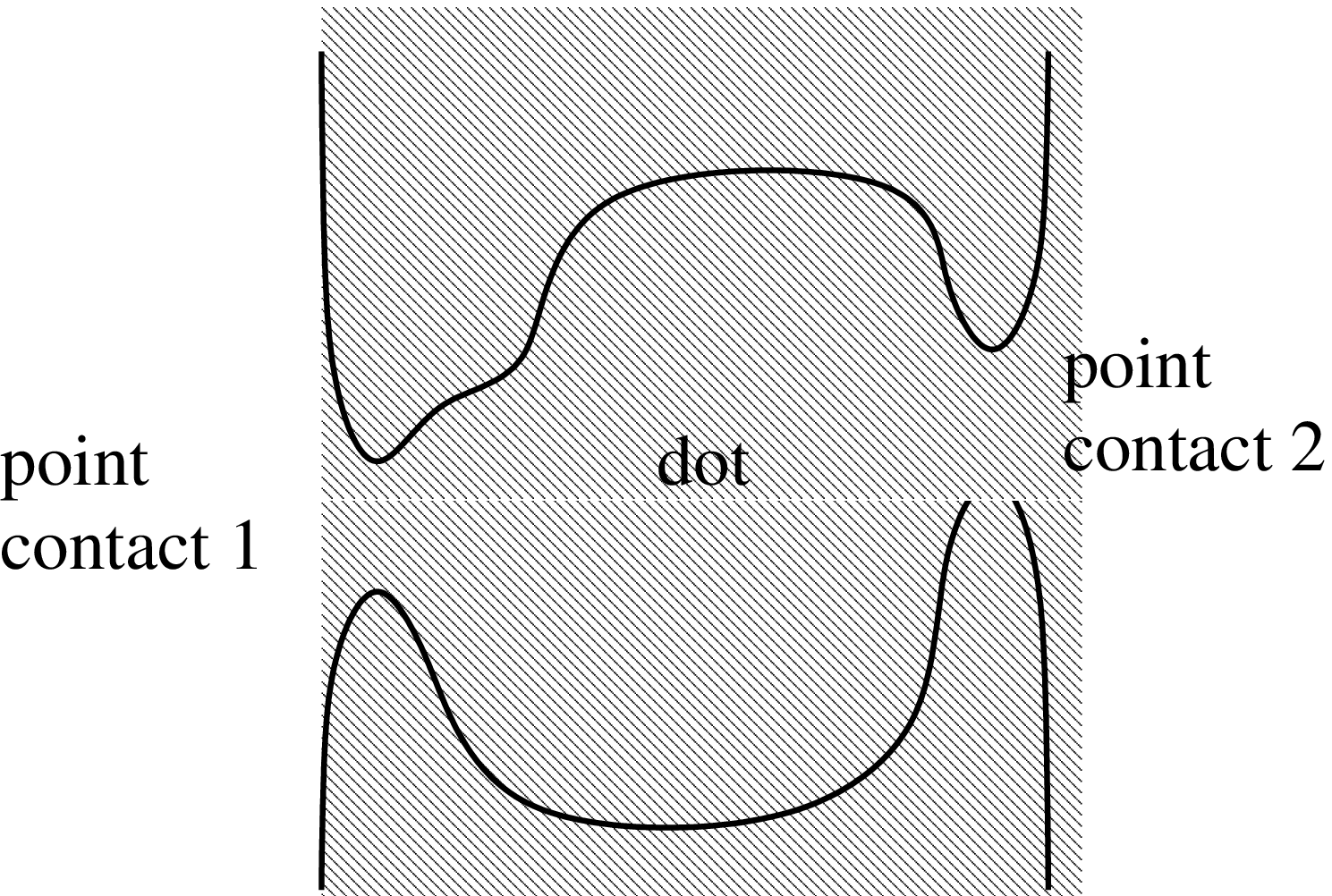}
\caption{\label{fig:1}
Schematic drawing of the system under consideration: a
quantum dot (center) coupled to source and drain reservoirs
(left and right) via point contacts. The two point contacts
have $N_1$ and $N_2$ propagating channels at the Fermi level.
Direct reflection at the point contact is described by energy-
independent reflection matrices $r_{{\rm c}1}$ and $r_{{\rm c}2}$,
so that their dimensionless conductances are $G_1 = (e^2/2 \pi \hbar)(
N_1 - \mbox{tr}\, r_{{\rm c}1}^{\protect\vphantom{\dagger}}
r_{{\rm c}1}^{\dagger})$ and $G_2 = (e^2/h)(
N_2 - \mbox{tr}\, r_{{\rm c}2}^{\protect\vphantom{\dagger}}
r_{{\rm c}2}^{\dagger})$, respectively. Interactions in the dot
are described by the charging energy
$E_{\rm c}$.}
\end{figure}
A schematic drawing of the system under consideration is shown in
Fig.\ \ref{fig:1}. It consists of a quantum dot coupled to two
electron reservoirs by means of point contacts. The point contacts
have $N_1$ and $N_2$ channels each (counting spin degeneracy).
The two contacts are characterized by
energy-independent reflection matrices $r_{{\rm c}1}$ and $r_{{\rm
    c}2}$, of dimension $N_1$ and $N_2$,
respectively. In this notation, the conductance of each contact
is
\begin{equation}
  G_i = \frac{e^2}{2 \pi \hbar} g_i,\ \
  g_i = N_i - \tr r_{{\rm c}i}^{\vphantom{\dagger}}
  r_{{\rm c}i}^{\dagger},\ \
  i = 1,2.
  \label{eq:gi}
\end{equation}
A contact with $N_i = g_i$, $i=1,2$ is called
ballistic or `ideal'; otherwise the contact is
called `nonideal'. A dot with $g_1+ g_2 \ll 1$ is referred
to as `closed'; an open dot has $g_1 + g_2 \gg 1$.
For future reference, we define the total number of channels in
the point contacts $N = N_1 + N_2$, the total dimensionless
conductance of the point contacts $g = g_1 + g_2$,
and the $N \times N$ combined reflection matrix of both contacts
\begin{equation}
  r_{\rm c} = \left( \begin{array}{ll} r_{{\rm c}1} & 0 \\
  0 & r_{{\rm c}2} \end{array} \right).
\end{equation}

The discussion of this section will be limited to the linear dc
conductance $G$ of the quantum dot. Our discussion of the literature
will be limited to the cases of a
`coherent' quantum dot, for which transport properties are expressed
in terms of its $N$-dimensional unitary
 scattering matrix $S(\varepsilon)$,
and an `incoherent' quantum dot, for which the dot itself is 
treated as a reservoir in equilibrium, with a self-consistent 
chemical potential.\cite{kn:flensberg1993,kn:matveev1995} In the
`incoherent' limit, coherent propagation across the dot, or 
backscattering from within, is not taken into account. 

Neglecting interactions for electrons
on the dot, for the `incoherent' case the conductance is
nothing but the classical series conductance $G_{\mathrm{cl}}=(e^2/\hbar) g_1 g_2/ g$ of
the two point contacts. For the `coherent' case, the
conductance is given by the Landauer formula, which we write as
\begin{eqnarray} \label{landauer}
  G_{\mathrm{0}} &=& \frac{e^2}{2\pi\hbar} \frac{N_1N_2}{N}
   \\ && \mbox{}
  - \frac{e^2}{2\pi\hbar}
  \int d\varepsilon \left( - \frac{\partial f(\varepsilon)}{\partial
  \varepsilon} \right) \mbox{tr}\, \Lambda S(\varepsilon)
  \Lambda S^{\dagger}(\varepsilon). \nonumber
\end{eqnarray}
Here $f(\varepsilon) = 1/(1 + \exp(\varepsilon/T))$ is the Fermi
distribution function and the matrix $\Lambda$ is defined as
\begin{equation}
  \Lambda_{ij}= \frac{\delta_{ij}}{N} \times \left\{
  \begin{array}{ll} N_2 & \mbox{for $j=1,\ldots,N_1$}, \\
  -N_1 & \mbox{for $j=N_1+1,\ldots,N$}, \end{array} \right.
  \label{eq:lambda}
\end{equation}
The introduction of
the matrix $\Lambda$ is a convenient way to
separate the classical conductance for ballistic contacts [the first
term in Eq.\ (\ref{landauer})] from the quantum
corrections and the effect of nonideal contacts [second term in
Eq.\ (\ref{landauer})].

The Landauer approach is an intrinsically single-particle picture, and
as such cannot capture any effects of electron-electron interaction by
itself. In a quantum dot, the most important part of the
interaction\cite{kn:aleiner1998a,kn:kurland2000,kn:aleiner2002}
is described by the Hamiltonian
\begin{equation}
  \hat H_{\mathrm{c}} = E_{\mathrm{c}}
  \left( \hat N_{\rm dot} - \cN \right)^2,
  \label{Hc}
\end{equation}
where the charging energy $E_{\rm c} = e^2/2C$ is related to the
geometric capacitance $C$ of the quantum dot, $\hat N_{\rm dot}$ is
the number of electrons on the quantum dot, and $\cN=CV_g/e$ is a
dimensionless gate voltage used to set the equilibrium value of
$N_{\rm dot}$.

The charging interaction (\ref{Hc}) is responsible for the phenomenon
of Coulomb blockade in closed dots at low temperatures: suppression of
the conductance, except if the gate voltage ${\cal N}$ is tuned to a
point of charge degeneracy. As a dot is opened, a situation best
realized in lateral semiconductor dots where point contacts may be
controlled electrostatically, strong quantum charge fluctuations lead
to the progressive diminishment of the Coulomb blockade. A theory of
this `weak Coulomb blockade' was developed by
Flensberg,\cite{kn:flensberg1993} Matveev,\cite{kn:matveev1995} and
Furusaki and Matveev,\cite{kn:furusaki1995a,kn:furusaki1995b} for
the `incoherent' case. For a quantum dot with two
nearly ideal single channel point contacts ($|r_{{\rm c}1}|,
|r_{{\rm c}2}| \ll 1$) and at temperatures $T \ll E_{\rm c}$,
the interactions were found to change
the conductance of the contacts according to $g_j \to g_j +
\delta g_j$, $j=1,2$, where
\begin{eqnarray}
  \delta g_j &=& - 2 \frac{\Gamma(3/4)}{\Gamma(1/4)}
  \sqrt{\frac{e^{\rm C} E_{\rm c}}{\pi T}}
  \,\tr r_{{\rm c}j}^{\vphantom{\dagger}} r_{{\rm c}j}^{\dagger},\ \
  j=1,2,
  \label{fm_res1}
\end{eqnarray}
up to corrections of order $|r_{\rm c}|^4$. Here ${\rm C} \approx
0.577$ is the Euler constant. As in the non-interacting case,
the conductance $G$ of the quantum dot
is simply the series conductance of the two point contacts,
\begin{eqnarray}\label{fm_res}
G &=&
  \frac{e^2}{2\pi\hbar}\left[ 1-\frac{\Gamma\left( {3}/{4}
    \right)}{\Gamma\left( {1}/{4}
    \right)}\sqrt{\frac{e^{\mathrm{C}}E_c}{\pi T}} \tr
    \rc \rcd
  \right].
\end{eqnarray}
The first term is the classical
conductance with $N_1=N_2=2$, accounting for spin degeneracy. The
second term shows the renormalization of the backscattering at the
contacts, see Eq.\ (\ref{fm_res1}). At low temperatures
the system enters into a non-perturbative regime. We refer to Refs.\
\onlinecite{kn:flensberg1993,kn:furusaki1995a,kn:furusaki1995b,kn:brouwer1999c}
for the expressions for the interaction corrections at different
numbers of channels in the point contacts.

A different approach to the same problem was taken by Golubev and
Zaikin\cite{kn:golubev2001}, and by Levy-Yeyati {\em et
al.}\cite{kn:levyyeyati2001} who described the interaction
Hamiltonian (\ref{Hc}) through an effective theory for the
electromagnetic environment of the junction. These authors also
calculated the interaction correction to the conductance of the
point contacts for an arbitrary transparency of the contacts using
$1/N$ as a small parameter. They found
\begin{eqnarray}
  \label{eq:gz1}
  \delta g_j &=& \lefteqn{- \frac{s_j g_j E_{\rm c}}{3 T},}
  \hphantom{- \frac{2 s_j g_j}{g} \ln \frac{E_{\rm c} g e^{1 + {\rm
  C}}}{2 \pi^2 T},\ } \ \, T \gg g E_{\rm c},  \\
  \delta g_j &=& - \frac{2 s_j g_j}{g} \ln \frac{E_{\rm c} g e^{1 + {\rm
  C}}}{2 \pi^2 T},\ \ T \ll g E_{\rm c},\ \ j=1,2, \nonumber
\end{eqnarray}
where $s_1$ and $s_2$ are
the so-called Fano factors of the contacts,
\begin{equation}
  s_j = \frac{1}{g_j} \tr r_{{\rm c}j}^{\vphantom{\dagger}}
  r_{{\rm c}j}^{\dagger} ( 1 - r_{{\rm c}j}^{\vphantom{\dagger}}
  r_{{\rm c}j}^{\dagger}),\ \ j=1,2.
  \label{eq:fano}
\end{equation}
The appearance of the Fano factors is quite common to corrections to
the conductance that are perturbative in the interaction, see Ref.\
\onlinecite{kn:matveev1993,kn:yue1994}. In fact, $s_i$ is the only
quadratic function of the product
$r_{{\rm c}i}^{\vphantom{\dagger}} r_{{\rm c}i}^{\dagger}$ that
vanishes both in the limits $r_{{\rm c}i} \to 0$ (ideal contacts)
and $r_{{\rm c}i} \to 1$ (completely closed contacts). Equation
(\ref{eq:gz1}) agrees with the large-$N$ generalization of Eq.\
(\ref{fm_res}) for nearly ideal
contacts.\cite{kn:flensberg1993,kn:brouwer1999c}

The results of Refs.\
\onlinecite{kn:flensberg1993,kn:matveev1995,kn:furusaki1995a,kn:furusaki1995b},
being non-perturbative in $E_{\rm c}$, are of conceptual importance,
but describe the incoherent case only. They neglect entirely the
possibility of coherent backscattering from within the dot, or
coherent transmission from one contact to another. Thus Eq.\
(\ref{fm_res}) misses quantum interference effects, such as weak
localization (the small negative magnetic-field dependent quantum
interference correction to the ensemble averaged conductance) and
conductance fluctuations, which one expects to be described by the
Landauer formula (\ref{landauer}) if the conductance of the
contacts is large enough. Interaction corrections to quantum
interference corrections cannot be captured by a simple
renormalization of the point contact conductances; one needs to
consider the total conductance $G$ of the quantum dot. Only at
temperatures for which relaxation processes in the dot are
dominant one expects the separate renormalizations of the point
contact conductances of Eqs.\ (\ref{fm_res1}) or (\ref{eq:gz1}) to
be a sufficient description.

The first results for the `coherent' case, concerning interaction
corrections to transport in open dots described by a unitary
scattering matrix, were obtained by Brouwer and
Aleiner~\cite{kn:brouwer1999c} (see also the review
Ref.~\onlinecite{kn:aleiner2002}).  They found $G = G_0 + G_{\rm BA}$
with
\begin{widetext}
\begin{eqnarray} \label{abg}
G_{\mathrm{BA}}&=&-\frac{e^2}{2\pi\hbar} \left(\frac{1}{\pi} \sin
{\frac{\pi}{ N}}\right) \int_{0}^{\infty} d \tau_1
  \int_0^{\infty} d \tau_2
\,\mathrm{tr}\, S^{\dagger}(-\tau_1) \Lambda S(\tau_2) \Lambda
\nonumber\\
  &&\mbox{} \times
\int_{t_0}^{\infty} d\sigma\,
\frac{(2 \sigma + \tau_2 + \tau_1)\pi^2 T^2}
 {\sinh[(\sigma+\tau_1)\pi T] \sinh[(\sigma+\tau_2)\pi T]}
   \left( \frac{
  \sinh[(\tau_1+\tau_2+\sigma+t_0) \pi T]
  \sinh[(\sigma - t_0) \pi T]
  }{
  \sinh[(  \tau_1 + t_0) \pi T]
  \sinh[(  \tau_2 + t_0) \pi T]
  } \right)^{1/N}.
\label{ABG_res}
\end{eqnarray}
\end{widetext}
Here $t_0=\pi/E_cNe^{\mathrm{C}}$ is a charge relaxation time
and $S(\tau)$ is the
Fourier transform of the scattering matrix $S(\varepsilon)$,
which is defined as
\begin{eqnarray}
  S(\tau) &=& \int \frac{d\varepsilon}{2\pi} S(\varepsilon) e^{-i
    \varepsilon \tau}, \\
  S^{\dagger}(\tau) &=&  \int \frac{d\varepsilon}{2\pi}
  S^{\dagger}(\varepsilon) e^{-i \varepsilon \tau}.
\end{eqnarray}
For a dot with two single-channel spin-polarized point contacts, $N_1
= N_2 = 1$, there is an additional interaction correction to the
conductance that depends explicitly on the gate voltage ${\cal N}$.
Equation (\ref{abg}) is obtained as a formal expansion in
the scattering matrix $S$. Such an expansion is controlled if $S$ is
subunitary, as is the case, {\em e.g.}, in an effective description of
inelastic processes in the dot.\cite{kn:brouwer1997c} Noting that only
times up to the `thermal time horizon' $\hbar/T$ contribute to the
interaction correction of Eq.\ (\ref{abg}), Brouwer and Aleiner argued
that their formal expansion is also justified for unitary $S$ if the
temperature is sufficiently high that the contribution from scattering
times $\lesssim \hbar/T$ is small. For a chaotic dot with mean dwell
time $\tau_{\rm d}$ and ideal point contacts, this translates to the
condition $T\gg
\hbar/\tau_{\rm d}$.

The result of Ref.\ \onlinecite{kn:brouwer1999c} captures the
Landauer part of the conductance exactly, whereas the interaction
correction (\ref{abg}) reproduces the singular behavior of Eq.\
(\ref{fm_res}) for the scattering matrix $S(\tau)=r_{\rm c}
\delta(\tau)+\ldots$, provided that the scattering from inside the
dot, represented by `$\ldots$', is from far beyond the `thermal
time horizon'. Using known statistical properties of the
scattering matrix for chaotic quantum dots, Brouwer and Aleiner
calculated the ensemble averages of the conductance and the
conductance fluctuations for $T \gg \hbar/\tau_{\rm d}$, and found
that both the weak localization correction to the conductance and
the conductance fluctuations are slightly enhanced by
interactions.

In this paper we question the arguments used to justify the formal
expansion in the scattering matrix used to obtain Eq.\
(\ref{abg}). As we will argue below, the interaction correction of
Eq.\ (\ref{abg}) is not the correct interaction correction to the
conductance of a coherent quantum dot at any temperature, despite
the fact that it is small if $T \gg \hbar/\tau_{\rm d}$. Our main
arguments will be given in the next sections, where we present two
calculations of the interaction correction to the conductance: The
first is a first-order-in-$E_{\rm c}$ calculation of the
conductance, which differs from Eq.\ (\ref{abg}) if that result is
expanded in $E_{\rm c}$. The second is a calculation to all orders
in $E_{\rm c}$, but to leading order in $1/N$. This calculation
confirms that Eq.\ (\ref{abg}) is found if the amplitude for
coherent scattering from the dot is small --- {\em i.e.},
transport through the dot is mainly incoherent (but not fully
incoherent)
 ---, but we find a different interaction
correction if transport is fully coherent.

The main
shortcoming of Eq.\ (\ref{abg}) can already be seen noting that, although
for $T \gg \hbar/\tau_{\rm d}$ most scattering processes have delay
time $\gg \hbar/T$ and thus do not contribute to the interaction
correction, the remaining interaction correction is dominated by
scattering processes with delay time $\lesssim \hbar/T$. Hence, it is
necessary that the theory describes this time range accurately; its
contribution being small is not sufficient to justify Eq.\
(\ref{abg}). That Eq.\
(\ref{abg}) does {\em not} describe delay times $\lesssim \hbar/T$
accurately can be seen by considering the example of a `dot'
that consists of a single ballistic channel in which there is an
interaction of the form (\ref{Hc}).
The channel is of length $L$, so the $2 \times 2$
scattering matrix describing
ballistic propagation from one end to the other is
\begin{equation}\label{ballistic}
S(\tau)=\begin{pmatrix}
0 & 1 \cr
1 & 0
\end{pmatrix}\delta(\tau-\tau_{\rm d}),
\end{equation}
where $\tau_{\rm d} = L/v_F$, $v_F$ being the Fermi velocity
inside the channel, is the time to propagate
through the channel. We assume $\tau_{\rm d} \gg t_0$. Then,
substituting Eq.\ (\ref{ballistic})
into (\ref{abg}) gives a conductance that is {\em larger} than
the conductance quantum $e^2/2 \pi \hbar$, $G = e^2/2 \pi \hbar
+ G_{\rm BA}$, with
\begin{eqnarray}
  G_{\rm BA} &=& \frac{e^2}{2\pi\hbar}
  \frac{\pi T \tau_{\rm d}}{\hbar} e^{-2\pi T \tau_{\rm d}/\hbar},
  \ \ T \gg \hbar/\tau_{\rm d}, \nonumber \\
  G_{\rm BA} &=& \lefteqn{\frac{e^2}{2 \pi \hbar}
  \frac{\pi \hbar}{8 T \tau_{\rm d}}, \ }
  \hphantom{\frac{e^2}{2\pi\hbar} \frac{\pi T \tau_{\rm d}}{\hbar}
  e^{-2\pi T \tau_{\rm d}/\hbar}, \ \ \, }
  T \ll \hbar/\tau_{\rm d}.
\end{eqnarray}
On the other hand it is well known that in the absence of
backscattering the conductance of such a system cannot deviate from
the quantized value $e^2/h$, see, {\em e.g.}, Refs.\
\onlinecite{kn:maslov1995,kn:ponomarenko1995,kn:safi1995}, so that
the interaction correction to the conductance must be zero for
all temperatures. Hence, although the error made by using Eq.\ (\ref{abg}) is
exponentially small for $\tau_{\rm d} \gg
\hbar/T$, it is substantial for $\tau_{\rm d} \lesssim \hbar/T$.

Very recently, Golubev and Zaikin extended their environmental
formalism to the case of a fully coherent dot.\cite{kn:golubev2004}
Again using $1/N$ as an
expansion parameter, they found $G = G_{0} + G_{\rm GZ}$, with
\begin{widetext}
\begin{eqnarray}
  \label{gz_res}
  G_{\rm GZ} &=&
  \frac{e^2}{2 \pi^2 \hbar} \mbox{Im}\,
  \int d\varepsilon d\omega
  \left( - \frac{\partial f(\varepsilon)}{\partial \varepsilon} \right)
  \kappa(\omega)
  \tr
  \left[\Lambda S(\varepsilon) \Lambda S^{\dagger}(\varepsilon + \omega)
  - \Lambda S(\varepsilon) S^{\dagger}(\varepsilon + \omega)
  S(\varepsilon) \Lambda S^{\dagger}(\varepsilon) \right]
  (1 - 2 f(\varepsilon + \omega)),
  \nonumber \\
\end{eqnarray}
\end{widetext}
where $\kappa(\omega)$ is an effective interaction kernel that
describes the fluctuations of the dot potential.
Since these fluctuations are small as $1/N$ if $N$ is large,
the correction (\ref{gz_res}) is a small correction to the Landauer
result $G_0$. Golubev and Zaikin performed the ensemble average over
$S$ and found that the leading interaction correction for the
coherent case is the same as in the incoherent limit ({\em i.e.},
the classical combination of the corrections Eq.~(\ref{eq:gz1})) as long as
$T \gg \hbar/\tau_{\rm d}$, and saturates when $T \lesssim
\hbar/\tau_{\rm d}$.
The agreement with the incoherent limit is no surprise,
since the leading-in-$N$ contributions to the conductance are
generally insensitive to the presence of quantum coherence.

The authors of Ref.\ \onlinecite{kn:golubev2004} did not analyze
the quantum interference corrections to the conductance and thus
attempted no comparison with the results of Refs.\
\onlinecite{kn:brouwer1999c,kn:aleiner2002}. Our full calculation, 
which is described in the next sections and which includes quantum
interference corrections,
agrees with Eq.\ (\ref{gz_res}) and extends it to the case of
general time dependent transport --- the focus of this paper.

Technically, our
formalism is very close to that of
Refs. \onlinecite{kn:brouwer1999c,kn:aleiner2002}. Within the same
formalism, we can obtain Eq.\ (\ref{abg}) using a formal expansion in
the scattering matrix $S$ and Eq.\ (\ref{gz_res}) as a formal
expansion in $1/N$. As we discussed above, expansion in powers of $S$
requires a relaxation mechanism, and Eq.\ (\ref{abg}) can be justified
only if relaxation is strong, so that, effectively, $|S(\epsilon)| \ll 1$.

\section{Formalism}
\label{sec:formalism}

\subsection{Definition of the problem} \label{sec:model}

The system under consideration --- a quantum dot coupled to source and
drain reservoirs via point contacts --- has been introduced at the
beginning of Sec.\ \ref{sec:previous}. There are two point contacts,
with $N_1$ and $N_2$ channels each (including spin degeneracy).
Electrons on the dot interact via the simple interaction Hamiltonian
(\ref{Hc}). We refer to Refs.\
\onlinecite{kn:aleiner1998a,kn:kurland2000,kn:aleiner2002} for a
microscopic justification of Eq.\ (\ref{Hc}) for quantum dots with a
large (internal) dimensionless conductance. In using Eq.\ (\ref{Hc})
we ignore those parts of the interaction describing superconducting
pairing and exchange.\cite{kn:aleiner2002}

\begin{figure}
\epsfxsize=0.7\hsize
\epsffile{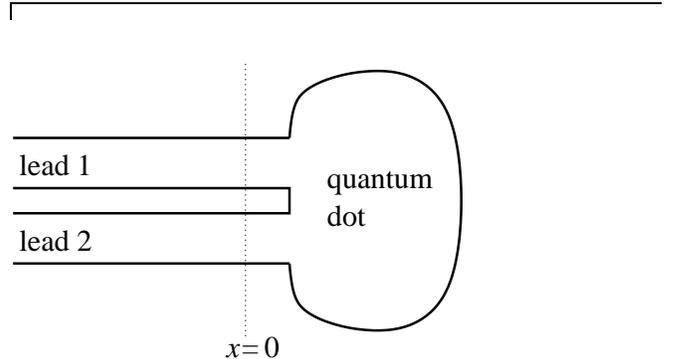}
\caption{\label{fig:2}
Representation of the source and drain reservoirs as one-dimensional
ideal leads. The lead-dot interface is at $x=0$. The one-dimensional
description (\protect\ref{eq:pc1d}) is valid from $x=-\infty$ up to
slightly beyond $x=0$.}
\end{figure}

Following Flensberg\cite{kn:flensberg1993} and
Matveev\cite{kn:matveev1995} we describe the electron dynamics in
the point contacts by a one-dimensional Hamiltonian. We locate the
lead-dot interface at $x=0$, taking dot and lead at $x>0$ and
$x<0$ respectively, see Fig.\ \ref{fig:2}.
Since there is no backscattering of electrons that
have left the dot through the point contacts and into the reservoirs
back into the dot, we may represent the reservoirs by extending the
one-dimensional description to all $x < 0$,
\begin{eqnarray}
  \label{eq:pc1d}
  \hat H &=&
  i v_F : \! \sum_{j=1}^{N} \int_{-\infty}^0 dx
  \left[ \hat \psi_{jL}^{\dagger}(x)
  \frac{\partial}{\partial x} \hat \psi_{jL}^{\vphantom{\dagger}}(x)
  \right. \\ && \ \ \left. \mbox{}
  - \hat \psi_{jR}^{\dagger}(x)
  \frac{\partial}{\partial x} \hat \psi_{jR}^{\vphantom{\dagger}}(x)
  \right] \! :
  +
  E_{\rm c} \left( \hat N_{\rm dot} - \cN \right)^2,
  \nonumber
\end{eqnarray}
Here the index $j$ labels the channels in the two point contacts,
including spin, $N$ is the total number of channels in the two
point contacts, $\hat \psi_{jL}$ and $\hat \psi_{jR}$ are annihilation
operators for left-moving and right-moving electrons in channel $j$,
respectively. Because of scattering by the dot, these are not
independent fields. 
In the one-dimensional Hamiltonian (\ref{eq:pc1d}), the kinetic
energy is linearized around the Fermi energy, which is appropriate
if the Fermi energy is not too close to a threshold at which a 
new channel in the point contact is opened.
The coordinate $x$ in the one-dimensional
Hamiltonian (\ref{eq:pc1d}) arises as the Fourier transform of the
one-dimensional Hamiltonian in momentum representation, see Ref.\
\onlinecite{kn:matveev1995}. Except for the immediate vicinity of 
the point contact, it cannot be identified with a distance to the
lead-dot interface.

It is convenient to follow Aleiner and
Glazman,\cite{kn:aleiner1998a} and write  the number of electrons
on the dot as
\begin{eqnarray}
  \label{eq:NdotNref}
  \hat N_{\rm dot} &=&
  N_{\rm ref}
  - \sum_{j}
  : \! \int_{-\infty}^{0} dx
  \left[\hat \psi_{jL}^{\dagger}(x)
  \hat \psi_{jL}^{\vphantom{\dagger}}(x)
  \nonumber \right. \\ && \left. \mbox{}
  + \hat \psi_{jR}^{\dagger}(x)
  \hat \psi_{jR}^{\vphantom{\dagger}}(x) \right]\! :.
\end{eqnarray}
Here $N_{\rm ref}$ is the total number of electrons in the system
(reservoirs and dot), which is time independent. Now the
interaction does not contain the dot degrees of freedom. Thus the
relation between the operators for right moving electrons entering
the quantum dot and left moving electrons exiting the quantum dot
is given by the noninteracting formula. In the general
time-dependent case one then has\cite{kn:polianski2003}
\begin{eqnarray}
  \hat \psi_{jL}^{\vphantom{\dagger}}(0,t) &=&
  \sum_{k=1}^{N} \int d\tau
  S_{jk}(t,t-\tau) \hat \psi_{kR}^{\vphantom{\dagger}}(0,t-\tau),
  \nonumber \\
  \hat \psi_{jL}^{{\dagger}}(0,t) &=&
  \sum_{k=1}^{N} \int d\tau
  \hat \psi_{kR}^{{\dagger}}(0,t-\tau)
  (S^{\dagger})_{kj}(t-\tau,t),
  \nonumber \\ \label{eq:psiSpsi}
\end{eqnarray}
where $S(t,t')$ is the scattering matrix of the dot.
Notice that the scattering matrix has two
time arguments. For time-independent transport, $S(t,t')$ is a
function of the time difference $t-t'$ only.

The relation (\ref{eq:psiSpsi}), together with the one-dimensional
Hamiltonian (\ref{eq:pc1d}), Eq.\ (\ref{eq:NdotNref}) for $\hat N_{\rm
dot}$, and the boundary conditions set by the reservoirs completely
define the problem. Notice that we never explicitly introduce the dot
degrees of freedom into the problem. If the scattering from the dot
is not fully coherent, Eq.\ (\ref{eq:psiSpsi}) has to be replaced by a
different boundary condition. We return to the case of
incoherent scattering in section~\ref{sec:relax}. A model for
partially coherent scattering is discussed in appendix \ref{app:c}.

\subsection{First order in $E_{\rm c}$}
\label{sec:fock}

It was shown by Matveev, Yue, and
Glazman\cite{kn:matveev1993,kn:yue1994} that, at first order in the
interaction, interaction corrections to conductance in the presence of
impurities may be understood in terms of the elastic scattering of
electrons from the Hartree-Fock potential created by the
impurities. This gave a simple and intuitive picture of the
renormalization of backscattering in the interacting one-dimensional
electron gas, without recourse to sophisticated formalism. We now
outline a similar calculation for transport through a quantum dot.

For linear transport, it is sufficient to consider the scattering off
the Hartree-Fock potential in equilibrium. With the replacement
(\ref{eq:NdotNref}) the Hartree-Fock potential acts in the leads, not
in the dot. In equilibrium there is no
Hartree potential as $\langle \hat N_{\mathrm{dot}} \rangle=\cN$. For
the Fock potential we find
\begin{equation}\label{fock}
  \hat H_{\rm F} =  \sum_{i,j=1}^N\int_{-\infty}^0 dx dy
  \left[ \hat \psi_{iL}^{\dagger}(x)V_{ij}(x,y)
  \hat \psi_{jR}^{\vphantom{\dagger}}(y)
  + \mbox{h.c.} \right],
\end{equation}
where ``h.c.'' denotes the hermitian conjugate,
plus forward scattering terms that do not affect the linear
conductance at lowest order in $E_{\rm c}$. The Fock potentials are
written in terms of the density matrices, which may be evaluated using
the boundary conditions (\ref{eq:psiSpsi})
\begin{eqnarray} \label{Fock}
V_{ij}(x,y) &=&-2E_{\mathrm{c}}\langle \hat \psi_{jR}^{\dagger}(y)\hat \psi_{iL}^{\vphantom{\dagger}}(x)  \rangle
  \nonumber \\ &=&
-2E_{\mathrm{c}}\int_{0}^{\infty}d\tau\, S_{ij}(\tau)
  \nonumber \\ && \mbox{} \times
  \frac{iT}{2\sinh\left[ \pi T\left(\tau-v_F[x+y]+i \lambda \right)
  \right]},
\end{eqnarray}
where $\lambda$ is a positive infinitesimal. Thus the effect of
the Coulomb interaction is to establish a non-local Fock potential
in the leads within a region of order the thermal length $v_F/T$
from the contacts. It is now straightforward to find the change in
the scattering matrix due to this potential
\begin{eqnarray}\label{deltaS}
\delta S(\epsilon) &=&
  i\int \frac{d\omega}{2\pi}\kappa(\omega)\left[
  (2 f(\epsilon-\omega) - 1) S(\epsilon-\omega)
  \right. \nonumber\\ &&  \left. \mbox{}
  + (2 f(\epsilon+\omega) -
  1)S(\epsilon)S^{\dagger}(\epsilon+\omega)S(\epsilon) \right],\ \ \ \
\end{eqnarray}
where
\begin{equation}
  \label{eq:kappaEc}
  \kappa(\omega)=-\frac{E_c}{(\omega-i 0^+)^2},
\end{equation}
$0^+$ being a positive infinitesimal.
The first term in (\ref{deltaS}) represents the effect of direct $R\to
L$ (in to out) scattering by the Fock potential (\ref{fock}). The
second corresponds to the dot scattering $R\to L$, followed by the
Fock potential scattering $L\to R$, and ending with the dot scattering
$R\to L$ once more. Substitution of (\ref{deltaS}) into the Landauer
formula (\ref{landauer}) immediate yields the result (\ref{gz_res})
for the interaction correction to conductance, with $\kappa(\omega)$
given by Eq.\ (\ref{eq:kappaEc}) above.

We may repeat the derivation for the incoherent case that is
considered in Refs.\
\onlinecite{kn:flensberg1993,kn:matveev1995,kn:furusaki1995a,kn:furusaki1995b,kn:golubev2001,kn:levyyeyati2001}.
In this case, the inclusion of the Fock potential leads to a
change of the reflection matrix $r_{\rm c}$ for direct reflection
at the contacts. Since electrons leaving the dot are incoherent
with those entering it, they do not contribute to the Fock
potential (\ref{Fock}). The renormalized reflection matrix is thus
\begin{eqnarray}\label{deltaSrc}
\delta  r_{{\rm c}}(\epsilon) &=&
  i\int \frac{d\omega}{2\pi}\kappa(\omega)\left[
  (2 f(\epsilon-\omega) - 1)  r_{{\rm c}}
  \right. \nonumber\\ &&  \left. \mbox{}
  + (2 f(\epsilon+\omega) -
  1) r_{{\rm c}} r_{{\rm c}}^{\dagger} r_{{\rm c}} \right],\ \ \ \
\end{eqnarray}
Taking into account the energy dependence of the renormalized
reflection matrix $r_{\rm c}$, we find that the interaction
correction to the dimensionless conductance of contact $j$, $j=1,2$,
becomes
\begin{eqnarray} \label{junction_res}
  \delta g_j
  &=& 2s_jg_j
  \mathrm{Im}\int d\varepsilon\int \frac{d\omega}{2\pi}
  \\ && \nonumber \mbox{} \times
  \left( - \frac{\partial f(\varepsilon)}{\partial
  \varepsilon} \right) \left( 1-2f(\varepsilon+\omega) \right)\kappa(\omega),
\end{eqnarray}
where $s_j$ was defined in Eq.\ (\ref{eq:fano}).
Although this looks very similar to (\ref{gz_res}) with the
scattering matrix
chosen as $S(\varepsilon) = r_{\rm c}$, the effect of using a non-unitary
scattering matrix is significant. Notice that for a unitary scattering
matrix, the trace in Eq.\ (\ref{gz_res}) vanishes as $\omega^2$, so
that the regulator in (\ref{eq:kappaEc}) plays no role. In
(\ref{junction_res}) on the other hand, the imaginary part of the
integrand comes from the region near $\omega=0$ and yields a finite,
regulator independent contribution to the conductance

\begin{equation}
\delta g_j=-s_jg_j\frac{E_c}{3T},\ \ j=1,2,
\end{equation}
which coincides with the high temperature limit of Eq.\
(\ref{eq:gz1}).\cite{kn:golubev2001} Recovering this same
result, together with the low-temperature limit of (\ref{eq:gz1}),
as the zero level spacing limit of scattering
from a coherent dot is subtle, see Sec.\ \ref{sec:steadystate}
and Ref. \onlinecite{kn:golubev2004}.

With this physical picture of the interaction corrections, we now proceed with the formal development of the theory.

\subsection{Effective action}
\label{sec:fictitiouslead}

We are interested in the current through the point contacts, at
the dot-lead interface $x=0$. The current in channel $j$ is
\begin{equation}
  \hat I_j = - e v_F \left[ \hat\psi_{jR}^{{\dagger}}(0)
  \hat \psi_{jR}^{\vphantom{\dagger}}(0) -
  \hat\psi_{jL}^{{\dagger}}(0) \hat \psi_{jL}^{\vphantom{\dagger}}(0)
  \right].
  \label{eq:Ij}
\end{equation}

Below, the expectation value of the current is calculated using standard
methods of nonequilibrium many-body theory.\cite{kn:rammer1986}
The current $\hat I(t)$ is related to the state of the
system at a reference time $t_{\rm ref}$ at which the Hamiltonian is
not time-dependent. The expectation value $I(t)$ is then found
from a thermodynamical average at the time $t_{\rm ref}$. An
important issue here is that the total number of electrons,
$N_{\rm ref}$, be kept constant --- especially when observables
at two different values of the gate voltage ${\cal N}$ are
compared.\cite{kn:aleiner2002} In Ref.\ \onlinecite{kn:aleiner2002}
this was implemented by taking this thermodynamical average in a
canonical ensemble. In our case, however, we can
take the average in a grand canonical ensemble, since, in the
non-equilibrium formalism, it is sufficient that the dynamics
conserves the number of electrons.

The one-dimensional dynamics of the Hamiltonian (\ref{eq:pc1d})
applies to the region $x < 0$ only; the dynamics in the quantum dot
is described by the boundary condition (\ref{eq:psiSpsi}). This
boundary condition is difficult to implement in the calculations,
and it is advantageous to reformulate the problem in terms of a
one dimensional Hamiltonian of the form (\ref{eq:pc1d}), where the
$x$ integration extends over the entire real axis. In this
formulation of the problem, the boundary condition (\ref{eq:psiSpsi}),
or its equivalent for a dot in which scattering is not fully
coherent, is replaced by an effective action $\cS$ that involves the
fermion operators at $x = \eta$, where $\eta$ is a positive
infinitesimal.
Such a formulation of the problem has first been given by Aleiner
and Glazman,\cite{kn:aleiner1998a} see also Ref.\
\onlinecite{kn:aleiner2002}. Here we will use an equivalent but
simpler effective action formulation.

\begin{figure}
\epsfxsize=0.8\hsize
\epsffile{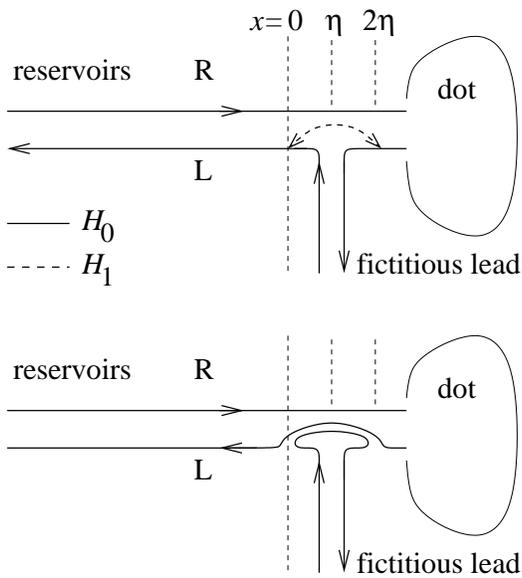}
\caption{\label{fig:3}
Inclusion of a fictitious lead at the point contact. Top panel:
With the Hamiltonian $H_0$, the fictitious lead couples ideally to
outgoing electrons (solid lines), so that all electrons that exit
the dot enter into the fictitious lead. The Hamiltonian $H_1$
describes scattering of electrons exiting the dot into the
physical lead and backscattering of electrons coming from the
fictitious lead (dotted line). The amplitude for this scattering
process in channel $j$ is $i r_j$, $j=1,\ldots,N$. Bottom panel:
If $r_j = 1$, the fictitious lead is decoupled again and the original
problem is restored.}
\end{figure}

In order to derive the effective action, a fictitious half-infinite
$N$-channel lead is side-coupled to the point contacts, at position
$x = \eta$, where $\eta$ is a positive infinitesimal.
The fictitious lead couples only to electrons moving out
of the dot, {\em i.e.}, to left movers. The total Hamiltonian
$\hat H$ is written as
\begin{equation}
  \hat H = \hat H_0 + \hat H_1,
\end{equation}
where $\hat H_0$ is the Hamiltonian of the full interacting system
(including the quantum dot, the reservoirs, and the fictitious lead),
with ideal coupling between the fictitious lead and the point contact,
see Fig.\ \ref{fig:3} (solid lines).
Hence, in the system described by Hamiltonian $\hat H_0$, all
electrons entering the point contact from the dot will exit through
the fictitious lead, whereas all electrons entering from the fictitious
lead will exit towards the source or drain reservoirs.
The Hamiltonian
$\hat H_1$ represents an impurity in the contact to the fictitious
lead which ensures that electrons coming from the dot in channel $j$
have amplitude $-i r_j$ to be transmitted coherently
towards the source or drain reservoirs (cf.\ dashed lines in Fig.\
\ref{fig:3}) and that electrons coming in from the fictitious lead have
amplitude $-i r_j$ to be reflected into the fictitious lead,
\begin{equation}
  \hat H_1 =
  \sum_{j=1}^{N} 2 v_F v_j \left[ \hat \psi_{jL}^{\dagger}(0)
  \hat \psi_{jL}^{\vphantom{\dagger}}(2 \eta) + \hat
  \psi_{jL}^{\dagger}(2 \eta)
  \hat \psi_{jL}^{\vphantom{\dagger}}(0) \right],
  \label{eq:Himp}
\end{equation}
where
\begin{equation}
  r_j = \frac{2 v_j}{1 + v_j^2},
  \ \ j=1,\ldots,N.
  \label{eq:wr}
\end{equation}
The parameters $r_j$ (or $v_j$) describe how well the fictitious
lead is coupled to the point contact. In the limit $r_j \to 1$,
$j=1,\ldots,N$, the fictitious lead is fully decoupled. We will
take the limit $r_j \to 1$ at the end of the calculation. In
writing Eq.\ (\ref{eq:Himp}), the same regularization of fermion
operators has been employed as in Ref.\
\onlinecite{kn:aleiner2002}.

The operator identity (\ref{eq:psiSpsi}) applies to left-moving
fermions at the moment when they exit the dot. With the inclusion
of the fictitious lead, this means that one should take $\psi_L$
at $x = 2 \eta$, not at $x=0$. Hence Eq.\ (\ref{eq:psiSpsi}) gives
a relation between $\hat \psi_L(2 \eta)$ and $\hat \psi_R(0)$,
\begin{eqnarray}
  \hat \psi_{jL}^{\vphantom{\dagger}}(2 \eta,t) &=&
  \sum_{k=1}^{N} \int d\tau
  S_{jk}(t,t-\tau) \hat \psi_{kR}^{\vphantom{\dagger}}(0,t-\tau).
  \nonumber \\
  \label{eq:psiESpsiR}
\end{eqnarray}

\begin{figure}
\epsfxsize=0.7\hsize
\epsffile{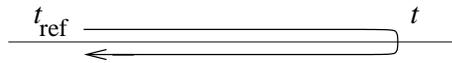}
\caption{\label{fig:4}
Integration contour for Eq.\ (\protect\ref{eq:Ipert2}).}
\end{figure}

The advantage of separating the Hamiltonian into the contributions
$\hat H_0$ and $\hat H_1$ is that the problem
described by the Hamiltonian $H_0$ alone is exactly solvable, see
Sec.\ \ref{sec:correlations}. The effect of $\hat H_1$ is then
treated in perturbation theory. Using Eq.\ (\ref{eq:psiESpsiR}),
the expectation value of the current at
time $t$ can be expressed
in terms of an effective action ${\cal S}$,
\begin{equation}
  I_j(t) =
  \left\langle T_{\rm c} e^{-i {\cal S}}
  \hat I_j(t) \right\rangle_0,
  \label{eq:Ipert2}
\end{equation}
where the current operator $\hat I_j$ is given by Eq.\ (\ref{eq:Ij})
above, ``c'' denotes the Keldysh contour, see Fig.\ \ref{fig:4},
$T_{\rm c}$ denotes contour ordering along the Keldysh contour,
the time-dependence of the operators is that of the interaction
picture with respect to $\hat H_0$, and the average $\langle \ldots
\rangle_0$ denotes an average with respect to $H_0$ at reference
time $t_{\rm ref}$. The effective action ${\cal S}$ is
\begin{eqnarray}
  {\cal S} &=&
  \label{eq:SeffA}
  \int_{\rm c} dt_1 \hat H_1(t_1) \\ &=&
  2 v_F \int_{\rm c} dt_1 \int d \tau
  \sum_{j=1}^{N} \sum_{k=1}^{N} v_j
  \nonumber \\ && \mbox{} \times
  \left[\psi^{\dagger}_{jL}(0,t_1)
  S_{jk}(t_1,t_1-\tau) \psi_{kR}(0,t_1-\tau)
  \right. \nonumber \\ && \ \ \left. \mbox{} +
  \psi^{\dagger}_{kR}(0,t_1-\tau) (S^{\dagger})_{kj}(t_1-\tau,t_1)
  \psi_{jL}(0,t_1) \right],\nonumber
\end{eqnarray}
where we dropped reference to the infinitesimal $\eta$.
It is important to note that in this effective action, the
contour-ordering occurs according to the `contour time' $t_1$,
not according to the `scattering delay time' $\tau$.

Both the current operator (\ref{eq:Ij}) and the effective
action (\ref{eq:SeffA}) do not contain any reference to the
electrons after they have passed through the point contact,
so that one may send the upper integration boundary in Eq.\
(\ref{eq:pc1d}) to infinity and set
\begin{eqnarray}
  \label{eq:pc1d2}
  \hat H_0 &=&  i v_F : \! \sum_{j=1}^{N} \int_{-\infty}^{\infty} dx
  \left[ \hat \psi_{jL}^{\dagger}(x)
  \frac{\partial}{\partial x} \hat \psi_{jL}^{\vphantom{\dagger}}(x)
  \right. \\ && \left. \mbox{}
  - \hat \psi_{jR}^{\dagger}(x)
  \frac{\partial}{\partial x} \hat \psi_{jR}^{\vphantom{\dagger}}(x)
  \right] \! :
  +
  E_{\rm c} \left( \hat N_{\rm dot} - {\cal N} \right)^2,
  \nonumber
\end{eqnarray}
where, as before, $\hat N_{\rm dot}$ is expressed in terms of the
fermion fields for $x < 0$ only, see Eq.\ (\ref{eq:NdotNref}). In the
system described by the Hamiltonian $\hat H_0$, the fermion
fields $\psi_R$ have chemical potentials $\mu_R$ corresponding to
the chemical potentials in the reservoirs, whereas the fermion
fields $\psi_L$ have the chemical potential $\mu_L$
of the fictitious lead. For the physically relevant case $r_j \to
1$, $j=1,\ldots,N$, the fictitious lead is fully decoupled from
the point contact. In that limit, all physical observables
become independent of the choice of $\mu_L$.

The effective action (\ref{eq:SeffA}) is different from the
effective action used in Refs.\
\onlinecite{kn:aleiner1998a,kn:brouwer1999c,kn:aleiner2002}: the
kernel in the latter action is a matrix $L$, which is a function
of the scattering matrix $S$ of the quantum dot.

\subsection{Relaxation inside quantum dot} \label{sec:relax}

If the fictitious lead used for the above derivation of the
effective action is fully coupled to the point contact ({\em
i.e.,}, if $r_j = 0$, $j=1,\ldots,N$), every electron exiting the
quantum dot will escape towards the fictitious lead, not towards
the (physical) reservoirs. Setting the chemical potential $\mu_L$
of the fictitious lead such that no net current is drawn, one finds
that each electron that escapes into the fictitious lead is
replaced by another one --- without phase
relationship.\cite{kn:buettiker1986,kn:buettiker1988} Hence,
coupling to the fictitious lead allows for complete phase and
energy relaxation of electrons exiting the dot, just before they
pass through the point contact to the reservoirs.

The simplest model for incomplete relaxation in the quantum dot is to
use the coupling to the fictitious lead with $0 < r_j < 1$. In this
case, the electrons are allowed to
escape into the fictitious lead and relax, but with finite probability
only. Hence imperfect coupling to the fictitious lead describes a
quantum dot with a finite probability that the electron exits the dot
without relaxation. Of course, this model is over-simplified, because
all relaxation is localized at the point contact. A more realistic,
but still phenomenological model for relaxation in the quantum dot
is discussed in appendix \ref{app:c}.

A connection can be made between the effective action (\ref{eq:SeffA})
for general $r_j$ and the 'fully incoherent' limit of the problem,
with nonideal contacts.\cite{kn:flensberg1993,kn:matveev1995,kn:furusaki1995a,kn:furusaki1995b,kn:levyyeyati2001,kn:golubev2001} Hereto, the
fictitious lead is interpreted as the ``quantum dot', the
scattering matrix $S_{ij}(t,t')$ is replaced by the identity matrix,
$S_{ij}(t,t') = \delta_{ij} \delta(t-t')$, and one identifies the
reflection matrix $r$ with the contact's reflection matrix $r_{\rm
  c}$. In this case, energy and
phase relaxation inside the dot is still complete, but a nonzero
value of the reflection probabilities $r_j^2$ allows one to describe
a nonideal coupling between dot and reservoirs.

\subsection{Correlation functions}
\label{sec:correlations}

In order to formulate a perturbation theory in the effective
action (\ref{eq:SeffA}), we need the contour-ordered fermion
correlation functions of the Hamiltonian $\hat H_0$. The
equilibrium correlators were calculated using the bosonization method
in Ref.\ \onlinecite{kn:aleiner2002}. The new elements in the nonequilibrium case are few. We discuss them now in physical terms, with the details being relegated to appendix \ref{app:boson}.

We specify the non-equilibrium conditions applied to the system through the chemical potentials $\mu_{jL}(t+x/v_F)$ and $\mu_{jR}(t-x/v_F)$  of the left and right moving electrons. Although the chemical potentials of the fictitious leads should not appear in physical quantities in the $r_j \to 1$ limit, we retain them in order to illustrate how this happens.

In the absence of scattering, i.e. setting $\hat{H}_1=0$, the
Hamiltonian is quadratic and a mean-field treatment is thus exact.
The average density of left moving electrons at the contact is
simply related to their chemical potential
\begin{equation} \label{rhoL_rel}
\langle\hat\rho_{jL}(0,t)\rangle=\frac{1}{2\pi v_F}\mu_{jL}(t),
\end{equation}
For right movers, however, the presence of the charging interaction in
the lead means that the `{\em electrochemical potential}' $\mu_{\rm
  c}$ is the relevant quantity
\begin{eqnarray} \label{rhoR_rel}
\langle\hat\rho_{jR}(0,t)\rangle &=&\frac{1}{2\pi v_F}
  \mu_{{\rm c}j}(t), \\
  \mu_{{\rm c}j}(t) &=& \mu_{jR}(t)+eV_{\mathrm{d}}(t),
  \label{eq:muc}
\end{eqnarray}
where $V_{\mathrm{d}}(t)$ was defined in (\ref{sc_pot}). The average current at the contact is
\begin{eqnarray} \label{current}I_{j}(t)&=&-ev_F\left(\langle \hat\rho_{jR}(0,t)\rangle-\langle\hat\rho_{j L}(0,t) \rangle\right)\\
&=&-\frac{e}{2\pi}\left[ \mu_{jR}-\mu_{jL}-2E_c\left( \langle\hat N_{\rm dot}(t)\rangle - {\cal N} \right)\right]\nonumber.\end{eqnarray}
Thus with an excess charge in the dot $\langle\hat N_{\rm dot}\rangle\neq {\cal N}$, there is a contribution from the interaction. We solve (\ref{rhoR_rel}) by writing the excess charge as an integral of the current
\begin{eqnarray} \label{rho_eqn}
\langle\hat\rho_{jR}(t)\rangle&=&\frac{1}{2\pi v_F}\mu_{jR}(t)\\
&&-\frac{E_c}{\pi}\sum_{k=1}^N\int^t dt' \langle \hat\rho_{kR}(t')\rangle-\langle\hat\rho_{k L}(t') \rangle\nonumber,
\end{eqnarray}
where we have introduced the convention that fields without a position label are taken to be evaluated at the origin. The solution of (\ref{rho_eqn}) is
\begin{widetext}
\begin{eqnarray} \label{rho_sol}\langle\hat\rho_{jR}(t)\rangle=\frac{1}{2\pi v_F}\left[\mu_{jL}(t)+\sum_{k=1}^N\int^{\infty}_0 d\tau i_{jk}(\tau)\left( \mu_{k R}(t-\tau)-\mu_{kL}(t-\tau) \right) \right],\end{eqnarray}
\end{widetext}
where we defined the kernel
\begin{equation}
  i_{mk}(\tau) = \delta_{mk} \delta(\tau-0^+) -
  \frac{E_{\rm c}}{\pi} e^{-E_{\rm c} N
  \tau/\pi} \theta(\tau),
    \label{eq:i}
\end{equation}
$0^+$ being a positive infinitesimal. The kernel $i_{jk}(\tau)$, which
appears frequently in the following development, relates the current
at the contacts to the chemical potentials in the leads, accounting
properly for the effect of the charging interaction. Note the
appearance of the $RC$ time $\pi/E_cN$ characteristic of an ideal
$N$-channel contact.

Combining the above results, we find that the current in channel $j$ is given by
\begin{eqnarray}
  I_j(t)&=&-\frac{e}{2\pi}\sum_{k=1}^N\int^{\infty}_0 d\tau
  i_{jk}(\tau)
  \nonumber \\ && \mbox{} \times
  \left[ \mu_{k R}(t-\tau)-\mu_{k L}(t-\tau) \right].
  \label{eq:currentzerosimple}
\end{eqnarray}
%
Note that in the dc
bias situation Eq.\ (\ref{eq:currentzerosimple}) simplifies to
\begin{equation} \label{dc_current}
I_j=-\frac{e}{2\pi}\sum_{k=1}^N\left( \delta_{jk}-\frac{1}{N} \right)\left( \mu_{k R}-\mu_{kL}\right),
\end{equation}
so that for the two-terminal set up
\begin{equation}
  I_j=-\frac{e}{2\pi}\Lambda_{jj}\left( \Delta\mu_1-\Delta\mu_2
  \right),
\end{equation}
where $\Delta\mu_{1,2}=\mu_{ R 1,2}-\mu_{L 1,2}$ are the chemical potential differences in the two contacts, and $\Lambda_{ij}$ is the matrix defined in Eq.\ (\ref{eq:lambda}).

Correlators of many fermion operators will be expressed in
terms of the single-fermion Green functions,
\begin{eqnarray}
  -i \langle T_{\rm c} \hat \psi_{mL}^{\vphantom{\dagger}}(s)
  \hat \psi^{\dagger}_{nL}(t) \rangle
  &=& G_{mnL}(s,t)  \\
  -i \langle T_{\rm c} \hat \psi_{mR}^{\vphantom{\dagger}}(t)
  \hat \psi^{\dagger}_{nR}(s) \rangle
  &=& G_{mnR}(t,s) .
\end{eqnarray}
These read
\begin{eqnarray} \label{GF}
   G_{mnL}(t,s) &=&
  -\frac{\delta_{mn} T e^{-i (\phi_{mL}(t) - \phi_{mL}(s))}}{2 v_F \sinh[\pi T(t - s
  - i \lambda \sign_{\rm c}(t-s))]}, \nonumber \\
  G_{mnR}(t,s) &=&
  -\frac{\delta_{mn} T e^{-i (\phi_{mR}(t) - \phi_{mR}(s))}}{2 v_F \sinh[\pi T(t - s
  - i \lambda \sign_{\rm c}(t-s))]}, \nonumber \\
\end{eqnarray}
where $\lambda$ is a positive infinitesimal and
\begin{widetext}
\begin{eqnarray}
  \label{eq:thetamu}
  \phi_{mL}(t) &=&
  \int_0^{\infty} d\tau \mu_{mL}(t-\tau),  \\
  \phi_{mR}(t) &=& \int_0^{\infty} d\tau \left[ \mu_{mL}(t-\tau)
  \vphantom{\sum_k^N}
  +
  \sum_{k=1}^{N} \int_{0}^{\infty} d\tau' i_{mk}(\tau')
  (\mu_{kR}(t-\tau-\tau') -
  \mu_{kL}(t-\tau-\tau')) \right], \nonumber
\end{eqnarray}
are the integrals of the electrochemical potentials
of right and left movers. It is straightforward to see that (\ref{GF}) is consistent with (\ref{rhoL_rel}) and (\ref{rho_sol}).
The single-fermion Green function can be written as a sum of
a Keldysh part, which does not depend on the contour positions,
and a delta function,
\begin{eqnarray}
  G_{mnL}(t,s) &=&
  \frac{1}{2} G^{\rm K}_{mnL}(t,s) - \frac{i}{2 v_F}\delta_{mn}
  \delta_{\rm c}(t,s),
  \nonumber \\
  G_{mnR}(t,s) &=&
  \frac{1}{2} G^{\rm K}_{mnR}(t,s) - \frac{i}{2 v_F}
 \delta_{mn} \delta_{\rm c}(t,s).
  \label{eq:GKdelta}
\end{eqnarray}
Here we abbreviated
\begin{eqnarray}
  \delta_{\rm c}(s,t) &=& \sign_{\rm c}(s-t) \delta(s-t).
  \label{eq:deltac}
\end{eqnarray}

Now the contour-ordered average of a product of fermion operators
is calculated as
\begin{eqnarray}
  \lefteqn{(-i)^{n+m}
  \left\langle T_{\rm c}
  \hat \psi_{l_1L}^{\vphantom{\dagger}}(s_1)
  \hat \psi_{k_1L}^{\dagger}(t_1)
  \ldots
  \hat \psi_{l_nL}^{\vphantom{\dagger}}(s_n)
  \hat \psi_{k_nL}^{\dagger}(t_n)
  \hat \psi_{k_1'R}^{\vphantom{\dagger}}(t_1')
  \hat \psi_{l_1'R}^{\dagger}(s_1')
  \ldots
  \hat \psi_{k_m'R}^{\vphantom{\dagger}}(t_m')
  \hat \psi_{l_m'R}^{\dagger}(s_m')
  \right\rangle } \nonumber \\
  &=&
  \sum_{P,Q} (-1)^{P+Q}
  \prod_{i=1}^{n} G_{l_i k_{P(i)}L}(s_i,t_{P(i)})
  \prod_{j=1}^{m} G_{k_j'l_{Q(j)}'R}(t_j',s_{Q(j)}')
  \prod_{i,j}
  \left(\frac{f(t_i-t_j') f(s_i-s_j')}{f(t_i-s_j') f(s_i-t_j')}
  \right)^{1/N},\ \ \
  \nonumber \\
  \label{eq:fermion}
\end{eqnarray}
where $P$ and $Q$ are permutations of the numbers $i=1,\ldots,n$
and $j=1,\ldots,m$, respectively, and the function $f(t-t')$ is
\begin{eqnarray}
  \ln f(t-t') &=&
  \ln \frac{\lambda E_{\rm C} N e^{C}}{\pi}
  - \frac{E_{\rm C} N}{\pi} \int_0^{\infty} d\zeta
  e^{-E_{\rm C} N \zeta/\pi} \ln \frac{\sinh[\pi T
  (t - t' + \zeta - i 0^+ \mbox{sign}_{\rm c}(t - t'))]}
  {\sinh(\pi T \zeta)}.
  \nonumber \\
  \label{eq:f}
\end{eqnarray}
In the last equation, $0^+$ is a positive infinitesimal
and ${\rm C} \approx 0.577$ is the Euler constant.
We separate
$\ln f(t-t')$ into a contour-independent (Keldysh)
part and a contour-dependent part,
\begin{equation}
  \ln f(t-t') = \ln |f(t-t')|
  + i N \kappa_0(t-t') \sign_{\rm c}(t-t'),
\end{equation}
where
\begin{equation}
  \kappa_0(t-t') = \frac{\pi}{N} \left(1 - e^{-E_{\rm c} N(t'-t)/\pi}
  \right) \theta(t'-t).
\end{equation}
Here $\theta(x) = 1$ if $x > 0$ and $0$ otherwise.

Note that the result (\ref{eq:fermion}) is compatible with the
symmetry of the Hamiltonian $H_0$. At first glance this is $U(N)\times
U(N)$, the two factors corresponding to the right and left
movers. Certainly this is the case for the non-interacting model with
$E_{\mathrm{c}}\to 0$, where (\ref{eq:fermion}) reduces to the usual
Wick's theorem result. The interaction, however, leads to the chiral
anomaly, as in the Schwinger model, that reduces the symmetry to
$SU(N)\times SU(N)\times U(1)$. Physically, this is because the interaction can scatter right movers to left movers and vice versa. This allows a second type of
correlation function to exist, characterized by the $SU(N)$-invariant
antisymmetric tensor $\varepsilon_{i_1i_2\ldots i_N}$. We will refer
to such correlators as anomalous. They describe the scattering of
right movers to left movers (and vice versa) by the
interaction.\cite{kn:matveev1995,kn:aleiner2002} Below we list the
first of these nontrivial correlators for $N=1$ and for $N=2$.  For
$N=1$ one has
\begin{eqnarray}
  -i \langle T_{\rm c}
  \hat \psi^{\vphantom{\dagger}}_{R}(t')
  \hat \psi^{\dagger}_{L}(s)\rangle &=&
  \frac{E_{\rm c} e^{\rm C} f(s-t')}{2 \pi^2 v_F f(0)}
  e^{- i (2 \pi {\cal N} + \phi_R(t') - \phi_L(s))}, \nonumber \\
  -i \langle T_{\rm c} \hat \psi^{\vphantom{\dagger}}_{L}(t)
  \hat \psi^{\dagger}_{R}(s')
  \rangle
  &=&  - \frac{E_{\rm c} e^{\rm C} f(t-s')}{2 \pi^2 v_F f(0)}
  e^{-i(\phi_L(t) - \phi_R(s') - 2 \pi {\cal N})}.
  \label{eq:psi4b}
\end{eqnarray}
For $N=2$ the relevant anomalous correlators are
\begin{eqnarray}
  \lefteqn{(-i^2)
  \left\langle T_{\rm c}
  \hat \psi_{1R}^{\vphantom{\dagger}}(0,t_1')
  \hat \psi_{1L}^{\dagger}(0,t_2)
  \hat \psi_{2R}^{\vphantom{\dagger}}(0,t_3')
  \hat \psi_{2L}^{\dagger}(0,t_4)
  \right\rangle_0} \nonumber \\
  &=&
  \left( \frac{E_{\rm c} e^{\rm C}}{\pi^2 v_F f(0)} \right)^2
  e^{-i (2 \pi {\cal N} +
  \phi_{R1}(t_1') -\phi_{L1}(t_2)  + \phi_{R2}(t_3')
  - \phi_{L2}(t_4))}
  \left[f(t_2,t_1') f(t_4,t_1') f(t_2,t_3') f(t_4,t_3')\right]^{1/2},
  \nonumber \\
  \lefteqn{(-i^2)
  \left\langle T_{\rm c}
  \hat \psi_{1L}^{\vphantom{\dagger}}(0,s_1)
  \hat \psi_{1R}^{\dagger}(0,s_2')
  \hat \psi_{2L}^{\vphantom{\dagger}}(0,s_3)
  \hat \psi_{2R}^{\dagger}(0,s_4')
  \right\rangle_0} \nonumber \\
  &=&
  \left( \frac{E_{\rm c} e^{\rm C}}{\pi^2 v_F f(0)} \right)^2
  e^{i (\phi_{L1}(s_1) - \phi_{R1}(s_2') + \phi_{L2}(s_3)
  - \phi_{R2}(s_4') - 2 \pi {\cal N}}
  \left[f(s_1,s_2') f(s_1,s_4') f(s_3,s_2') f(s_3,s_4')\right]^{1/2}.
\end{eqnarray}
\end{widetext}
Notice that the `vacuum-angle' describing the $U(1)$ symmetry breaking is just $2\pi\cN$.

For the perturbation theory calculation of the current,
we need contour-ordered correlators of the form
$$
  \langle T_{\rm c}
  \hat I_j(t) \hat A(t_1') \ldots \hat A(t_n') \rangle,
$$
where the symbols $\hat A(t')$ represent creation or
annihilation operators for left or right moving fermions.
There is a simple relation between such correlators and
the corresponding correlator without current operator,
\begin{eqnarray}
  \lefteqn{\langle T_{\rm c}
  \hat I_j(t) \hat A_1(t_1') \ldots \hat A_n(t_n') \rangle}
  \nonumber \\
  &=&
  \langle T_{\rm c}
  \hat A_1(t_1') \ldots \hat A_n(t_n') \rangle
  \sum_{m=1}^{n} F_m(t,t_m').
  \label{eq:currentgeneral}
\end{eqnarray}
Here the $F_m$ depend on whether the corresponding operator
$A_m$ is a creation or annihilation operator for left or
right moving fermions.
For an annihilation operator for left-moving and right-moving
fermions in channel $k$, one has
\begin{eqnarray}
  F_{j,kL}(t,t_1) &=&
  - \frac{e T}{2 i} \int_0^{\infty} d\tau i_{jk}(t-t_1-\tau)
  \coth(\pi T \tau)
  \nonumber \\ && \mbox{}
   - \frac{e}{2} \sign_{\rm c}(t-t_1)
  i_{jk}(t-t_1)
\end{eqnarray}
respectively. Factors $F_m$ for the creation operators
for left-moving and right-moving fermions in channel $k$ are
$-F_{j,kL}$ and $-F_{j,kR}$, respectively.

\section{Calculation of the current}
\label{sec:current}

With the help of the correlators listed in the previous
section, the perturbation expansion (\ref{eq:Ipert2}) can
be evaluated. In this section, we first describe a full
evaluation of $I(t)$ up to second order in the effective
action (first order for $N=1$). We then reorganize the
perturbation theory, and calculate $I(t)$ to all orders in
${\cal S}$, but in a perturbation expansion in $1/N$, which
turns out to correspond to a loop expansion.\cite{kn:zinnjustin1993}

\subsection{Expansion in effective action}
\label{sec:currentA}

To {\em zeroth order} in the effective action,
one finds that the current in channel $j$ is given by
\begin{eqnarray}
  \label{eq:I0}
  I_{0,j}(t) &=&
  \frac{1}{2} i e v_F \left[ G^{\rm K}_{jjR}(t,t) - G^{\rm K}_{jjL}(t,t)
  \right] \\ &=&
  \frac{e}{2 \pi \hbar}
  \int dt_1 \sum_{k=1}^{N}
  i_{jk}(t-t_1) \left[ \mu_{kL}(t_1) - \mu_{kR}(t_1) \right],
  \nonumber
\end{eqnarray}
in agreement with Eq.\ \eqref{eq:currentzerosimple} and with
$i_{jk}(t-t_1)$ defined in Eq.\ (\ref{eq:i}) above.

To {\em first order} in the effective action one has
\begin{eqnarray}
  I_{j,1}(t) &=&
  -i \left\langle {\cal S}
  \hat I_{j}(t) \right\rangle_0.
\end{eqnarray}
Substituting Eq.\ (\ref{eq:SeffA}) for the effective action
and using the fermion correlation functions of the previous section,
one finds $I_{j,1}(t) = 0$, except for $N=1$. In that case, one
has
\begin{widetext}
\begin{eqnarray}
  I_1(t) &=&
  4 v_F v \mbox{Im}\, \int_{\rm c} dt_1 \int d\tau
  e^{-2 \pi i {\cal N}} S(t_1,t_1-\tau) [-F_{L}(t,t_1) + F_{R}(t,t_1-\tau)]
  \left( \frac{ -i E_{\rm c} e^{\rm C} f(\tau)}{2 \pi^2 v_F f(0)} \right)
  e^{-i (\phi_R(t_1-\tau) - \phi_L(t_1))}. \nonumber \\
\end{eqnarray}
In this equation, it is important to note that all
contour-dependence should depend on $t_1$ only; the time
difference $\tau$ should not enter into the contour signs. We also
note that the argument of $f(\tau)$ is always positive, so that
$f(\tau)$ does not depend on the contour position, the only the
contour-dependent part of the function $F$ that contributes to the
integration, and that the contribution from right moving fermions
is zero by causality, so that
\begin{eqnarray}
  I_1(t) &=&
  - \frac{2 e v E_{\rm c} e^{\rm C}}{\pi^2 f(0)}
  \int dt_1 i(t,t_1) \int d\tau f(\tau) \mbox{Re}\,
  S(t_1,t_1-\tau) e^{- i (\phi_R(t_1-\tau) - \phi_L(t_1))
  - 2 \pi i {\cal N}}
  \label{eq:IoscN1}
\end{eqnarray}
One verifies that for a time-independent situation ---
$S(t,t-\tau)$ independent of $t$ and chemical potentials
$\mu_{L}$ and $\mu_R$ independent of time --- the current is
zero.

The {\em second order} calculation proceeds similarly.
We first restrict our attention to the case $N > 2$.
Using matrix notation, we then find
\begin{eqnarray}
  I_{j,2}(t) &=&
  - \frac{e}{2} (2 v_F)^2 \int_{\rm c} dt_1 ds_1 \int d\tau_1 d\sigma_1
  \left( \frac{f(\tau_1) f(\sigma_1)}{f(t_1-s_1+\sigma_1) f(s_1-t_1+\tau_1)}
  \right)^{1/N}
  \nonumber \\ && \mbox{} \times
  \mbox{tr}\, v^2 [i_j(t,t_1) \mbox{sign}_{\rm c}(t-t_1)
  - i_j(t,s_1) \mbox{sign}_{\rm c}(t-s_1)]
  \nonumber \\ && \mbox{} \times
  G_L(s_1,t_1)
  S(t_1,t_1-\tau_1) G_{R}(t_1-\tau_1,s_1-\sigma_1)
  S^{\dagger}(s_1-\sigma_1,s_1).
\end{eqnarray}
Here $i_j$ is an $N \times N$
diagonal matrix, closely related to the kernel defined in Eq.\
(\ref{eq:i}),
\begin{equation}
  (i_j)_{mn}(\tau)
  =
  \delta_{mn} i_{jm}(\tau).
\end{equation}

At this point, we write the Green functions $G_R$ and $G_L$ as the
sum of the Keldysh component and a delta function, see Eq.\
(\ref{eq:GKdelta}). Doing this, there will be four terms: one
term with delta functions for both $G_L$ and $G_R$, two terms with
one delta function and one Keldysh Green function, and one term
with two Keldysh Green functions. The first term is easily shown
to vanish. In the second and third term, the interaction function
$f$ cancels, and one finds
\begin{eqnarray}
  I_{j,2,0}(t) &=&
  2 i v_F e \int dt_1
  \mbox{tr}\, v^2 i_{jL}(t,t_1)
  \nonumber \\ && \mbox{} \times
  \left[
  G^{\rm K}_{L}(t_1,t_1)
   - \int d\tau_1 d\sigma_1S(t_1,t_1-\tau_1)
  G^{\rm K}_{R}(t_1-\tau_1,t_1-\sigma_1)
  S^{\dagger}(t_1-\sigma_1,t_1)
  \right].
  \label{eq:I21}
\end{eqnarray}
Here the last index ``0'' indicates that, apart from the difference
between $i_{jk}(\tau)$ and $\delta_{jk} \delta(\tau)$,
Eq.\ (\ref{eq:I21}) --- together with the
zeroth-order result of Eq.\ (\ref{eq:I0}) --- reproduces
precisely what one would obtain from the scattering approach for
non-interacting fermions.
Finally, in the remaining term, the interaction contribution is
retained. In fact, it is just because of the contour-dependence
of the function $f$ that the one finds a nonzero result. We
then find
\begin{eqnarray}
  I_{j,2,{\rm ee}}(t) &=&
  - 2 e i v_F^2 \int dt_1 \int^{t_1} ds_1 \int d\tau_1
  d\sigma_1
  \left| \frac{f(\tau_1) f(\sigma_1)}{f(t_1-s_1+\sigma_1)
  f(s_1-t_1+\tau_1)}
  \right|^{1/N} \sin [\kappa_0(s_1-t_1+\tau_1)]
  \nonumber \\ && \mbox{} \times
  \mbox{tr}\, v^2 i_j(t,t_1)
  \left[ G^{\rm K}_{L}(s_1,t_1) S(t_1,t_1-\tau_1)
  G^{\rm K}_{R}(t_1-\tau_1,s_1-\sigma_1) (S^{\dagger})(s_1-\sigma_1,s_1)
  \right. \nonumber \\ && \ \ \ \left. \mbox{}
  - G^{\rm K}_{L}(t_1,s_1) S(s_1,s_1-\sigma_1)
  G^{\rm K}_{R}(s_1-\sigma_1,t_1-\tau_1) (S^{\dagger})(t_1-\tau_1,t_1)
  \right].
  \label{eq:I22}
\end{eqnarray}

For $N=2$ there is an additional oscillating contribution to the current,
\begin{eqnarray}
  I_{j,{\rm osc}}(t) &=&
  8 e  v_1 v_2
  \left( \frac{2 E_{\rm c} e^{\rm C}}{\pi^2 f(0)} \right)^2
  \mbox{Im}\,
  e^{-2 \pi i {\cal N}}
  \int dt_1 \int^{t_1} dt_2 dt_1' dt_2'
  \sin(\phi(t_2,t_1'))
  \nonumber \\ && \mbox{} \times
  \det[i_j(t,t_1) e^{i \phi_L(t_1)} S(t_1,t_1')
  e^{-i \phi_R(t_1')}, e^{i \phi_L(t_2)} S(t_2,t_2')
  e^{-i \phi_R(t_2')}]
  \nonumber \\ && \mbox{} \times
  |f(t_1,t_1') f(t_1,t_2') f(t_2,t_1') f(t_2,t_2')|^{1/2},
  \label{eq:IoscN2}
\end{eqnarray}
where we defined
\begin{eqnarray}
  \det(B(1),\ldots,B(N)) &=& \frac{1}{N!}
  \sum_{P}  \sum_{k_1,\ldots,k_N} (-1)^P
  B(1)_{k_1,P(k_1)} \ldots B(N)_{k_N,P(k_N)},
\end{eqnarray}
\end{widetext}
where $B(j)$ is an $N \times N$ matrix, $j=1,\ldots,N$.
 Equations (\ref{eq:IoscN1}), (\ref{eq:I22}),
and (\ref{eq:IoscN2}) represent the interaction corrections to the
current.

This result is found to agree precisely with the linear dc
conductance (\ref{abg}) calculated in Refs.\
\onlinecite{kn:brouwer1999c} and \onlinecite{kn:aleiner2002} in
the dc bias limit, to first order in the bias. To see this, we set
$\mu_{R1}-\mu_{R2} = - e V$, expand to linear order in the bias
voltage, take $S(t,t-\tau)$ to depend on $\tau$ only, set
$\mu_{jL} = 0$ for all channels $j=1,\ldots,N$ in the fictitious
lead, use the approximations $|f(\tau)| \approx \lambda \pi T/ v_F
\sinh[\pi T |\tau + t_0|]$, $\kappa_0(\tau) = \pi
\theta(-\tau-t_0)/N$ valid for $\tau \ll t_0$ and $\tau \gg t_0$,
and set $v_j=1$, so that  $r_j=1$ (see Eq.~\ref{eq:wr}), which
corresponds to detaching the fictitious lead and thus restoring
the original system. The oscillating current corrections
(\ref{eq:IoscN1}) and (\ref{eq:IoscN2}) reproduce the
periodic-in-${\cal N}$ interaction corrections to the dot's
`electrochemical capacitance' $C_{\rm c} = dQ/dV$ found in Ref.\
\onlinecite{kn:aleiner1998a} if we set $\mu_{jL} = 0$, choose a
time-dependent chemical potential $\mu_{jR}(t) = e V t$ for all
channels, and expand in $V$. [In order to recover the non-periodic
interaction correction to the capacitance reported in Ref.\
\onlinecite{kn:aleiner2002} from Eqs.\ (\ref{eq:I21}) and
(\ref{eq:I22}) one has to include time-dependencies in both
$\mu_{L}$ and $\mu_{R}$ in order to ensure that no charge escapes
through the fictitious lead.]

However, truncating at second order in
the effective action ${\cal S}$ is not a satisfactory description for
a fully coherent quantum dot as there is no a priori reason why the
effective action is small in this case.
This will be made explicit in the next
subsection, where we report a calculation to all orders in ${\cal S}$
(but for $N$ large). However, that Eqs.\ (\ref{eq:IoscN1}),
(\ref{eq:I22}), and (\ref{eq:IoscN2}) cannot be the
correct interaction correction for a fully coherent dot becomes clear
when one realizes that
the current in fact depends on the chemical potentials $\mu_{jL}$,
$j=1,\ldots,N$, of the fictitious reservoir, despite the fact that one
has taken the limit $r_j \to 1$, $j=1,\ldots,N$. (Note that this
dependence on the fictitious
chemical potentials is not manifest in the linear response calculation
of Refs.\ \onlinecite{kn:aleiner1998a,kn:brouwer1999c,kn:aleiner2002}
because $\mu_{jL} = 0$, $j=1,\ldots,N$, in equilibrium.)

Of course, truncating the perturbation expansion at second order
in ${\cal S}$ is justified if there is a reason why ${\cal S}$ is
small. This is the case,
{\em e.g.,} if the $r_j$ are small, $j=1,\ldots,N$. Then the
fictitious lead is strongly coupled to the point contact, and
serves as a source of relaxation. In that case the chemical
potentials $\mu_{jL}$ must be chosen such that no current flows
through the fictitious lead at any point in time. A more realistic
(but still phenomenological) model of relaxation is to couple to
quantum dot itself (rather than the point contact) to a fictitious
lead with many weakly coupled
channels.\cite{kn:buettiker1986,kn:buettiker1988,kn:brouwer1995,%
kn:baranger1995,kn:brouwer1997c,kn:alves2002}
As shown in appendix \ref{app:c}, this model results in an
effective action ${\cal S}$ similar to (\ref{eq:SeffA}), but with a
sub-unitary scattering matrix $S$. If relaxation is strong, $S$ is
small, and expanding in the action is justified. For this situation,
we recover the interaction corrections found in Refs.\
\onlinecite{kn:aleiner1998a,kn:brouwer1999c,kn:aleiner2002}.

\subsection{Large number of channels}
\label{sec:currentB}

We have not been able to do a full calculation to all orders in ${\cal
 S}$. However, by organizing each order in the effective action to the
(formal) power of $N$ it carries, we have been able to calculate the
current to all orders in ${\cal S}$ while expanding in $1/N$.
The reason that an expansion in $1/N$ is possible is that
the correlator (\ref{eq:fermion}) of interacting fermions
admits a systematic expansion around the non-interacting
correlator, by expanding
\begin{equation}
  f^{1/N} = 1 + \frac{1}{N} \ln f + \ldots.
  \label{eq:Nf}
\end{equation}
Counting the power of $1/N$ is done keeping the $RC$ time
$\pi/E_{\rm c} N$ and the dwell time $\tau_{\rm d}$ constant, so
that the only factors $1/N$ arise from the expansion
(\ref{eq:Nf}) and from explicit summations over the channel
indices.
Oscillating contributions with an explicit dependence on the
dimensionless gate voltage ${\cal N}$, involve $N$th
order in perturbation theory and cannot be considered
in this method.

Non-oscillating contributions to the current occur to even order
in the effective action only. The non-oscillating current
contribution of order $2n$ in the action reads
\begin{widetext}
\begin{eqnarray}
  \label{eq:Ijn}
  I_{j,n}(t) &=& \frac{1}{(2n)!} (-i)^{2n}
  \langle {\cal S}^n I_j \rangle_0
  \nonumber \\ &=&
  \frac{1}{(n!)^2} (-2 i v_F)^{2n}
  \sum_{k_1,l_1,k_1',l_1'} \ldots \sum_{k_n,l_n,k_n',l_n'}
  \int_{\rm c} dt_1 \ldots dt_n
  ds_1 \ldots ds_n \int d\tau_1 \ldots d\tau_n d\sigma_1 \ldots d\sigma_n
  v_{k_1} \ldots v_{k_N}
  \nonumber \\ && \mbox{} \times
  v_{l_1} \ldots v_{l_N}
  S_{k_1,k_1'}(t_1,t_1-\tau_1) \ldots S_{k_n,k_n'}(t_n,t_n-\tau_n)
  (S^{\dagger})_{l'_1,l_1}(s_1-\sigma_1,s_1) \ldots
  (S^{\dagger})_{l'_n,l_n}(s_n-\sigma_n,s_n)
  \nonumber \\ && \mbox{} \times
  \left\langle T_{\rm c}
  \hat \psi_{k_1L}^{\dagger}(t_1)
  \hat \psi_{k_1'R}^{\vphantom{\dagger}}(t_1-\tau_1)
  \hat \psi_{l_1'R}^{\dagger}(s_1-\sigma_1)
  \hat \psi_{l_1L}^{\vphantom{\dagger}}(s_1)
  \ldots
  \right. \nonumber \\ && \ \ \left. \mbox{} \times
  \hat \psi_{k_nL}^{\dagger}(t_n)
  \hat \psi_{k_n'L}^{\vphantom{\dagger}}(t_n-\tau_n)
  \hat \psi_{l_n'R}^{\dagger}(s_n-\sigma_n)
  \hat \psi_{l_nL}^{\vphantom{\dagger}}(s_n)
  \hat I_j (t) \right\rangle_0.
\end{eqnarray}
\end{widetext}
As before, the time differences denoted by Greek symbols do not
affect the contour ordering for the fermion operators.

\begin{figure}
\epsfxsize=0.9\hsize \epsffile{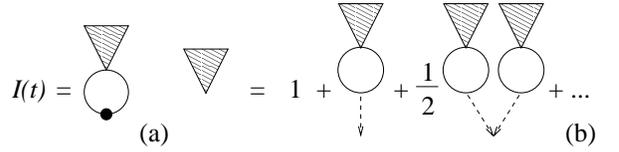} \caption{\label{fig:tree}
(a) Diagrammatic representation of leading-in-$1/N$ contribution
to the current. The current vertex $i$ is denoted with a filled
circle, the `bubble' of single-fermion Green function with a solid
loop, and the tree structure with a hatched triangle. The defining
equation for the tree structure is shown in (b), where the dotted
arrow represents the interaction line.}
\end{figure}

The average of the product of $4n$ fermion operators and one current
operator in Eq.\ (\ref{eq:Ijn}) admits a standard diagrammatic
representation if the factors $1/N \ln f$ in the expansion
(\ref{eq:Nf}) of the general average (\ref{eq:fermion}) are considered
`interaction lines'. Without `interaction lines', the average is
simply given by Wick's theorem: Denoting each single-fermion Green
function by a solid line, Wick's theorem expresses the average as a
sum over partitions in `bubbles' of alternating single-fermion
Green functions for right moving and left moving fermions. One
bubble contains the current vertex $i$. In our
formal expansion in $1/N$, each bubble contributes a factor $N$, as
it contains a trace over the channel index. Each `interaction
line', on the other hand, contributes a factor $1/N$. Only diagrams
in which all `bubbles' are connected by interaction lines contribute
to the current. This means that, to leading order in $1/N$, the
relevant diagrams have a `tree' structure, see Fig.\
\ref{fig:tree}: Every `bubble' has
precisely one connection to the current vertex via intermediate
bubbles and interaction lines. Subleading contributions to the
current contain `loops'.

To {\em leading order} in $1/N$, we only have to examine the
`tree' diagrams, plus the current contribution that is zeroth
order in the action, see Eq.\ (\ref{eq:I0}). Combining these,
we find
\begin{eqnarray}
  I_j(t) &=& \frac{e}{2 \pi \hbar} \int dt_1 \sum_{k=1}^{N}
  i_{jk}(t-t_1)[\mu_{Lk}(t_1) - \mu_{Rk}(t_1)]
  \nonumber \\ && \mbox{}
  + e
  \int dt_1 \tr i_{j}(t-t_1) r^2 j_-(t_1),
  \label{eq:Ij1}
\end{eqnarray}
where we used Eq.\ (\ref{eq:wr}) to express $v$ in terms of the
reflection matrix $r$. In Eq.\ (\ref{eq:Ij1}), we abbreviated the
contribution of the bubble by
\begin{eqnarray}
  j_-(t) &=& \frac{1}{2} i v_F \left[
  G^{\rm K}_{L}(t,t) -
  \int d\tau d\sigma S(t,t-\tau)
  \right. \nonumber \\ && \left. \mbox{} \times
  \tilde G^{\rm K}_{R}(t-\tau,t-\sigma)
  S^{\dagger}(t-\sigma,t) \right]
  \label{eq:IL}
\end{eqnarray}
where
\begin{equation}
  \tilde G^{\rm K}_{R}(t,s) = G^{\rm K}_{R}(t,s)
  e^{- i (\tilde \phi(t) - \tilde \phi(s))}.
\end{equation}
The phase $\tilde \phi(t)$ satisfies the self-consistent equation
\begin{eqnarray}
  \tilde \phi(t) &=& 2 \int dt_1
  \kappa_0(t_1-t)
  \tr r^2 j_-(t_1).
  \label{eq:tphi}
\end{eqnarray}
The quantity $j_-$ is nothing but the Landauer expression for the
difference of the current flowing into the leads coming from the
quantum dot [second term in Eq.\ (\ref{eq:IL})] and coming from
the fictitious reservoir [first term in Eq.\ (\ref{eq:IL})].

The overall phase factor that appears in $\tilde G^K_{mR}(t,s) $ can
be written in terms of the electrochemical potential $\mu_{\rm c}$
\begin{eqnarray*}
  \phi_{mR}(t)+\tilde \phi(t)-\phi_{mR}(s)+\tilde \phi(s)
  &=&
  \int^t_s dt' \mu_{{\rm c}m}(t').
\end{eqnarray*}
Here $\mu_{{\rm c}m} = \mu_{m} + e V_{\rm d}$
[the plus sign arises because the interaction is in the leads, not the
dot, {\em cf.}\ Eq.\ (\ref{eq:muc})] and the dot potential is
\begin{equation} \label{eq:Ivd}
CV_{\mathrm{d}}(t)=\int^t \sum_{k=j}^N dt' I_j(t'),
\end{equation}
where $C$ is the dot's geometrical capacitance and we used the
specific form of $\kappa_0(t'-t)$ and $i_{jk}(\tau)$. Note that in the
limit $r_j\to 1$ the $\mu_{mL}(t)$ drop out of
(\ref{eq:Ij1}). Equation (\ref{eq:Ivd}) is just the same as (\ref{sc_pot}),
written in terms of the currents. Hence, to leading order in $1/N$,
our formalism
recovers the self-consistent theory for the effect of the
Coulomb interaction on transport through quantum dots.

\begin{figure}[t]
\epsfxsize=0.4\hsize
\epsffile{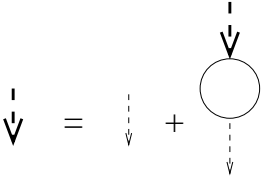}
\caption{\label{fig:rpa} Diagrammatic representation of the
effective interaction $\kappa$ (thick arrow) in terms of the
bare interaction line $\ln f$ (thin arrow) and a `bubble'
of single-fermion Green functions (solid loop).}
\end{figure}

\begin{figure}
\epsfxsize=0.95\hsize
\epsffile{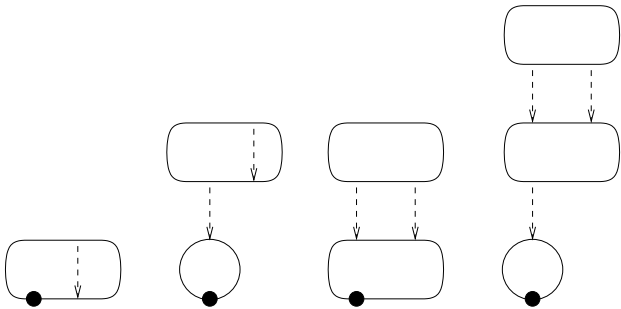}
\caption{\label{fig:oneloop} Diagrammatic representation of
subleading-in-$1/N$ corrections to the current}
\end{figure}

To {\em subleading order} in $1/N$, one has to solve diagrams
with one loop. The `interaction lines' in these diagrams are
replaced by an effective interaction, which is an RPA-like
series, see Fig.\ \ref{fig:rpa}. Denoting the effective
interaction by $\kappa(t,s)$, one finds that $\kappa(t,s)$
satisfies the self-consistent equation
\begin{widetext}
\begin{eqnarray}
  \kappa(t,s) &=& \kappa_0(t,s) - 2 v_F \int dt_1 d\tau_1 d\sigma_1
  (\kappa_0(t,t_1 - \tau_1) - \kappa_0(t,t_1-\sigma_1))
  \nonumber \\ &&
  \mbox{} \times
  \tr r^2 S(t_1,t_1 - \tau_1) \tilde G^{\rm
  K}_{R}(t_1-\tau_1,t_1-\sigma_1)
  S^{\dagger}(t_1-\sigma_1,t_1)  \kappa(t_1,s).
  \label{eq:kappaone}
\end{eqnarray}
Note that $\kappa(t,s) = 0$ if $t \ge s$ and that
$\kappa(t,s)$ is real.

The one-loop interaction correction to the current can be represented
by four diagrams, see Fig.\
\ref{fig:oneloop}. The first and second diagram in Fig.\
\ref{fig:oneloop} represent a correction to $j_-$,
\begin{eqnarray}
  j_{-}(t) &=&
  \frac{1}{2} i v_F \left[
  G^{\rm K}_{L}(t,t) -
  \int d\tau d\sigma S(t,t-\tau)
  \tilde G^{\rm K}_{R}(t-\tau,t-\sigma)
  S^{\dagger}(t-\sigma,t) \right]
  \nonumber \\ && \mbox{} +
  v_F^2 \mbox{Im}\, \int ds_1 d\tau_1 d\sigma_1
  \kappa(s_1,t-\tau_1)
  S(t,t-\tau_1) \tilde G^{\rm K}_{R}(t-\tau_1,s_1-\sigma_1)
  S^{\dagger}(s_1-\sigma_1,s_1)
  \nonumber \\ && \mbox{} \times
  \left[ (1 - r^2)
  G^{\rm K}_{L}(s_1,t)
  + r^2 \int d\tau_2 d\sigma_2
  S(s_1,s_1-\tau_2)
  \tilde G^{\rm K}_{R}(s_1-\tau_2,t-\sigma_2)
  S^{\dagger}(t-\sigma_2,t) \right],
  \label{eq:IRR}
\end{eqnarray}
whereas the third and fourth diagram represent an additional
renormalization of the distribution function for right moving
fermions, which can be represented by a change in the
relation between to $\tilde G_R$ and $G_R$,
\begin{eqnarray}
  \tilde G_R^{\rm K}(t,s) &=& G_R^{\rm K}(t,s)
  e^{-i(\tilde \phi(t) - \tilde \phi(s)) + \alpha(t,s)}.
  \label{eq:tG}
\end{eqnarray}
For the calculation of $\alpha(t,s)$, the regularization of the
fermion operators at the impurity site is important, which makes
the actual calculation rather cumbersome. The final result
reads
\begin{eqnarray}
  \alpha(t,s) &=& - \frac{1}{2} v_F^2 \int dt_1 dt_2 (\kappa(t_1-t) -
  \kappa(t_1-s))(\kappa(t_2-t) - \kappa(t_2-s))
  \tr \left\{ G^{\rm K}_{L}(t_1,t_2) G^{\rm K}_{L}(t_2,t_1)
  \vphantom{\int}
  \right. \nonumber \\ && \left.
  - \left[ (1 - r^2) G^{\rm K}_{L}(t_1,t_2)
  + r^2 \int d\tau_1 d\sigma_2 S(t_1,t_1-\tau_1)
  \tilde G^{\rm K}_{R}(t_1-\tau_1,t_2-\sigma_2)
  S^{\dagger}(t_2-\sigma_2,t_2) \right]
  \right. \nonumber \\ && \left. \times
  \left[(1 - r^2) G^{\rm K}_{L}(t_2,t_1)
  + r^2 \int d\tau_1 d\sigma_2 S(t_2,t_2-\tau_2)
  \tilde G^{\rm K}_{R}(t_2-\tau_2,t_1-\sigma_1)
  S^{\dagger}(t_1-\sigma_1,t_1) \right]
  \right\}.
  \label{eq:alpha}
\end{eqnarray}
\end{widetext}
Equations (\ref{eq:Ij1}), (\ref{eq:tphi}), (\ref{eq:kappaone}),
(\ref{eq:IRR}),
(\ref{eq:tG}), and (\ref{eq:alpha}) give the solution of the
problem up to sub-leading order in $1/N$.

Note that the leading
corrections to the
theory are in the form of a modified expression for the
current $j_-$; Equation (\ref{eq:tphi}) and, hence,
the relation (\ref{eq:Ivd}) between $I_j(t)$ and the self-consistent
potential $V_{\rm d}$ remains unchanged.

The interaction correction to $j_-$, contained in the
second and third line of Eq.\ (\ref{eq:IRR}), consists of a term
proportional to $r^2$ and a term proportional to $1-r^2$.
When expanding in the action and truncating the expansion
at second order, as is done in Sec.\ \ref{sec:currentA} and
in Refs.\ \onlinecite{kn:brouwer1999c}
and \onlinecite{kn:aleiner2002}, one recovers
the first term only, with prefactor unity instead of $1-r^2$.
However, upon setting $r^2 \to 1$, which is the relevant limit
for a coherent quantum dot with a unitary scattering matrix,
the prefactor of that first term vanishes, and
the interaction correction to
$j_-$ is given by the second term instead. It is this
replacement of the first term by the second term in the
interaction correction that is the essential difference
between the theories of Refs.\ \onlinecite{kn:brouwer1999c}
and \onlinecite{kn:aleiner2002} on the one hand and that of
this work and
Ref.\ \onlinecite{kn:golubev2004} on the other hand.

\section{Steady state transport}
\label{sec:steadystate}

In this section we analyze the results of the previous section
for the case of steady state transport through the
quantum dot. We assume that the chemical potentials in the reservoirs
and the gate voltages defining the shape of the
dot are time independent. This implies that the scattering matrix
$S(t,t-\tau)$ is independent of $t$ and the Green functions
$G_L(t,t-\tau)$ and $G_R(t,t-\tau)$ are independent of $t$.
Current conservation then implies that there
is no net current into the dot, $\sum_{j} I_j = 0$ -- see Eq.~(\ref{dc_current}). This allows us
to focus our calculation on the weighted difference of currents in
the left and right leads,
\begin{equation}
  I = \sum_{j=1}^{N} \Lambda_{j} I_{j},
\end{equation}
where $\Lambda$ was defined in Eq.\ (\ref{eq:lambda}) above.
We will be using the Fourier transform of the scattering matrix
and the Green functions,
\begin{eqnarray}
  S(\varepsilon) &=& \int_0^{\infty} dt S(t) e^{i \varepsilon t}, \\
  S^{\dagger}(\varepsilon) &=& \int_{-\infty}^{0} dt S^{\dagger}
  (t) e^{i \varepsilon t}, \\
  G^{\rm K}_{mn R,L}(\varepsilon) &=& \int dt G^{\rm K}_{mnR,L}(t)
  e^{i \varepsilon t}
  \nonumber \\ &=&
  \frac{i}{v_F}\delta_{mn}(2 f_{R,L}(\varepsilon) - 1),
\end{eqnarray}
where $f$ is the distribution function.

\subsection{Fully coherent dot}

We first consider the case $r^2 = 1$, corresponding to a fully
coherent dot.

We first consider the function $\tilde \phi$. By Eq.\
(\ref{eq:tphi}), $\tilde \phi$ depends on $\mbox{tr}\, j_-$ only.
For the time independent case, one has
\begin{eqnarray}
  \mbox{tr}\,
  j_- = - \frac{N}{2 \pi} \frac{\partial \tilde \phi}{\partial t}.
\end{eqnarray}
Calculating the derivative $\partial \tilde \phi/\partial t$ from
Eq.\ (\ref{eq:tphi}) one then finds
\begin{eqnarray}
  \frac{\partial \tilde \phi}{\partial t}
  &=& - \frac{N}{\pi} \int dt_1 \frac{\partial \kappa_0(t-t_1)}{\partial
  t} \frac{\partial \tilde \phi}{\partial t} \nonumber \\
  &=& \frac{\partial \tilde \phi}{\partial t},
\end{eqnarray}
which implies that the derivative $\partial \tilde \phi/\partial
t$ is left undetermined by the self-consistency equation
(\ref{eq:tphi}); any time-independent value for $\partial \tilde
\phi/\partial t$ is a solution of the self-consistency equation.
This is no problem for steady state transport; in the previous
section we have seen that $\tilde \phi$ is determined once the
time-dependence of bias and gate voltages is taken into account.
In this section, we will set $\tilde \phi = 0$ from now on.

The effective interaction function $\kappa$ can be solved from
Eq.\ (\ref{eq:kappaone}) using Fourier transform,
\begin{widetext}
\begin{eqnarray}
  \kappa(\omega) &=&
  \kappa_0(\omega) \left[\kappa_0(\omega)
  - \frac{i}{\pi}
  \int d\varepsilon
  \tr S(\varepsilon-\omega)
  (f_{R}(\varepsilon - \omega)
  - f_{R}(\varepsilon))
   S^{\dagger}(\varepsilon) \right]^{-1}. \nonumber
\end{eqnarray}
Here $\kappa_0(\omega)$ is the Fourier transform of $\kappa_0(\tau)$,
\begin{eqnarray}
  \kappa_0(\omega) &=& \int d\tau \kappa_0(\tau) e^{i \omega \tau}.
\end{eqnarray}
This effective interaction is the same as that obtained by Golubev and
Zaikin.\cite{kn:golubev2004}
We further need to calculate the function $\alpha(t-s)$, see Eq.\
(\ref{eq:tG}). Fourier transforming Eq.\ (\ref{eq:alpha}) one has
\begin{eqnarray}
  \label{eq:b}
  \alpha(\tau) &=&
  -  \frac{1}{2 \pi^2} \int d\omega
  (1 - \cos(\omega \tau)) |\kappa(\omega)|^2
  \\ && \mbox{} \times
  \left[ \omega N \coth(\omega/2 T) -
  \sum_{\pm} \int d\varepsilon
  \tr S(\varepsilon\pm\omega) f_R(\varepsilon\pm\omega)
  S^{\dagger}(\varepsilon\pm\omega)
  S(\varepsilon) (1-f_R(\varepsilon)) S^{\dagger}(\varepsilon)
  \right].\nonumber
\end{eqnarray}
Finally, with these results the current through the dot is
\begin{eqnarray}
  I &=&
   -\frac{e}{2 \pi \hbar} \tr \Lambda \mu_R
  + \frac{e}{2 \pi \hbar} \int d\varepsilon \tr \Lambda
  S(\varepsilon) f_R(\varepsilon)
  S^{\dagger}(\varepsilon)
  \nonumber \\ && \mbox{}
  + \frac{e}{2 \pi^2 \hbar} \mbox{Im}\,
  \int d\omega d\varepsilon \kappa(\omega) \tr \Lambda
  \left[ S(\varepsilon) f_R(\varepsilon + \omega)
  S^{\dagger}(\varepsilon + \omega)
  S(\varepsilon) (1-f_R(\varepsilon))
  S^{\dagger}(\varepsilon)
  \right. \nonumber \\ && \ \ \ \left. \mbox{}
  + S(\varepsilon) (1-f_R(\varepsilon + \omega))
  S^{\dagger}(\varepsilon + \omega)
  S(\varepsilon) f_R(\varepsilon)
  S^{\dagger}(\varepsilon)
  \right]
  \nonumber \\ && \mbox{}
  + \frac{e}{16 \pi^4 \hbar}
  \int d\omega d\varepsilon |\kappa(\omega)|^2
  \tr \Lambda S(\varepsilon)[2 f_R(\varepsilon)
  - f_R(\varepsilon - \omega) - f_R(\varepsilon+ \omega)]
  S^{\dagger}(\varepsilon)
  \nonumber \\ && \ \ \mbox{} \times
  \left[ \omega N \coth \frac{\omega}{2 T} -
  \sum_{\pm} \int d\varepsilon'
  \tr S(\varepsilon'\pm\omega) f_R(\varepsilon'\pm\omega)
  S^{\dagger}(\varepsilon'\pm\omega)
  S(\varepsilon) (1-f_R(\varepsilon')) S^{\dagger}(\varepsilon')\right].
 \label{eq:Irfinal}
\end{eqnarray}
In order to find the linear conductance, we set $\mu_R =
- \Lambda e V$, $\mu_L = 0$, expand in $V$,
and find $I = G V$, with
\begin{eqnarray}
  \label{eq:lincond}
  G &=&
  \frac{e^2}{2 \pi \hbar} \tr \Lambda^2
  - \frac{e^2}{2 \pi \hbar}
  \int d\varepsilon \left( - \frac{\partial f(\varepsilon)}
  {\partial \varepsilon} \right)
  \tr \Lambda S(\varepsilon) \Lambda S^{\dagger}(\varepsilon)
  \\ && \mbox{} +
  \frac{e^2}{2 \pi^2 \hbar} \mbox{Im}\,
  \int d\varepsilon d\omega
  \left( - \frac{\partial f(\varepsilon)}{\partial \varepsilon} \right)
  \kappa(\omega)
  \tr
  \left[\Lambda S(\varepsilon) \Lambda S^{\dagger}(\varepsilon + \omega)
  - \Lambda S(\varepsilon) S^{\dagger}(\varepsilon + \omega)
  S(\varepsilon) \Lambda S^{\dagger}(\varepsilon) \right]
  (1 - 2 f(\varepsilon + \omega)). \nonumber
\end{eqnarray}
The same equation was derived previously by Golubev and Zaikin using
a different method.\cite{kn:golubev2004}

Equation (\ref{eq:Irfinal}) gives the current for one particular
realization of the scattering matrix $S(\varepsilon)$. In order to
find the average and variance of the current or the conductance
for an ensemble of quantum dots, we need to average
$S(\varepsilon)$ over the appropriate ensemble of scattering
matrices. Details on the averaging procedure can be found in
Appendix \ref{app:a}. For large $N$, the average of a product of
traces may be calculated as the product of the averages. This
means that the interaction kernel $\kappa(\omega)$ may be averaged
separately, with the result
\begin{eqnarray}
  \kappa(\omega) &=&
  \frac{\pi}{g i \omega} \frac{1 + i \omega \tau_{\rm d}}{i \omega
  \tau_{\rm d}
  [1 + (\pi/E_{\rm c} g \tau_{\rm d})(1 + i \omega \tau_{\rm d})]}.
  \label{eq:kappaomega}
\end{eqnarray}
In Eq.\ (\ref{eq:kappaomega}), $\tau_{\rm d}$ is the mean dwell
time for electrons entering the dot. In terms of the dot's mean
level spacing $\Delta$ and the total conductance $g = g_1 + g_2$
of the point contacts, one has
\begin{equation}
  \tau_{\rm d} = \frac{2 \pi}{g \Delta}.
  \label{eq:taud}
\end{equation}
(If levels are spin degenerate, $\Delta$ is half the mean spacing
between spin-degenerate levels.) Performing the
ensemble average for the current, one then
finds, to leading and sub-leading order in $N$ and in the absence
of time-reversal symmetry,
\begin{eqnarray}
  \langle I \rangle &=&
  - \frac{e}{2 \pi \hbar} \tr \Lambda \mu_R
  + \frac{e}{2 \pi \hbar g} \int d\varepsilon
  [g \tr \Lambda \rc f(\varepsilon) \rcd
  + \tr \Lambda (1 - \rc \rcd)\,
  \tr f(\varepsilon) (1 - \rcd \rc)] \nonumber \\ && +
  \frac{e}{2 \pi^2 \hbar g^2}\mbox{Im}\,
  \int d\varepsilon d\omega \kappa(\omega)
  \frac{(i \omega \tau_{\rm d})^2}{(1 + i \omega \tau_{\rm d})^2}
  \label{eq:avgI}
   \\ && \mbox{} \times
  \left[ g
  \tr \Lambda r_{c} f(\varepsilon + \omega) r_{c}^{\dagger}
  (1 - r_{c} r_{c}^{\dagger})\, \tr (1 - f(\varepsilon))
  (1 - r_{c}^{\dagger} r_{c})
  \right. \nonumber \\ && \ \ \left. \mbox{}
  + g
  \tr \Lambda r_{c} (1-f(\varepsilon + \omega)) r_{c}^{\dagger}
  (1 - r_{c} r_{c}^{\dagger})\, \tr f(\varepsilon)
  (1 - r_{c}^{\dagger} r_{c})
  \right. \nonumber \\ && \ \ \left. \mbox{}
  - \tr \Lambda(1 - \rc \rcd)\,
  \tr f(\varepsilon + \omega) (1 - \rcd \rc) \rcd \rc\,
  \tr (1 - f(\varepsilon)) (1 - \rcd \rc)
  \right. \nonumber \\ && \ \ \left. \mbox{}
  - \tr \Lambda(1 - \rc \rcd)\,
  \tr (1-f(\varepsilon + \omega)) (1 - \rcd \rc) \rcd \rc\,
  \tr f(\varepsilon) (1 - \rcd \rc) \right], \nonumber
\end{eqnarray}
{}From Eq.\ (\ref{eq:avgI}) we conclude that
there is no interaction correction to the
average current for ideal point contacts. Setting $\mu_R = - e V
\Lambda$ and expanding Eq.\
(\ref{eq:avgI}) to first order in the bias voltage $V$, we find
the ensemble average of the linear dc conductance,
\begin{eqnarray}
  \langle G \rangle &=& \frac{e^2}{2 \pi \hbar}
  \frac{g_1 g_2}{g}
  \left\{ 1
    -
  \mbox{Im}\, \int d\omega \frac{(g_2 s_1 + g_1 s_2)
  \tau_{\rm d}[(\omega/T) -
  \sinh(\omega/T)]}{2 g^2
  (1 + i \omega \tau_{\rm d})
  [1 + (\pi/E_{\rm c} g \tau_{\rm d})(1 +  i \omega \tau_{\rm d})]
  \sinh^2(\omega/2 T)} \right\}.
\end{eqnarray}
\end{widetext}
Performing the frequency integral in the limit $\hbar/\tau_{\rm d}
\ll T \ll
E_{\rm c} g$ one finds\cite{kn:golubev2004}
\begin{equation}
  \langle G \rangle = \frac{e^2}{2 \pi \hbar}
  \frac{g_1 g_2}{g}
  \left[1 - \frac{2(s_1 g_2 + s_2 g_1)}{g^2} \ln \frac{E_{\rm
  c} g e^{1 + {\rm C}}}{2 \pi^2 T} \right],
\end{equation}
which is the same result as for a quantum dot without coherent
scattering from inside the dot, see Eq.\ (\ref{eq:gz1}). In the
low temperature limit $T \ll \hbar/\tau_{\rm d} \ll E_{\rm c} g$
one finds a saturation of the interaction
correction,\cite{kn:golubev2004}
\begin{equation}
  \langle G \rangle = \frac{e^2}{2 \pi \hbar}
  \frac{g_1 g_2}{g} \left[1
  - \frac{2(g_2 s_1 + g_1 s_2)}{g^2}
  \ln \frac{E_{\rm c} g \tau_{\rm d}}{\pi} \right].
\end{equation}

For the quantum interference corrections we have performed
calculations for the linear conductance with ideal point contacts
only. Weak localization is a small difference $\delta G$ of the
ensemble averaged conductance with and without magnetic field.
We find that the interaction correction to weak localization
is zero, so that $\delta G$ is given by the non-interacting
theory,\cite{kn:jalabert1994,kn:baranger1994}
\begin{equation}
  \delta G = - \frac{\nu_{s} e^2 N_1 N_2}{2 \pi
  \hbar N^2},
\end{equation}
where $\nu_s$ denotes the spin degeneracy.

\begin{figure}
\epsfxsize=0.9\hsize
\epsffile{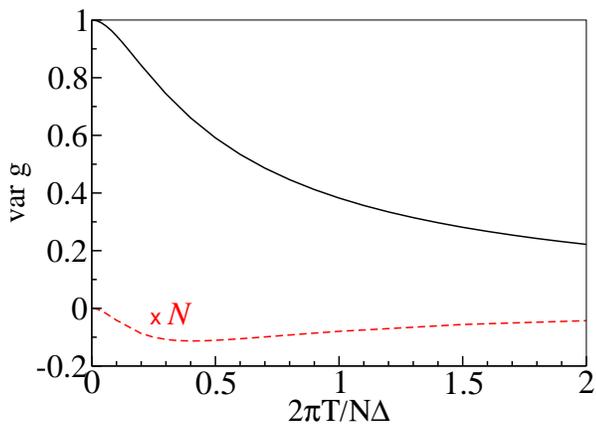}
\caption{\label{fig:variance}
The interaction correction to the variance of the conductance,
made dimensionless by writing $G = (\nu_s e^2/2 \pi \hbar)
(N_1 N_2/N^2) g$. The solid curve shows the non-interacting
contribution to the conductance fluctuations; the dashed curve
shows the interaction correction for $\pi/E_{\rm c} N \tau_{\rm d}
 = 0.1$, multiplied by $N$ to make it fit on the same scale as the
non-interacting contribution.}
\end{figure}

Closed form expressions for the interaction correction to
the conductance fluctuations could be obtained for the limiting cases
$T \ll \hbar/\tau_{\rm d} \ll N E_{\rm c}$ and $\hbar/\tau_{\rm d}
\ll T \ll N E_{\rm c}$,
for which we find, in the absence of time-reversal symmetry
\begin{eqnarray}
  \mbox{var}\, G &=&
  \left( \frac{\nu_s e^2 N_1 N_2}{2 \pi \hbar N^2} \right)^2
  \left[ 1 -  \frac{2}{3} \left(\pi T \tau_{\rm d} \right)^2
  \right. \nonumber \\ && \left. \mbox{} \times
  \left(1 + \frac{4 \pi}{E_{\rm c} N^2 \tau_{\rm d}}
  \ln \frac{E_{\rm   c} N \tau_{\rm d}}{\pi}  \right) \right].
  \nonumber \\
\end{eqnarray}
and
\begin{eqnarray}
  \mbox{var}\, G &=&
  \left( \frac{\nu_s e^2 N_1 N_2}{2 \pi \hbar N^2} \right)^2
  \frac{\pi}{6 T \tau_{\rm d}}
  \left( 1 -
  \frac{2 \pi}{5 T N \tau_{\rm d}}
  \right),\ \ \ \
\end{eqnarray}
respectively. For intermediate temperatures, the final result
still contains two energy integrations, which are easily
integrated numerically. Figure \ref{fig:variance}
shows the interaction correction to the
variance as a function of temperature for the value
$\pi/E_{\rm c} N \tau_{\rm d} = 0.1$.
In the presence of time-reversal symmetry, the variance of the
conductance a factor two higher than in the absence of
time-reversal symmetry.
We conclude that, at zero temperature, there is no interaction
correction to the conductance fluctuations, to leading order in
$1/N$. At small but finite temperatures, interactions lead to a
small decrease of the conductance fluctuations.

It is legitimate to ask how one can distinguish between the interaction
correction to weak localization and the conductance fluctuations
and the second order interaction corrections to the classical
[{\em i.e.,} of order $N$] conductance or the second order quantum 
interference corrections in the non-interacting theory, as both are 
of order $1/N$. With respect to the comparison to the second order 
interaction correction to the classical conductance, an answer is 
readily given: this interaction correction does not depend on the 
magnetic field and has no mesoscopic fluctuations as a function of,
{\em e.g.}, Fermi energy or magnetic field. A formal method
to distinguish the interaction corrections discussed above from
second-order quantum interference corrections follows after 
introducing the number of `orbital
channels' $N^{\rm o} = N/\nu_s$. Then one finds that the 
interaction correction is a $1/N$ correction, while the second-order
quantum interference correction is of order $N/(N^{\rm o})^2$.
For the conductance fluctuations in the absence of time-reversal
symmetry this problem does not arise, as there are no second
order quantum interference corrections in this
case.\cite{kn:baranger1994,kn:jalabert1994,kn:beenakker1997}

The absence of an interaction correction to weak localization, and to
the conductance fluctuations at zero temperature, is perhaps one of
the most striking predictions of our theory. In
Section~\ref{sec:conclusion}, we will give an argument to explain why
such a renormalization leaves these ensemble averaged quantities
unchanged, using the observation that the form of the conductance
correction derives from the renormalized scattering matrix
Eq.~(\ref{deltaS}).

\subsection{Incoherent dot}

The opposite limit of a quantum dot without coherent reflection
from inside the dot can be obtained from the general
formalism of Sec.\ \ref{sec:current} by setting
$S(t,t-\tau) \to \delta(\tau)$ and interpreting $r$ as the
reflection matrix $r_c$ of the point contacts. In this case,
the distribution function of the left moving fermions represents
the distribution function of electrons leaving the quantum dot.
Self-consistent determination of the corresponding chemical
potential $\mu_L$ is problematic, however,
because that would require knowledge of the
average dwell time inside the dot, which is not contained in this model.

For small $r_{\rm c}$, the results of Sec.\ \ref{sec:currentA} can
be used. In this case one finds
\begin{eqnarray} \label{eq:Ir}
  I &=& - \frac{e}{2 \pi \hbar} \tr \Lambda (1 - r_c r_c^{\dagger})
  \mu_R \\ && \mbox{} +
  \frac{1}{2} e i v_F^2 \int_0^{\infty} d\tau \sin(\kappa_0(-\tau))
  \left| \frac{f(0)^2}{f(\tau) f(-\tau)} \right|^{1/N}
  \nonumber \\ && \mbox{} \times
  \tr \Lambda
  r_c r_c^{\dagger}
  (G^{\rm K}_{R}(\tau) G^{\rm K}_{L}(-\tau) -
  G^{\rm K}_{L}(\tau) G^{\rm K}_{R}(-\tau)).
  \nonumber
\end{eqnarray}
For $N=2$ there is an additional oscillating interaction correction
to the current. Equation (\ref{eq:Ir}) is the non-equilibrium
generalization of the original linear response theory of Refs.\
\onlinecite{kn:flensberg1993,kn:furusaki1995a,kn:furusaki1995b}.
For the linear conductance one then finds
\begin{eqnarray}
  G &=& \frac{e^2}{2 \pi \hbar}\left[ \frac{N_1 N_2}{N} -
  \mbox{tr}\, \Lambda^2 \rc \rcd
  \left( 1 + \frac{2}{N}
  \ln \frac{E_{\rm c} N e^{1 +{\rm C}}}{2 \pi^2 T}
  \right) \right]
  \nonumber \\
  \label{eq:GThigh}
\end{eqnarray}
if $\ln(E_{\rm c}/\pi T) \ll N/2$ but $T \ll E_{\rm c} N$ and
\begin{eqnarray}
  G & = &
  \frac{e^2}{2 \pi \hbar}\left[ \frac{N_1 N_2}{N} -
  \frac{1}{2 \pi^{1/2}}
  \Gamma \left(1- \frac{1}{N} \right)
  \Gamma \left(\frac{1}{N}-\frac{1}{2} \right)
  \nonumber \right. \\ && \left. \mbox{} \times
  \left( \frac{E_{\rm c} N e^{\rm C}}{\pi^2 T} \right)^{2/N}
  \mbox{tr}\, \Lambda^2 \rc \rcd
  \cos \frac{\pi}{N}
  \right]
\end{eqnarray}
if $\ln(E_{\rm c}/\pi T) \gg N/2$ (but temperature larger than a
suitable lower limit\cite{kn:furusaki1995a,kn:furusaki1995b,kn:aleiner2002}).

For large $N$, one recovers the same expressions as Golubev and
Zaikin, see Eq.\ (\ref{eq:gz1}) and Ref.\ \onlinecite{kn:golubev2001},
starting from the results of Sec.\ \ref{sec:currentB}.

\section{Time-dependent transport:
Adiabatic approximation and linear response}

\label{sec:adiabatic}

In this section we will consider the case of a fully coherent dot
only. We will limit ourselves to a theory of adiabatic and linear
response transport --- the response to either a slowly varying
internal potential or to a small bias voltage, but not both ---,
and consider the average and variance of the current only.

We will present our final results in terms of the Fourier
transform of the scattering matrix,
\begin{eqnarray}
  S(t;\varepsilon) &=&
  \int d\tau e^{i \varepsilon \tau} S(t+\tau/2,t-\tau/2),
  \nonumber \\
  S^{\dagger}(t;\varepsilon) &=&
  \int d\tau e^{i \varepsilon \tau} S^{\dagger}(t+\tau/2,t-\tau/2).
  \label{eq:Stransform}
\end{eqnarray}
In the adiabatic approximation, the Fourier transform
$S(t;\varepsilon)$ satisfies the unitarity condition
\begin{eqnarray}
  (1 - (i/2) D_{t,\varepsilon}) S(t;\varepsilon)
  S^{\dagger}(t;\varepsilon)
  = 1,
\end{eqnarray}
where $D_{t,\varepsilon} A B = (\partial A/\partial t)
(\partial B/\partial \varepsilon) - (\partial A/\partial \varepsilon)
(\partial B/\partial t)$. To leading order in the adiabatic
approximation, the Fourier transform $S(t;\varepsilon)$ is equal
to the `frozen' scattering matrix, taken by fixing the internal
potentials at the value they have at time $t$. However, the
`frozen' scattering matrix is unitary, whereas $S(t;\varepsilon)$
is not. Since our final expression for the current will be
manifestly of first order in combined linear response and adiabatic
approximation, we can neglect the difference between
$S(t;\varepsilon)$ and the `frozen' scattering matrix in our
final expressions.

We will set the chemical potential $\mu_L$ of the electrons in the
fictitious reservoir equal to zero for all times. In the adiabatic
approximation and for linear transport, the Keldysh Green function
$G^{\rm K}_{mR}(t,s)$ depends on the time difference $t-s$ only,
$m=1,\ldots,N$, and reads
\begin{eqnarray}
  G^{\rm K}_{mnR}(t+\tau/2,t-\tau/2) &=&
  - \frac{T}{v_F}\delta_{mn}
  \\ && \nonumber \mbox{} \times
  {\rm P}\, \frac{e^{-i (\mu_{mR}(t) + e V_{\rm d}(t))
  \tau}}
  {\sinh(\pi T \tau)},
\end{eqnarray}
where $V_{\rm d}$ is the dot potential, see Eq.\ (\ref{eq:Ivd}).

Calculating the interaction function $\kappa$ and the interaction
correction $\alpha$ in the adiabatic limit is problematic. The reason
is that both quantities involve long time scales during which one
cannot assume that the derivative of the scattering matrix is constant
(as one does in the adiabatic approximation).  However, since $\kappa$
and the interaction correction $\alpha$ contain traces, they are
self-averaging, and we can replace them by their ensemble
averages. Taking the ensemble average in the defining equation
(\ref{eq:kappaone}) for $\kappa(t,s)$, using
the known probability distributions of time-dependent scattering
matrices,\cite{kn:polianski2003} the equation for $\kappa(t,s)$
becomes
\begin{eqnarray}
  \kappa(t,s) &=& \kappa_0(t-s)
  + \frac{E_{\rm C} N}{\pi} \int_{t} dt_2 \int_{t}^{t_2} dt_2'
   \kappa(t_2,s)
  \nonumber \\ && \mbox{} \times
  e^{-E_{\rm C} N(t_2'-t)/\pi} e^{-(t_2-t_2')/\tau_{\rm d}}.
\end{eqnarray}
This equation has no reference to the distribution functions in the
lead or the time dependence of the scattering matrix. Hence we find
that the solution is of the equilibrium form $\kappa(t-s)$, the
Fourier transform of which is given in Eq.\ (\ref{eq:kappaomega}).
For the interaction correction $\alpha(t+\tau/2,t-\tau/2)$, we note
that it is an even function of $\tau$. This implies its ensemble
average must vanish if we consider the adiabatic approximation and
linear response only.

We are now in a position to calculate the self-consistent
potential $V_{\rm d}$ and the current $I_j$. The self-consistent
equation for the potential $V_{\rm d}$ reads
\begin{widetext}
\begin{eqnarray}
  e \frac{\partial V_{\rm d}}{\partial
  t} &=&
  - \frac{E_{\rm c}}{\pi}
  \int d\varepsilon
  \left( - \frac{ \partial f(\varepsilon - \mu_{\rm c})}{\partial
  \varepsilon} \right)
  \tr \left\{ i \frac{\partial \mu_{\rm c}}{\partial t}
  S(t;\varepsilon) \frac{\partial S^{\dagger}(t;\varepsilon)}
  {\partial \varepsilon}
  + i S(t;\varepsilon) \frac{\partial S^{\dagger}(t;\varepsilon)}
  {\partial t}
  \right. \nonumber \\ && \ \ \left. \mbox{}
  + \frac{1}{\pi} \mbox{Re}\,
  \int d\omega \kappa(\omega)
  (2 f(\varepsilon + \omega - \mu_{\rm c}) - 1)
  \left( \frac{\partial \mu_{\rm c}}{\partial t}
  \frac{\partial}{\partial \varepsilon}
  + \frac{\partial}{\partial t} \right)
  [S(t;\varepsilon) S^{\dagger}(\varepsilon + \omega)]
  \right\},
  \label{eq:mueq1}
\end{eqnarray}
where the distribution function $f$ is evaluated with respect to the
electrochemical potential
\begin{equation}
  \mu_{\rm c} =  e V_{\rm d} + \frac{1}{N} \sum_{k} \mu_{kR}.
\end{equation}
The first terms give the mean-field contribution to $V_{\rm d}$.
The last term gives the interaction correction. Upon taking the
ensemble average, which is appropriate if $V_{\rm d}$ enters as a
variable in the expression for the current, the last term
vanishes.

For the current we find
\begin{eqnarray}
  I_j(t) &=&
  \frac{C}{N} \frac{\partial V_{\rm d}}{\partial t}
   +  \frac{e}{2 \pi} \int d\varepsilon
  \left( -
  \frac{\partial f(\varepsilon  - \mu_{\rm c})}{\partial
  \varepsilon} \right)
  \tr \left(\delta_{jk} - \frac{1}{N} \right)
  \nonumber \\ && \mbox{} \times
   \left\{j(t;\varepsilon)
  + \frac{1}{\pi} \mbox{Im}\, \int d\omega \kappa(\omega)
  \left(2 f(\varepsilon + \omega - \mu_{\rm c})-1 \right)
    S(t;\varepsilon) S^{\dagger}(t;\varepsilon+\omega)
    (j(t;\varepsilon+\omega) - j(t;\varepsilon))
  \right\},
  \label{eq:currentad}
\end{eqnarray}
where we abbreviated
\begin{eqnarray}
  j(t;\varepsilon) &=&
  \mu_R - S(t;\varepsilon) \left( \mu_R
  - i \frac{\partial \mu_{\rm c}}{\partial t}
  \frac{\partial}{\partial \varepsilon}
  - i \frac{\partial}{\partial t} \right)
  S^{\dagger}(t;\varepsilon).
\end{eqnarray}
In the expressions for $I_j(t)$ and $\partial V_{\rm d}(t)/\partial
t$ one may replace
the Fourier transformed scattering matrix $S(t;\varepsilon)$ by the
`frozen' scattering matrix. Then a time derivative
$\partial/\partial t$ is replaced by $(\partial X/\partial t)(
\partial/\partial X)$, where $X$ is a (set of) parameter(s) that
characterizes the potentials in the dot.

In Eq.\ (\ref{eq:currentad}) the terms proportional to $\mu_R$ give
the dc conductance of the dot, {\em cf.}
Eq.\ (\ref{eq:lincond}). The first line of Eq.\
(\ref{eq:currentad}) is nothing but the Landauer formula with a
self-consistent potential in the dot; the last lines give the
correction from interactions beyond the mean-field level. The
self-consistent potential $V_{\rm d}$ was omitted from Eq.\
(\ref{eq:lincond}) because it cannot be determined from the
steady state formulation of the problem only. The terms
proportional to the time derivative give the emissivity $-dQ/dX$
of the quantum
dot. Again, the first two lines of Eq.\ (\ref{eq:currentad})
correspond to the emissivity according to the self-consistent
(Hartree) theory of B\"uttiker and coworkers \cite{kn:buettiker1994}
whereas the remaining lines of that equation give the correction from
interactions beyond the mean-field level. The governing
equation (\ref{eq:mueq1}) for the self-consistent potential
$V_{\rm d}(t)$ is the same as that of the Hartree theory of
Ref.\ \onlinecite{kn:buettiker1994}. It depends both on
(changes in) the dot potential and the chemical potential in the
lead. After taking the ensemble average there are no
interaction corrections to the Hartree theory for $V_{\rm d}$.

Although the above expression for the current was derived in
linear response and in the adiabatic approximation, it can be used
to calculate the electrochemical capacitance $C_{\rm c}$ of the
quantum dot, the derivative of the charge on the dot to a uniform
shift if chemical potential $\mu_R$. This is possible because, for
the Hamiltonian $H_0$, slowly increasing the chemical potential of
all channels simultaneously corresponds to a time-independent
small shift of the electrochemical potential, see Eq.
(\ref{eq:thetamu}). Hence, the time derivatives of the phases
carried by the bare Green functions $G_R$ and $G_L$ differ only by
a small amount, and linear response is meaningful. Considering the
rate of change of the charge on the dot for the case in which
there is no explicit time dependence of the dot scattering matrix,
one finds from Eq.\ (\ref{eq:currentad})
\begin{eqnarray}
  \frac{d Q}{d t}
  &=& - C \frac{\partial V_{\rm d}}{\partial t},
\end{eqnarray}
where $\partial V_{\rm d}/\partial t$ can be calculated from Eq.\
(\ref{eq:mueq1}). Upon changing all chemical potentials for incoming
electrons simultaneously by an amount $-e V(t)$, we
find\cite{kn:buettiker1993a}
\begin{eqnarray}
  C_{\rm c} &=& \frac{dQ}{dV} \nonumber \\ &=&
  \left[C^{-1} + (e^2 dn/d\varepsilon)^{-1} \right]^{-1},
\end{eqnarray}
where $dn/d\varepsilon$ is the density of states of the dot,
\begin{eqnarray}
  \frac{dn}{d\varepsilon} &=&
  \frac{1}{2 \pi} \int d\varepsilon
  \left( - \frac{\partial f(\varepsilon - \mu_{\rm c})}{\partial
  \varepsilon} \right)
  \mbox{tr}\,
  \left[ i S(\varepsilon)
  \frac{\partial S^{\dagger}(\varepsilon)}{\partial \varepsilon}
  + \frac{1}{\pi} \mbox{Re}\,
  \int d\omega \kappa(\omega)
  (2 f(\varepsilon + \omega - \mu_{\rm c}) - 1)
  \frac{\partial}{\partial \varepsilon}
  S(\varepsilon) S^{\dagger}(\varepsilon + \omega)
  \right].\ \ \
  \nonumber \\
\end{eqnarray}
\end{widetext}
Here, the first term is the Hartree expression for the density of
states of an open quantum
dot;\cite{kn:buettiker1993a,kn:buettiker1993b} the second term is
the interaction correction.

\begin{figure}
\epsfxsize=0.9\hsize
\epsffile{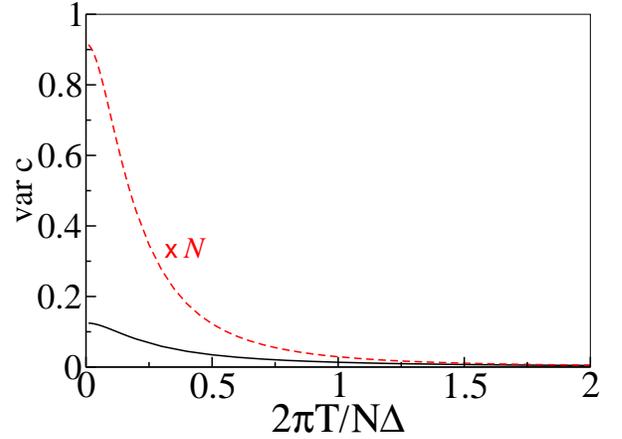}
\caption{\label{fig:capvariance}
The variance of the mean-field contribution to the
capacitance, made dimensionless by writing
$C_{\rm c} = 2 \pi e^2 c/(E_{\rm c}^2 N^2 \tau_{\rm d})$ (solid curve)
and the interaction correction to the variance of the electrochemical
capacitance (dashed), as a function of temperature. The interaction
correction to the variance has been multiplied by $N$ in order to
fit on the same vertical scale.}
\end{figure}

The ensemble average of the interaction correction to the density
of states, and hence of the interaction correction to the
electrochemical capacitance, is zero. It has, however, mesoscopic
fluctuations. Calculating the interaction correction to the
variance of the density of states, we find, for zero temperature,
\begin{equation}
  \mbox{var}\, (dn/d\varepsilon)
  = \frac{\nu_{\rm s}^2 \tau_{\rm d}^2}{2 \pi^2}
  \left(1 + \frac{22}{3 N} \right),
  \label{eq:vardnde}
\end{equation}
from which one concludes, for $\hbar/\tau_{\rm d} \ll E_{\rm c} N$,
\begin{equation}
  \mbox{var}\, C_{\rm c} = \frac{\pi^2 e^4 \nu_{\rm s}^2}
  {2 E_{\rm c}^4 N^4 \tau_{\rm d}^2}
  \left(1 + \frac{22}{3 N} \right). \label{eq:varcc}
\end{equation}
Equations (\ref{eq:vardnde}) and (\ref{eq:varcc}) are valid in the
absence of time-reversal symmetry. With time-reversal symmetry,
the variance is a factor two larger. The first
term is the variance of the mean-field contribution to the
capacitance,\cite{kn:fyodorov1997,kn:brouwer1997b}
the second term is the interaction correction.
Figure \ref{fig:capvariance} shows the temperature dependence of
the variance of the electrochemical capacitance.

If the quantum dot is operated as a quantum pump, two shape-defining
parameters $X_1$ and $X_2$ are varied periodically in time, whereas
the bias voltages are kept zero. We
consider a harmonic time dependence for the parameters $X_1$
and $X_2$,
\begin{eqnarray}
  X_1(t) &=& \delta X_1 \sin (\omega t), \nonumber \\
  X_2(t) &=& \delta X_2 \sin (\omega t + \phi).
\end{eqnarray}
The charge pumped in one cycle
$Q_{\rm pump}$ is defined as the integral of the current in one
of the point contacts. Integrating over one cycle, one then
finds, for small pumping amplitudes\cite{kn:brouwer1998}
\begin{eqnarray}
  Q_{\rm pump} &=&   (\pi \delta X_1 \delta X_2 \sin \phi)^2
  \nonumber \\ && \mbox{} \times
  \left( \frac{\partial}{\partial X_1}
  \frac{dQ_{\rm pump}}{dX_2} -
  \frac{\partial}{\partial X_2} \frac{dQ_{\rm pump}}{dX_1} \right).
  \label{eq:Qpump}
  \ \ \ \ \
\end{eqnarray}
[For large amplitudes $\delta X_1$ and $\delta X_2$ one has to
average Eq.\ (\ref{eq:Qpump}) over the area enclosed in the
two-dimensional space spanned by the parameters $X_1$ and
$X_2$.]

Since the total current is conserved,
we may calculate $Q_{\rm pump}$ as the integral of $I(t) =
\sum_{j} I_{j}(t) \Lambda_{jj}$ over one period, {\em i.e.,}
$(N_2/N)$ times the
integral over the current through the first point contact minus
$(N_1/N)$ times the current in the second point contact. The
ensemble average $\langle Q_{\rm pump} \rangle$ is zero for
symmetry reasons.
When calculating the variance of the pumped charge, we note that,
up to corrections
of order $1/N^2$, one may replace the average of a product of
traces by the product of the averages, with the exception of the
two traces that contain the matrix $\Lambda$. Their average is
zero, and one needs to consider the average of the product. This
means that all terms proportional to $\partial V_{\rm d}/\partial
t$ can be dropped, and one may set
\begin{widetext}
\begin{eqnarray}
  \frac{dQ_{\rm pump}}{dX} &=&
  - \frac{e}{2 \pi} \int d\varepsilon
  \left( - \frac{\partial f(\varepsilon)}{\partial \varepsilon}
  \right)
  \tr \left\{ i \Lambda S(\varepsilon) \frac{\partial
  S^{\dagger}(\varepsilon)}{\partial X}
  \right. \nonumber \\ && \left. \mbox{}
  + \frac{1}{\pi} \mbox{Re}\,
  \int d\omega \kappa(\omega)
  (2 f(\varepsilon + \omega) -1 ) \mbox{tr}\,
  \Lambda S(\varepsilon)
  \left[\frac{\partial
  S^{\dagger}(\varepsilon + \omega)}{\partial X} -
  S^{\dagger}(\varepsilon +
  \omega) S(\varepsilon) \frac{\partial
  S^{\dagger}(\varepsilon)}{\partial X}
  \right]
  \right\}.
\end{eqnarray}
\end{widetext}
Using the results of appendix \ref{app:a} to perform the ensemble
average, we find that, at
zero temperature and for small pumping
amplitudes, the variance of the pumped charge is given by
\begin{eqnarray}
  \mbox{var}\, Q_{\rm pump} = \frac{4 N_1 N_2 \nu_{\rm s}^4}{N^4}
  \left(1 + \frac{2}{N} \right)
  (\delta X_1 \delta X_2 \sin \phi)^2,\ \ \
  \label{eq:varqpump}
\end{eqnarray}
irrespective of the presence or absence of a time-reversal
symmetry breaking magnetic field. Here we have used the
standard convention (within random matrix theory) to relate
the effect of the shape-defining parameters $X_1$ and $X_2$ to
a parametric motion of the scattering matrix.
The first term in Eq.\ (\ref{eq:varqpump})
is the leading non-interacting contribution
to the variance;\cite{kn:brouwer1998,kn:shutenko2000} the
second term is the interaction correction. For temperatures
$T \gg \hbar/\tau_{\rm d}$ we find
\begin{eqnarray}
  \mbox{var}\, Q_{\rm pump} &=&
  (\pi \delta X_1 \delta X_2 \sin \phi)^2 \nonumber \\ &&
  \mbox{}
  \times
  \frac{N_1 N_2 \nu_{\rm s}^4}{12 T N^4 \tau_{\rm d}}
  \left(1 + \frac{\pi}{T N \tau_{\rm d}} \right),
\end{eqnarray}
where, again, the first term is the leading non-interacting
contribution to the variance of the pumped
current\cite{kn:shutenko2000,kn:vavilov2001a} and the second
term is the interaction correction.

\section{Conclusion}
\label{sec:conclusion}

In this paper we have presented a systematic theory to evaluate
the effects of electron-electron interactions on quantum transport
through open quantum dots, accommodating arbitrary bias and time
dependence of the dot's scattering matrix $S(t,t')$. Our result
takes the form of an expression for the sample-specific current
through a quantum dot in terms of $S(t,t')$. This expression has
been obtained using a systematic expansion in $1/N$, where $N$ is
the total number of channels in the point contacts connecting the
dot to the electron reservoirs. To leading order in $1/N$, our
theory reproduces what one finds if the interactions were treated
in the Hartree approximation. The subleading-in-$1/N$ corrections
represent contributions to the current through the dot that cannot
be described by means of a self-consistent potential.

Upon taking the average over an ensemble of chaotic
quantum dots, we have been able to calculate the interaction
corrections to weak localization, universal conductance fluctuations,
capacitance fluctuations, and to the pumped current in case the
quantum dot is operated as a quantum pump.
For a quantum dot with ideal point contacts, we found that the
interaction correction to the
weak localization correction to the conductance vanishes, as well as the
interaction correction to the conductance fluctuations at zero
temperature. There is a negative interaction correction to the
conductance fluctuations at finite temperature. There are
positive interaction corrections to the capacitance fluctuations
and to the fluctuations of the pumped current, but no corrections
to their average. All interaction corrections are small as $1/N$
in comparison to the non-interacting ({\em i.e.,} Hartree) contributions.

The problem of `weak Coulomb blockade' in open quantum dots has
been addressed previously in the
literature.\cite{kn:brouwer1999c,kn:aleiner2002,kn:golubev2004}
Our findings for the quantum interference corrections, as well as
our expression for the sample-specific dc conductance and
capacitance, differ from those of Refs.\
\onlinecite{kn:brouwer1999c,kn:aleiner2002}. To explain the
difference, we have shown that the formal expansion in the
scattering matrix, used in Refs.\
\onlinecite{kn:brouwer1999c,kn:aleiner2002}, cannot be used to
describe a coherent quantum dot. We believe that the expressions
presented here, which were obtained using a systematic expansion
in $1/N$, are correct. Our sample-specific expression agrees with
that obtained in Ref.\ \onlinecite{kn:golubev2004}, although Ref.\
\onlinecite{kn:golubev2004} does not consider the quantum
corrections to the conductance.

The absence of an interaction correction to weak localization and
to the conductance
fluctuations at zero temperature can be understood using a
simple argument: Since the charging interaction cannot lead to
dephasing\cite{kn:brouwer1999c} --- it corresponds to a time-dependent
{\em uniform} shift of the dot's potential ---, its only effect is a
renormalization $S(\varepsilon) \to S'(\varepsilon)$
of the scattering matrix $S(\varepsilon)$ of the dot.
To leading order in perturbation theory, such a
renormalization was calculated explicitly in Sec.\ \ref{sec:fock} and
in Ref. \onlinecite{kn:golubev2004}.
Like the probability distribution functional $P[S(\varepsilon)]$ of the
`bare' energy-dependent scattering matrix $S(\varepsilon)$, the probability
distribution functional $P[S'(\varepsilon)]$ of the renormalized scattering
matrix $S'(\varepsilon)$ is invariant under left and right
multiplication of $S'(\varepsilon)$
with energy-independent unitary matrices $U$ and
$V$,\cite{kn:wigner1951a,kn:wigner1951b,kn:wigner1952,kn:brouwer1999a}
\begin{equation}
  P[S'(\varepsilon)] = P[U S'(\varepsilon) V].
  \label{eq:wigner}
\end{equation}
In the presence of time-reversal symmetry, $U =
V^{\rm T}$. The statistical invariance
(\ref{eq:wigner}), together with the same invariance for
$P[S(\varepsilon)]$, implies that the distributions of
both $S(\varepsilon)$ and $S'(\varepsilon)$ for a fixed energy
$\varepsilon$ are those of the circular ensemble from random matrix
theory.\cite{kn:mehta1991,kn:beenakker1997}
Since the weak localization correction to the conductance and
the conductance fluctuations at zero temperature do not depend on statistical
correlations between scattering matrices $S(\varepsilon)$ at different energies, we conclude
that these are given by the circular ensemble
averages,\cite{kn:baranger1994,kn:jalabert1994} and, hence,
 that there is no
interaction correction. There is an interaction correction, however,
to the conductance fluctuations at a finite temperature, which do involve
such correlations. The above argument does not apply to a quantum dot with
nonideal leads, for which the invariance property (\ref{eq:wigner})
does not hold, or to a quantum dot with relaxation, for which the
scattering matrix is sub-unitary and the invariance property
(\ref{eq:wigner}) is not sufficient to fix the distribution.

How reliable is the expansion in $1/N$ and how significant are the
interaction corrections, as they are a factor $1/N$ smaller than
the Hartree contributions? Recalling that $N$ is the total number
of channels in the point contacts connecting the dot and the
electron reservoirs, not counting spin degeneracy, one has $N=4$
for a dot with two spin-degenerate single-mode point contacts.
Although $N=4$ cannot be classified as `large', we believe that
for $N=4$ the $1/N$ expansion should give a reliable estimate of
the interaction correction, whereas, on the other hand, the
interaction corrections should still be significant (if they are
nonzero). Measurements of the conductance distribution in
single-mode quantum dots have shown excellent agreement with the
(non-interacting) random matrix theory down to the lowest
temperatures.\cite{kn:huibers1998} This experimental observation
is in agreement with our theory since, to leading order in $1/N$
and at zero temperature, we find no interaction correction to the
average and variance of the conductance. The argument following
Eq.\ (\ref{eq:wigner}) is extremely suggestive and may hint that
the agreement with random matrix theory is not restricted to the
approximations of this paper, though we caution that higher orders
in the interaction are presumably not simply described by an
effective $S'(\varepsilon)$ due to inelastic processes. As long as
the Fermi Liquid picture holds, however, one may expect this
description to be restored at zero temperature.

The interaction correction we calculated here is the result of the
capacitive interaction only. For large $N$, it is a factor $\sim 1/N$
smaller than the leading non-interacting (Hartree) contribution to the
transport properties and their quantum interference corrections. We
have not considered other contributions to the electron-electron
interaction. These were omitted from the Hamiltonian of the quantum
dot because they are much smaller than the capacitive interaction if
the dimensionless conductance of the dot is
large.\cite{kn:aleiner1998a,kn:kurland2000,kn:aleiner2002} The
interactions we omitted will give additional corrections to the
dot's transport properties, {\em e.g.,} because they lead to
dephasing. To the best of our knowledge,
a microscopic theory of the effect these residual
interactions on the transport properties of a chaotic quantum dot
still has to be developed.

In addition to the interaction corrections studied here, which
are not periodic in the dimensionless gate voltage ${\cal N}$,
there exist periodic-in-${\cal N}$ interaction corrections to the
dot's capacitance and
conductance.\cite{kn:aleiner1998a,kn:brouwer1999c,kn:aleiner2002}
 These occur to $N$th order in
perturbation theory, which rules out a large-$N$ approach.
Periodic-in-gate-voltage interaction corrections have been
studied using an expansion in the scattering
matrix.\cite{kn:aleiner1998a,kn:aleiner2002} Although we have
shown that such an expansion is not justified for aperiodic
interaction corrections, we must leave the question of its
validity for the periodic-in-${\cal N}$ corrections unanswered.

Before closing, we would like to make two remarks on the relation
of our work to that of Golubev and Zaikin.\cite{kn:golubev2004}
First, Golubev and Zaikin consider both the case of a
`voltage biased' and a `current biased' quantum dot. The latter
case corresponds to a quantum dot in series with a large ohmic
resistor. In Ref.\ \onlinecite{kn:golubev2004} it is shown
that the interaction corrections
in the two cases are qualitatively different. Our calculation, in which
the chemical potentials of the electrons coming in through the
point contacts are kept fixed, corresponds to the `voltage
biased' case. Second, Golubev and Zaikin argue that the sole
effect of interactions as a renormalization of the reflection
matrix of the point contacts $r_{\rm c}$ only. While this is
true for an incoherent
dot,\cite{kn:matveev1993,kn:yue1994,kn:kindermann2003,kn:bagrets2004}
our calculation
shows that this is not correct for a coherent dot: When
quantum interference corrections are taken into account, the
renormalization of the scattering matrix is more complicated,
and involves more than one energy. This is manifested, {\em e.g.},
in the existence of an interaction correction to the conductance
fluctuations at finite temperature.

\acknowledgments

We would like to thank I.\ Aleiner, K.\
Matveev, and L.\ Glazman for
discussions. This work was supported by the NSF under grant no.\
DMR-0334499 and by the Packard Foundation.

\appendix

\section{Bosonization}
\label{app:boson}

In this section we calculate the contour-ordered correlation
functions of the fermion fields in the Hamiltonian $\hat H_0$ of
Eq.\ (\ref{eq:pc1d2}). The bosonization method allows for an exact
calculation of the fermion correlation functions for the
Hamiltonian $\hat H_0$, despite the fact that $\hat H_0$ is
quadratic in the fermion operators:\cite{kn:haldane1981}  $\hat
H_0$ is rewritten in terms boson fields $\hat \varphi_{jL}(x)$ and
$\hat \varphi_{jR}(x)$. In the bosonized language $\hat H_0$ is
diagonalized easily, and the fermion correlators can be found from
the boson correlation functions.

The boson fields $\hat \varphi_{jL}(x)$ and $\hat \varphi_{jR}(x)$ are
related to the fermion fields $\hat \psi_{jL}(x)$ and $\hat
\psi_{jR}(x)$ as
\begin{eqnarray}
  \hat{\psi}_{jL} &=&
  \frac{\hat{\eta}_j}{\sqrt{2 \pi \lambda v_F}} e^{-i
  \hat{\varphi}_{jL}},\ \ j=1,\ldots,N, \\
  \hat{\psi}_{jR} &=&
  \frac{\hat{\eta}_j}{\sqrt{2 \pi \lambda v_F}}
  \lefteqn{e^{ i \hat{\varphi}_{jR}},}
  \hphantom{e^{-i \hat{\varphi}_{jL}},\ } \ j=1,\dots, N.
\label{eq:psiphi}
\end{eqnarray}
They obey the commutation rules
\begin{eqnarray}
  \left[ \hat{\varphi}_{jL}(x), \hat{\varphi}_{Li}(y)\right] &=&
  -i \pi \delta_{ij}
  \sign(x-y), \nonumber \\
  \left[ \hat{\varphi}_{jR}(x), \hat{\varphi}_{Ri}(y)\right] &=&
  i \pi \delta_{ij}
  \sign(x-y), \nonumber \\
  \left[ \hat{\varphi}_{jR}(x), \hat{\varphi}_{Li}(y)\right] &=&
  -i \pi \delta_{ij}.
  \label{eq:commboson}
\end{eqnarray}
In Eq.\ (\ref{eq:psiphi}) $\hat{\eta}_j = \hat{\eta}_j^{\dagger}$ is a Majorana
fermion, $\ \left\{\hat{\eta}_j, \hat{\eta}_i\right\}=2 \delta_{ij}$,
and $\lambda$ is a cut-off time that is taken to zero at the end
of the calculation.
Since the Majorana fermions do not enter into the Hamiltonian, their
average is given by
\begin{equation}
  \langle T_{\rm c} \hat{\eta}_j(t), \hat{\eta}_i(s)\rangle =
  \delta_{ij}\sign_{\rm c} \left(t-s\right),
  \label{eq:etacorr}
\end{equation}
where the sign $\sign_{\rm c}$ is defined with respect
to the ordering along the integration contour c.

The (normal-ordered) densities of left and right moving fermions
are proportional to the derivatives of the boson fields,
\begin{eqnarray}
  :\! \hat{\psi}_{jL}^{\dagger}(x)
  \hat{\psi}_{jL}^{\vphantom{\dagger}}(x)\! :\,
    &=& \frac{1}{2 \pi} \frac{\partial}{\partial x}
  \hat{\varphi}_{jL}(x),\\
  :\! \hat{\psi}_{jR}^{\dagger}(x)
  \hat{\psi}_{jR}^{\vphantom{\dagger}}(x)\! :\,
    &=& \frac{1}{2 \pi} \frac{\partial}{\partial x} \hat{\varphi}_{jR}(x).
\end{eqnarray}
Hence, the number of electrons on the quantum dot is simply expressed
as the sum of left-moving and right-moving boson fields at
$x=0$,
\begin{equation}
  \hat N_{\rm dot} = - \frac{1}{2 \pi} \sum_{j=1}^{N}
  (\hat{\varphi}_{jL}(0) + \hat{\varphi}_{jR}(0)).
\end{equation}
There are no boundary terms at $x=-\infty$ because the interaction
extends over the entire lead. The reference electron number $N_{\rm
  ref}$
has been absorbed in the definition of normal-ordering of the
fermion operators.
Then, once expressed in terms of the
 boson fields, the Hamiltonian $\hat H_0$ becomes quadratic,
\begin{eqnarray}
  \hat H_0 &=&
  \frac{v_F}{4\pi}
\sum_{j=1}^{N}\int_{-\infty}^{\infty}dx
\left[
\left(\frac{\partial \hat{\varphi}_{jL}}{\partial x}\right)^2 +
\left(\frac{\partial \hat{\varphi}_{jR}}{\partial x}\right)^2
\right] \label{eq:50.6}
 \nonumber \\ && \mbox{}
  + \frac{E_{\rm c}}{4\pi^2}
\left[ \sum_{j=1}^{N} \left(
\hat{\varphi}_{jL}(0) +
\hat{\varphi}_{jR}(0) + 2\pi{\cal N} \right)
\right] ^2. \nonumber \\
\end{eqnarray}

The Hamiltonian is supplemented with the condition that the
chemical potential of left-moving fermions moving towards the
dot-lead interface at $x = 0$ is equal to $\mu_{jL}(t)/2 \pi v_F$ and
that the density of right-moving fermions is equal to $\mu_{jR}(t)/2
\pi v_F$. We generate the electron densities corresponding to these
chemical potentials by the inclusion of a time-dependent
forward-scattering potential in the Hamiltonian. This potential is
located at a distance $a$ from the dot-lead interface
at $x=0$. (The precise value of $a$ is not important for the
subsequent considerations, as long as $a > 0$.)
In boson language it has the form
\begin{eqnarray}
  \hat H'(t) &=& \frac{v_F}{2 \pi}
  \sum_{j=1}^{N} m_{jL}(t)
  \left. \frac{\partial \hat \varphi_{jL}(x)}{\partial x} \right|_{x \to
  a}
  \nonumber \\ && \mbox{}
  - \frac{v_F}{2 \pi}
  \sum_{j=1}^{N} m_{jR}(t)
  \left. \frac{\partial \hat \varphi_{jR}(x)}{\partial x} \right|_{x \to
  - a},
\end{eqnarray}
where $m_{jL}$ and $m_{jR}$ are the integrals of the chemical
potentials
$\mu_{jL}$ and $\mu_{jR}$, respectively,
\begin{eqnarray}
  m_{jL}(t) &=& \int^{t+a/v_F}_{-\infty} dt' \mu_{jL}(t'),\ \
  j=1,\ldots,N, \nonumber \\
  m_{jR}(t) &=& \int^{t+a/v_F}_{-\infty} dt' \mu_{jR}(t'),\ \
  j=1,\ldots,N.\ \ \
\end{eqnarray}
We assume that the system is in equilibrium, $\mu_{jL}(t) =
\mu_{jR}(t) = 0$, $j=1,\ldots,N$ for times $t$ smaller than a
reference time $t_{\rm ref}$.

In order to find correlation functions of fermion creation
and annihilation operators, expressions of the form
$T_{\rm c} e^{a}$ need to be averaged over the boson fields,
where $a$ is an arbitrary linear combination of the boson
fields $\varphi_{jL}$ and $\varphi_{jR}$, $j=1,\ldots,N$, at
different times. Since the Hamiltonian is quadratic in the
boson fields, such an average can be calculated as
\begin{equation}
  \langle \exp(\hat{a}) \rangle =
\exp\left\{\langle\hat{a}\rangle + \frac{1}{2}
\left[\langle \hat {a}^2 \rangle - \langle \hat{a} \rangle^2
\right] \right\},
  \label{eq:expavg}
\end{equation}
which means that only contour-ordered correlators of up to
two boson fields at different times are needed. Hence, it remains
to calculate the correlation functions of the boson fields.

For this calculation, it is necessary that we
transform the boson fields such that the charge mode is
separated from modes that do not charge the quantum dot.
The corresponding basis change reads
\begin{eqnarray}
  \varphi_{\mu L} = \sum_{j} o_{\mu j} \varphi_{j L},\ \
  \varphi_{\mu R} = \sum_{j} o_{\mu j} \varphi_{j R},
\end{eqnarray}
where the transformed fields carry a Greek index and the original
fields carry a roman index. Here the $o_{\mu j}$ are real numbers
satisfying
\begin{eqnarray}
  \sum_{j} o_{\mu j} o_{\nu j} &=& \delta_{\mu \nu},
  \ \
  \sum_{\mu} o_{\mu j} o_{\mu i} = \delta_{ij}.
  \nonumber
\end{eqnarray}
and
\begin{eqnarray}
  o_{1 j} &=& N^{-1/2},\ \ j=1,\ldots,N.
\end{eqnarray}
With this choice, the transformed fields $\hat \varphi_{1L}$ and
$\hat \varphi_{1R}$ describe a mode that charges the
quantum dot, whereas all other fields $\hat \varphi_{\mu L}$ and
$\hat \varphi_{\mu R}$, $\mu=2,\ldots,N$, do not involve charging
of the dot.
The transformed boson fields $\hat{\varphi}_{\mu L}$ and $\hat{\varphi}_{\mu
  R}$ satisfy the same boson commutation relations as the original
fields $\hat{\varphi}_{j L}$ and $\hat{\varphi}_{j R}$,
\begin{eqnarray}
  \left[ \hat{\varphi}_{\mu L}(x), \hat{\varphi}_{\nu L}(y)\right] &=&
  -i \pi \delta_{\mu\nu}
  \sign(x-y), \nonumber \\
  \left[ \hat{\varphi}_{\mu R}(x), \hat{\varphi}_{\nu R}(y)\right] &=&
  i \pi \delta_{\mu\nu}
  \sign(x-y), \nonumber \\
  \left[ \hat{\varphi}_{\mu R}(x), \hat{\varphi}_{\nu L}(y)\right] &=&
  -i \pi \delta_{\mu\nu}.
  \label{eq:commboson2}
\end{eqnarray}
In terms of the transformed fields, the boson Hamiltonian $\hat H_0$
and the time-dependent perturbation $\hat H'$ read
\begin{eqnarray}
  \hat H_0 &=&
    \frac{v_F}{4\pi}
\sum_{\mu=1}^{N}\int_{-\infty}^{\infty}dx
\left[
\left(\frac{\partial \hat{\varphi}_{\mu L}}{\partial x}\right)^2 +
\left(\frac{\partial \hat{\varphi}_{\mu R}}{\partial x}\right)^2
\right] \label{eq:50.6b}
 \nonumber \\ && \mbox{}
  + \frac{E_{\rm c} N}{4\pi^2}
\left( \hat{\varphi}_{L,1}(0) +
\hat{\varphi}_{R,1}(0)
+\frac{2 \pi {\cal N}}{\sqrt{N}}
\right)^2, \ \ \ \ \\
  \hat H' &=&
  \frac{v_F}{2 \pi}
  \sum_{\mu=1}^{N} m_{\mu L}(t)
  \left. \frac{\partial \hat \varphi_{\mu L}(x)}{\partial x} \right|_{x \to
  a}
  \nonumber \\ && \mbox{}
  - \frac{v_F}{2 \pi}
  \sum_{\mu=1}^{N} m_{\mu R}(t)
  \left. \frac{\partial \hat \varphi_{\mu R}(x)}{\partial x} \right|_{x \to
  - a},
\end{eqnarray}
where
\begin{equation}
  m_{\mu L}(t) = \sum_{j} o_{\mu j} m_{j L}(t),\ \
  m_{\mu R}(t) = \sum_{j} o_{\mu j} m_{j R}(t).
\end{equation}

In the transformed basis there is no charging interaction for the
modes $\mu=2,\ldots,N$. Solving the
Heisenberg evolution equation for the fields $\hat \varphi_{\mu L}$
and $\hat \varphi_{\mu R}$,
\begin{eqnarray*}
  \frac{\partial \hat \varphi_{\mu L}(x,t)}{\partial t}
  &=& v_F m_{\mu L}(t) \delta(x-a) +
  v_F \frac{\partial \hat \varphi_{\mu L}(x,t)}{\partial x}, \\
  \frac{\partial \hat \varphi_{\mu R}(x,t)}{\partial t}
  &=& v_F m_{\mu R}(t) \delta(x+a)
  - v_F \frac{\partial \hat \varphi_{\mu L}(x,t)}{\partial x},
\end{eqnarray*}
we obtain an expression for $\hat \varphi(x,t)$ in terms
of boson fields at the reference time $t_{\rm ref}$,
\begin{eqnarray}
  \hat \varphi_{\mu L}(x,t) &=&
  \hat \varphi_{\mu L}(x + v_F(t-t_{\rm ref}),t_{\rm ref})
  \nonumber \\ && \mbox{}
  +  m_{\mu L}(t + (x-a)/v_F)
  \theta(a-x),
  \nonumber \\
  \label{eq:psiLR}
  \hat \varphi_{\mu R}(x,t) &=&
  \hat \varphi_{\mu R}(x - v_F(t-t_{\rm ref}),t_{\rm ref})
  \nonumber \\ && \mbox{}
  - m_{\mu R}(t - (x+a)/v_F)
  \theta(a+x),\ \ \
\end{eqnarray}
(Recall that $m_{\mu L}(t) = m_{\mu R}(t) = 0$ for $t < t_{\rm ref}$.)
For the transformed mode $\mu=1$, the charging interaction is
important. In this case, the Heisenberg equation of motion reads
\begin{widetext}
\begin{eqnarray*}
  \frac{\partial \hat \varphi_{1 L}(x,t)}{\partial t}
  &=& v_F m_{1 L}(t) \delta(x-a)
  + v_F \frac{\partial \hat \varphi_{1 L}(x,t)}{\partial x}
  + \frac{E_{\rm c} N}{\pi}
  \left( \hat{\varphi}_{1 L}(0) + \hat{\varphi}_{1 R}(0)
  + \frac{2 \pi {\cal N}}{\sqrt{N}} \right)
  \theta(-x), \\
  \frac{\partial \hat \varphi_{1 R}(x,t)}{\partial t}
  &=& v_F m_{1 R}(t) \delta(x+a)
  - v_F \frac{\partial \hat \varphi_{1 L}(x,t)}{\partial x}
  - \frac{E_{\rm c} N}{\pi}
  \left( \hat{\varphi}_{1 L}(0) + \hat{\varphi}_{1 R}(0)
  + \frac{2 \pi {\cal N}}{\sqrt{N}} \right) \theta(-x).
\end{eqnarray*}
and has the solution
\begin{eqnarray}
  \hat{\varphi}_{1 L}(x,t) &=&
  \hat \varphi_{1 L}(x + v_F(t-t_{\rm ref}),t_{\rm ref}) +
  m_{1 L}(t + (x-a)/v_F)
  \theta(a-x)
  \nonumber \\ && \mbox{}
  - \frac{E_{\rm c} N}{\pi} \int_{t_{\rm ref}}^{t} dt'
  \theta(-x) \theta(t'- t - x/v_F)
  e^{-E_{\rm c} N(t-t')/\pi}
  \nonumber \\ && \mbox{} \times
  \left[ \hat\varphi_{1 R}(-v_F(t'-t_{\rm ref}),t_{\rm ref})
  -
  m_{1 R}(t' - a/v_F)
  + \hat \varphi_{1 L}(v_F(t-t_{\rm ref}),t_{\rm ref}) +
  m_{1 L}(t' - a/v_F)
  + \frac{2 \pi {\cal N}}{\sqrt{N}} \right], \nonumber \\
  \hat{\varphi}_{1 R}(x,t) &=&
  \hat \varphi_{1 R}(x - v_F(t-t_{\rm ref}),t_{\rm ref}) -
  m_{1 R}(t + (x+a)/v_F)\theta(a+x)
  \nonumber \\ && \mbox{}
  - \frac{E_{\rm c} N}{\pi} \int_{t_{\rm ref}}^{t} dt'
  \left[ \theta(x) \theta(t - x/v_F - t')
  e^{-E_{\rm c} N (t-x/v_F-t')/\pi}
  + \theta(-x) e^{-E_{\rm c} N (t-t')/\pi} \right]
  \nonumber \\ && \mbox{} \times
  \left[ \hat\varphi_{1 R}(-v_F(t'-t_{\rm ref}),t_{\rm ref})
  \vphantom{\frac{M}{M}}
  - m_{1 R}(t' - a/v_F)
  + \hat \varphi_{1 L}(v_F(t-t_{\rm ref}),t_{\rm ref}) +
  m_{1 L}(t' - a/v_F)
  + \frac{2 \pi {\cal N}}{\sqrt{N}} \right]. \nonumber \\
  \label{eq:psi1LR}
\end{eqnarray}

According to Eqs.\ (\ref{eq:psiLR}) and (\ref{eq:psi1LR}), the
effect of the time-dependent
potentials is to shift the boson fields by a real number. Such a
shift affects the average of the boson fields,
\begin{eqnarray}
  \langle \hat \varphi_{\mu L}(0,t) \rangle
  &=&
  m_{\mu L}(t - a/v_F), \\
  \langle \hat \varphi_{\mu R}(0,t) \rangle
  &=& -
  m_{\mu R}(t - a/v_F)
  - \frac{2 \pi {\cal N}}{\sqrt{N}} \delta_{\mu,1}
  + \frac{E_{\rm c} N}{\pi} \delta_{\mu,1}
  \int_{t_{\rm ref}}^{t} dt' e^{-E_{\rm c}N(t-t')/\pi}
  \left[ m_{1 R}(t' - a/v_F) -
  m_{1 L}(t' - a/v_F) \right],
  \nonumber
\end{eqnarray}
but not the
connected two-point correlators. Hence, the connected
two-boson correlation functions are the same as in equilibrium.
They can be obtained by analytical continuation from the
imaginary-time correlators obtained in Refs.\
\onlinecite{kn:aleiner1998a,kn:aleiner2002} or they can be
calculated
directly from the solution of the equation of motion. In the
latter method one first calculates the spectral
density $A(t,t')$.
Since Eqs.\ (\ref{eq:psiLR}) and (\ref{eq:psi1LR})
express all boson operators in terms of the operators at the
same reference time $t_{\rm ref}$, the spectral density can
be found from the equal-time commutation relations
(\ref{eq:commboson2}). We only need the spectral density at
$x=0$,
\begin{eqnarray}
  A_{\mu L, \nu L}(t,t') &=&
  \langle [\hat \varphi_{\mu L}(0,t),
  \hat \varphi_{\nu L}(0,t') ]_{-} \rangle \nonumber \\
  &=&  - i \pi \delta_{\mu\nu} \sign(t-t'), \nonumber \\
  A_{\mu L, \nu R}(t,t') &=&
  \langle [\hat \varphi_{\mu L}(0,t)
  \hat \varphi_{\nu R}(0,t') ]_{-} \rangle \nonumber \\
  &=& i \pi \delta_{\mu \nu}
  \left[1 - \frac{E_{\rm c} N}{\pi} \delta_{\mu,1}
  \int_{0}^{\infty} d\zeta e^{-E_{\rm c} N \zeta/\pi}
  \sign(t'-t-\zeta) \right], \nonumber \\
  A_{\mu R,\nu L}(t,t') &=&
  \langle [\hat \varphi_{\mu R}(0,t)
  \hat \varphi_{\nu L}(0,t') ]_{-} \rangle \nonumber \\
  &=& -i \pi \delta_{\mu \nu}
  \left[1 - \frac{E_{\rm c} N}{\pi} \delta_{\mu,1}
  \int_{0}^{\infty} d\zeta e^{-E_{\rm c} N \zeta/\pi}
  \sign(t-\zeta-t') \right], \nonumber \\
  A_{\mu R,\nu R}(t,t') &=&
  \langle [\hat \varphi_{\mu R}(0,t),
  \hat \varphi_{\nu R}(0,t') ]_{-} \rangle \nonumber \\
  &=&
  -i \pi \delta_{\mu\nu} \sign(t-t').
\end{eqnarray}
The two-boson correlation functions are then found
using the fluctuation-dissipation theorem,
see, {\em e.g.,} Ref.\ \onlinecite{kn:abrikosov1963}. The resulting
expressions for the averages and the connected correlation functions
of the original boson fields $\hat{\varphi}_{Lj}$ and $\hat{\varphi}_{Rj}$,
$j=1,\ldots,N$ are
\begin{eqnarray}
  \langle \hat{\varphi}_{jL}(t) \rangle
  &=&
  \int^{t} dt' \mu_{jL}(t') \\
  \langle \hat{\varphi}_{jR}(t) \rangle &=&
  - \int^{t} dt' \mu_{jL}(t')
  - \frac{2 \pi {\cal N}}{N}
  - \int^{t} dt' \sum_{k=1}^{N} i_{jk}(t-t')
  [\mu_{k R}(t'') - \mu_{k L}(t'')].
  \label{eq:muRmuL}
\end{eqnarray}
where $i_{ik}$ was defined in Eq.\ (\ref{eq:i}),
and
\begin{eqnarray}
  \langle T_{\rm c} \hat{\varphi}_{iL}(t')
  \hat{\varphi}_{jL}(t'') \rangle
  &=& \delta_{ij} \ln\left(\frac{\lambda \pi T/i}{\sinh
  \left[ \pi T\left(t' - t'' - i \lambda \sign_{\rm c}(t'-t'')
  \right) \right]}\right) + A
  + \left( \delta_{ij} - \frac{1}{N} \right) B
  \nonumber \\
  &=&
  \langle T_{\rm c} \hat{\varphi}_{iR}(t')
  \hat{\varphi}_{jR}(t'') \rangle,
  \nonumber\\
  \langle T_{\rm c}
  \hat{\varphi}_{iL}(t') \hat{\varphi}_{jR}(t'') \rangle
  &=&  \delta_{ij} \frac{i \pi}{2} \sign_{\rm c} (t'-t'')
    - \frac{1}{N} \ln f(t'',t') - A
  \nonumber \\
  &=&
  \langle T_{\rm c}
  \hat{\varphi}_{iR}(t'') \hat{\varphi}_{jL}(t') \rangle,
\end{eqnarray}
\end{widetext}
where $A$ and $B$ are positive constants that are taken to infinity at
the end of the calculation and $f$ is given in Eq.\ (\ref{eq:f}).
{}Calculation of the fermion correlators of Sec.\
\ref{sec:correlations} using the rule (\ref{eq:expavg}) is
straightforward now.

\section{Ensemble average}
\label{app:a}

In this section we present some material relevant for performing
the averages over an ensemble of microscopically different but
macroscopically equivalent quantum dots.

We consider the statistical distribution of the scattering matrix for
an ensemble of chaotic quantum dots. The dots are placed in a magnetic
field. The magnetic field strength is described by the dimensionless
parameter $\alpha$, which is proportional to the magnetic flux $\Phi$
through the dot,
\begin{equation}
  \alpha = \frac{e \Phi}{h c} \left[ \frac{\kappa \hbar v_F l}{L^2
  \Delta} \right]^{1/2},
\end{equation}
where $\Delta$ is the mean spacing between spin-degenerate levels,
$L$ its linear
size, $l$ the elastic mean free path inside the dot, and $\kappa$ a
numerical constant of order unity. One has $\kappa = 4\pi/15$ for
a diffusive sphere of radius $L$ and $\kappa = \pi/2$ for a diffusive
disk of the same radius \cite{kn:frahm1995b}.
If the electron motion inside the dot is
ballistic, but with diffusive boundary scattering, $l$ is replaced
by $5L/8$ and $\pi L/4$ for the cases of the sphere and the disk,
respectively. The shape of the dot is controlled by two gate
voltages, which are described by the dimensionless parameters
$X_1$ and $X_2$. The normalization of the parameters $X_1$ and $X_2$
is the same as in Ref.\ \onlinecite{kn:polianski2003}.

Two types of scattering processes contribute to the scattering
matrix: direct reflection at the contacts, which is described by
the energy-independent reflection matrix $r_{\rm c}$, and scattering from
inside the quantum dot. The matrix $r_{\rm c}$ is a property of
the contacts and is not subject to mesoscopic fluctuations; it is
only the contribution that involves scattering from inside the
dot that has fluctuations. In order to manifestly separate the two
contributions, $S$ is parameterized as\cite{kn:friedman1985}
\begin{equation}
  S = r_{\rm c} + (1 - r_{\rm c} r_{\rm c}^{\dagger})^{1/2}
  S_0 (1 + r_{\rm c}^{\dagger} S_0)^{-1}
  (1 - r_{\rm c}^{\dagger} r_{\rm c})^{1/2},
  \label{eq:src}
\end{equation}
where $S_0$ has the same statistics as the scattering matrix of
a chaotic quantum dot ideal contacts.

For electrons with spin (and without spin-orbit scattering), the
scattering matrix takes the form
\begin{equation}
  S = S^{\rm o} \otimes \openone_2,
\end{equation}
where $\openone_2$ is the $2 \times 2$ unit matrix (acting on the
spin degrees of freedom) and $S^{\rm o}$ is a unitary matrix of
size $N^{\rm o} = N/2$ representing scattering of the orbital
degrees of freedom. A similar decomposition applies to the
matrices $r_{\rm c}$ and $S_0$. The results below are for the
distribution $P(S^{\rm o}_0)$ of the orbital scattering matrix
$S_0^{\rm o}$, in the limit of large $N$. For simplicity of
notation we will drop the subscript ``0'' and the superscript
``o''.

For large $N$, only
moments of the distribution that contain as many factors
of $S$ as of its complex conjugate $S^*$ and in which the
indices of factors $S$ and $S^*$ are pairwise equal
are nonzero.\cite{kn:brouwer1996a} This implies that all odd moments
of $P(S)$ are zero. Since the
probability of $S$ is close to Gaussian for large $N$, with small
non-Gaussian corrections, $P(S)$ is well characterized by its
moments. The second moment characterizes the Gaussian part of
the distribution; Higher moments represent the non-Gaussian
corrections to $P(S)$.

In order to specify $P(S)$, it is sufficient to consider
generic moments in which all (pairs of) indices are different. Moments
in which two or more pairs of indices coincide are fully determined
by the invariance of the ensemble under basis changes in the
leads.\cite{kn:brouwer1996a} For the general time-dependent case, we
need the second moment $\langle S_{ij}(t,t-\tau)
S_{ij}(s,s-\sigma)^*$ for $i \neq j$ only, which was calculated
in Ref.\ \onlinecite{kn:polianski2003},
\begin{widetext}
\begin{eqnarray}
  \langle S_{ij}(t,t-\tau) S_{ij}(s,s-\sigma)^*
  \rangle &=& \frac{1}{N \tau_{\rm d}}
  \delta(\tau-\sigma) \theta(\tau)
  \\ && \mbox{} \times
  \exp \left[ - \int_0^{\tau} d\tau'
  \left( \frac{1}{\tau_{\rm d}}
  + (\alpha(t-\xi) - \alpha(s-\xi))^2
  + 2 \sum_{j} (X_j(t-\xi) - X_j(s-\xi))^2 \right) \right].
  \nonumber
\end{eqnarray}

For the time-independent case, we need to go calculate the
corrections to the Gaussian distribution.
Using the short-hand notation
$S(1) = S[\varepsilon(1),\alpha(1),X_1(1),X_2(1)]$, the
relevant moments are
\begin{eqnarray}
  \langle S_{ij}(1) S_{ij}(2)^* \rangle
  &=& W_1(1,2), \\
  \langle S_{ij}(1) S_{kj}(2)^* S_{kl}(3)
  S_{il}(4)^*  \rangle &=&
  W_2(1,2,3,4), \\
  \langle S_{ij}(1) S_{ij}(2)^* S_{kl}(3)
  S_{kl}(4)^*  \rangle &=&
  W_1(1,2) W_1(3,4) + W_{1,1}(1,2,3,4), \\
  \langle S_{ij}(1) S_{kj}(2)^* S_{kl}(3)
  S_{ml}(4)^* S_{mn}(5) S_{in}(6)^* \rangle
  &=&
  W_{3}(1,2,3,4,5,6),
\end{eqnarray}
where all indices $i$, $j$, $k$, $l$, $m$, and $n$ are different.
In addition to the moments listed above, we need moments in
which the indices of a factor $S_{ij}$ or $S^{*}_{ij}$
are interchanged. These follow from the above
moments using the identity $S_{ij}(\varepsilon,X_1,X_2,\alpha)
= S_{ji}(\varepsilon,X_1,X_2,-\alpha)$. The function $W_1$ was
calculated in Refs.\ \onlinecite{kn:efetov1995,kn:frahm1995a};
the function $W_2$ was calculated in Ref.\
\onlinecite{kn:polianski2003}. Repeating the calculations of
Ref.\ \onlinecite{kn:polianski2003} for $W_{1,1}$ and $W_{3}$, one
finds, for large $N$,
\begin{eqnarray}
  W_1(1,2) &=& \frac{1}{F_1(1,2)} \\
  W_2(1,2,3,4) &=& - \frac{F_2(1,2,3,4)}{F_1(1,2) F_1(3,2) F_1(3,4)
  F_1(1,4)}, \\
  W_{1,1}(1,2,3,4) &=& \frac{1}{F_1(1,2)^2 F_1(3,4)^2}
  \left[
  \frac{F_2(1,2,3,4) F_2(1,4,3,2)}{F_1(1,4) F_1(3,2)}
  + \frac{F_2(1,2,1,4) F_2(1,4,3,4)}{F_1(1,4)^2}
  \right. \nonumber \\ && \left. \mbox{}
  + \frac{F_2(1,2,3,2) F_2(3,2,3,4)}{F_1(3,2)^2}
  - \frac{F_3(1,2,1,4,3,4)}{F_1(1,4)}
  - \frac{F_3(1,2,3,4,3,2)}{F_1(3,2)}
  \right], \\
  W_{3}(1,2,3,4,5,6) &=&
  \frac{1}{F_1(1,2) F_1(3,2) F_1(3,4) F_1(5,4) F_1(5,6) F_1(1,6)}
  \left[
  \frac{F_2(1,2,3,4) F_2(1,4,5,6)}{F_1(1,4)}
  \right. \nonumber \\ && \left. \mbox{}
  + \frac{F_2(1,2,5,6) F_2(3,4,5,2)}{F_1(5,2)}
  + \frac{F_2(3,4,5,6) F_2(1,2,3,6)}{F_1(3,6)}
  - F_3 (1,2,3,4,5,6)
  \right],
\end{eqnarray}
where
\begin{eqnarray}
  F_n(1,\ldots,n) &=&
  N \left[1 -  i \sum_{m=1}^{n}
  (-1)^m \varepsilon(m) \tau_{\rm d} \right]
  + \left[ \sum_{m=1}^{n} (-1)^m \alpha(m) \right]^2
  + 2 \sum_{j=1}^{2} \left[ \sum_{m=1}^{n} (-1)^m X_j(m) \right]^2.
  \ \ \
\end{eqnarray}

With the help of these moments, the ensemble averages of Secs.\
\ref{sec:steadystate} and \ref{sec:adiabatic} can be done using the
diagrammatic technique of Ref.\ \onlinecite{kn:brouwer1996a}.
\end{widetext}

\section{Quantum dot with relaxation}
\label{app:c}

A highly simplified model for relaxation is to consider the model
of Sec.\ \ref{sec:fictitiouslead} for $r < 1$. In this appendix,
we describe a more realistic (but still phenomenological) model
for relaxation in the quantum dot.

\begin{figure}[t]
\epsfxsize=0.8\hsize
\epsffile{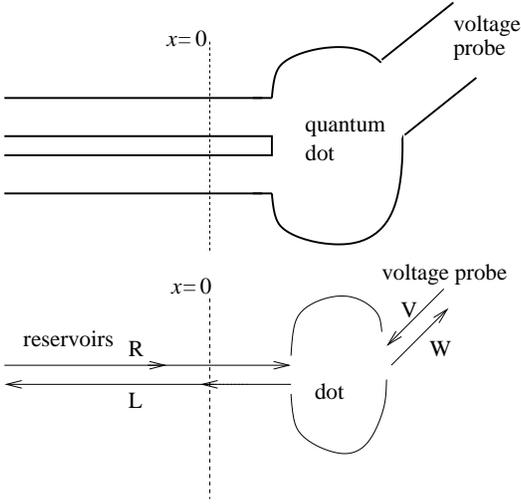}
\caption{Top panel: schematic drawing of a quantum dot with a
voltage probe. Bottom panel: layout of labeling of fermion fields
``R'', ``L'', ``V'', and ``W''.
\label{fig:geometry5}}
\end{figure}

In this model, the quantum dot itself (rather than the point
contact) is coupled to a fictitious lead, see Fig.\
\ref{fig:geometry5}. This fictitious lead is connected to a
reservoir, the chemical potential of which is chosen such that no
net current flows through this lead. For that reason, the
fictitious lead is referred to as `voltage probe'. This model was
originally proposed by
B\"uttiker,\cite{kn:buettiker1986,kn:buettiker1988} and has been
applied by various authors to describe relaxation effects in
chaotic quantum dots.\cite{kn:brouwer1995,kn:baranger1995,
kn:brouwer1997c,kn:huibers1998,kn:alves2002} In order to model
spatially distributed relaxation, the voltage probe has $M \gg 1$
channels, each coupled to the quantum dot via a tunnel barrier
with transmission probability $\Gamma_{V} \ll
1$.\cite{kn:brouwer1997c} The relaxation rate is set by the
product $M \Gamma_{V}$.

\begin{figure}
\epsfxsize=0.95\hsize \epsffile{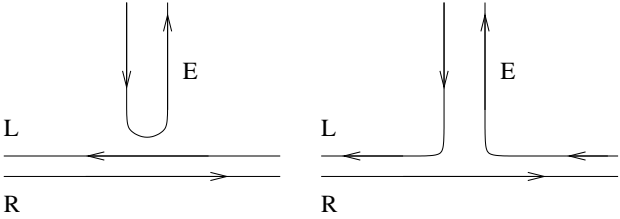}
\caption{\label{fig:5} Left diagram: Schematic drawing of the
fictitious lead (E) coupling of which to the left-moving fermions
through the Hamiltonian (\protect\ref{eq:Hperfect}) gives the
second term of the effective action (\protect\ref{eq:SeffV}).
Right diagram: Definition of the fermion fields R, E, and L after
the change of variables of Eq.\ (\protect\ref{eq:LE}).}
\end{figure}

Electrons in the physical point contact are labeled ``R'' and
``L'' as before. We label electrons entering the dot from the
voltage probe by ``V'', and electrons exiting the dot towards the
voltage probe by ``W'', see Fig.\ \ref{fig:5}. The scattering
matrix of the dot now has dimension $(N+M)$, and relates the
fields $\psi_R$ and $\psi_V$ to the fields $\psi_E$ and $\psi_W$,
\begin{widetext}
\begin{eqnarray}
  \hat \psi_{jL}^{\vphantom{\dagger}}(0,t) &=&
  \sum_{k=1}^{N} \int d\tau
  S_{jk}(t,t-\tau) \hat \psi_{kR}^{\vphantom{\dagger}}(0,t-\tau)
  + \sum_{k=1}^{M} \int d\tau
  s_{jk}(t,t-\tau) \hat \psi_{kV}^{\vphantom{\dagger}}(t-\tau), \nonumber \\
  \hat \psi_{jW}^{\vphantom{\dagger}}(t) &=&
  \sum_{k=1}^{M} \int d\tau
  \tilde S_{jk}(t,t-\tau) \hat \psi_{kV}^{\vphantom{\dagger}}(t-\tau)
  + \sum_{k=1}^{N} \int d\tau
  \tilde s_{jk}(t,t-\tau) \hat \psi_{kR}^{\vphantom{\dagger}}(0,t-\tau).
  \nonumber \\
  \label{eq:psiEVpsi}
\end{eqnarray}
The fields $\psi_V$ and $\psi_W$ are evaluated at the interface
between dot and the voltage probe reservoir.

In order to derive an effective action for the quantum dot with
voltage probe attached to it, we again attach
a fictitious lead at the location of the point contact, as described
in Sec.\ \ref{sec:fictitiouslead}. The Hamiltonian is written as
the sum $\hat H = \hat H_0 + \hat H_1$, where $\hat H_0$ and $\hat
H_1$ represent ideal coupling of the fictitious lead and an impurity
at the contact to the fictitious lead, respectively. For $\hat H_1$
we take Eq.\ (\ref{eq:Himp}) with $r_j = 1$, $j=1,\ldots,N$. We then
use Eq.\ (\ref{eq:psiEVpsi}) to eliminate the fields $\psi_L$ from
the expression for $\hat H_1$ and arrive at the effective action
\begin{eqnarray}
  {\cal S} &=& 2 v_F \int_{\rm c} dt_1 \int d\tau
  \sum_{mn}
  \left[ \hat \psi^{{\dagger}}_{mL}(0,t_1) S_{mn}(t_1,t_1-\tau)
  \hat \psi^{\vphantom{\dagger}}_{nR}(0,t_1-\tau)
  + \mbox{h.c.} \right]
  \nonumber \\ && \mbox{}
  +
  2 v_F \int_{\rm c} dt_1 \int d\tau
  \sum_{mp}
  \left[ \hat \psi^{{\dagger}}_{mL}(0,t_1) s_{mp}(t_1,t_1-\tau)
  \hat \psi^{\vphantom{\dagger}}_{Vp}(t_1-\tau)
  + \mbox{h.c.} \right].
\end{eqnarray}

The following formal manipulations closely follow those of
Ref.\ \onlinecite{kn:aleiner1998a}.
Since the current $I_j$, $j=1,\dots,N$, does not depend on the
fields $\hat \psi_{Vk}$, $k=1,\ldots,M$, we can integrate these out.
This results in a new effective action
\begin{eqnarray}
  {\cal S}' &=& 2 v_F \int_{\rm c} dt_1 \int d\tau_1
  \sum_{mn}
  \left[ \hat \psi^{{\dagger}}_{mL}(t_1) S_{mn}(t_1,t_1-\tau_1)
  \hat \psi^{\vphantom{\dagger}}_{nR}(t_1-\tau_1) +
  \mbox{h.c.} \right]
  \label{eq:Sprime}
  \\ && \mbox{}
  + 4 v_F^2 \int_{\rm c} dt_1 dt_2 \int d\tau_1 d\tau_2
  \sum_{mnp}
  \hat \psi^{{\dagger}}_{mL}(t_1) s_{mp}(t_1,t_1-\tau_1)
  G_{V}(t_1-\tau_1,t_2-\tau_2)
  (s^{\dagger})_{pn}(t_2',t_2)
  \hat \psi^{\vphantom{\dagger}}_{nL}(t_2),
  \nonumber
\end{eqnarray}
where
\begin{equation}
  G_{V}(t_1-\tau_1,t_2-\tau_2) = -i
  \langle T_{\rm c} \hat \psi^{\vphantom{\dagger}}_{Vp}(t_1-\tau_1)
  \hat \psi^{{\dagger}}_{Vp}(t_2-\tau_2) \rangle
\end{equation}
is the contour-ordered Green functions for fermions coming
in from the voltage probe. (Note that the contour ordering is
with respect to the contour times $t_1$ and $t_2$.) We
assume that the change
of potentials (either external or internal) is slow in
comparison to the {\em minimum} of the relaxation time and
the escape rate into the physical leads. Then, using the
gradient expansion we find
\begin{eqnarray}
  \int d\tau_1 d\tau_2' s(t_1,t_1-\tau_1)
  G_V(t_1-\tau_1,t_2-\tau_2) s^{\dagger}
  (t_2-\tau_2,t_2)  &=&
  G_V(t_1,t_2) + \delta G_V(t_1,t_2),
  \label{eq:GVsubst}
\end{eqnarray}
where $\delta G_V(t_1,t_2)$ is defined as
\begin{eqnarray}
  \delta G_V(t_1,t_2) &=&
  - \int d\tau_1 d\tau_2
  S(t_1,t_1-\tau_1) G_V(t_1-\tau_1,t_2-\tau_1) S^{\dagger}
  (t_2-\tau_2,t_2)
  +
  \frac{1}{2 \pi v_F} \int d\varepsilon
  e^{-i \varepsilon (t_1-t_2)}
  \frac{\partial f_V(\varepsilon)}{\partial \varepsilon}
  \nonumber \\ && \ \ \mbox{} \times
  \tr
  \left[ \left( \frac{\partial S(t;\varepsilon)}{\partial t}
  + \frac{\partial \mu_V}{\partial t}
  \frac{\partial S(t;\varepsilon)}{\partial \varepsilon} \right)
  S^{\dagger}(t;\varepsilon)
  + \left( \frac{\partial s(t;\varepsilon)}{\partial t}
  + \frac{\partial \mu_V}{\partial t}
  \frac{\partial s(t;\varepsilon)}{\partial \varepsilon} \right)
  s^{\dagger}(t;\varepsilon) \right].
  \label{eq:deltaG}
\end{eqnarray}
Here $f_V(\varepsilon)$ is the distribution function in the
voltage probe, and $S(t;\varepsilon)$ and $s(t;\varepsilon)$ are
the Fourier transforms of the scattering matrices $S(t,t-\tau)$
and $s(t,t-\tau)$, respectively, see Eq.\ (\ref{eq:Stransform}).
The second term in Eq.\
(\ref{eq:deltaG}) does not depend on the contour positions of
$t_1$ and $t_2$. Upon substitution of Eq.\ (\ref{eq:GVsubst})
into the effective action (\ref{eq:Sprime}), one finds
\begin{eqnarray}
  {\cal S}' &=& 2 v_F \int_{\rm c} dt_1 \int d\tau_1
  \sum_{mn}
  \left[ \hat \psi^{{\dagger}}_{mL}(t_1) S_{mn}(t_1,t_1-\tau_1)
  \hat \psi^{\vphantom{\dagger}}_{nR}(t_1-\tau_1) +
  \mbox{h.c.} \right]
  \nonumber \\ && \mbox{}
  + 4 v_F^2 \int_{\rm c} dt_1 dt_2
  \sum_{mn} \hat \psi^{{\dagger}}_{mL}(t_1) G_{V}(t_1,t_2)
  \hat \psi^{\vphantom{\dagger}}_{nL}(t_2)
  + 4 v_F^2 \int_{\rm c} t_1 dt_2
  \sum_{mn}
  \hat \psi^{{\dagger}}_{mL}(t_1)
  \delta G_{V}(t_1,t_2)
  \hat \psi^{\vphantom{\dagger}}_{nL}(t_2).
  \label{eq:SeffV}
\end{eqnarray}
\end{widetext}

At this point, we recognize that the second term in Eq.\
(\ref{eq:SeffV}) is the same one as the effective action that one
would have obtained by coupling the left-moving fermions to a
fictitious lead with $N$ channels, with chemical potential $\mu_V$.
This situation is shown schematically in Fig.\ \ref{fig:5}, where the
fermions in that fictitious lead are labeled with the subscript
``E''. The coupling that corresponds to the second term in Eq.\
(\ref{eq:SeffV}) is that of the Hamiltonian
\begin{eqnarray}
  \hat H_2 &=& 2 v_F \sum_{m} v_m \left[\hat \psi^{\dagger}_{nL}(0)
  \hat \psi^{\vphantom{\dagger}}_{mE}(0)
  \right. \nonumber \\ && \ \ \left. \mbox{}
  + \hat \psi^{\dagger}_{mE}(0) \hat \psi^{\vphantom{\dagger}}_{mL}(0)\right].
  \label{eq:Hperfect}
\end{eqnarray}
In order to eliminate the Hamiltonian (\ref{eq:Hperfect}) we
introduce new fermion fields according to\cite{kn:aleiner2002}
(see Fig.\ \ref{fig:5})
\begin{eqnarray}
  \mbox{old} && \mbox{new} \nonumber \\
  \hat \psi_{mL}(x) &\to& \hat \psi_{mL}(x) \theta(-x) + \hat \psi_{mE}(-x)
  \theta(x),
  \nonumber \\
  \hat \psi_{Em}(x) &\to& i[\hat \psi_{mE}(-x) \theta(-x) - \hat \psi_{mL}(x)
  \theta(x),\ \
  \label{eq:LE}
\end{eqnarray}
where the Heaviside function $\theta(x)$ is defined such that
$\theta(0) = 1/2$. For the new fermion fields, the ``L'' fields
have the chemical potential $\mu_V$ of the voltage probe,
whereas the ``E'' fields have
the chemical potential $\mu_L$ of the fictitious lead.
Substituting Eq.\ (\ref{eq:LE})
into the remaining terms of Eq.\ (\ref{eq:SeffV}), one finds the
effective action
\begin{widetext}
\begin{eqnarray}
  {\cal S} &=& v_F \int_{\rm c} dt_1 \int d\tau_1
  \sum_{mn}
  \left[ (\hat \psi^{{\dagger}}_{mL}(t_1) +
  \hat \psi^{{\dagger}}_{mF}(t_1)) S_{mn}(t_1,t_1-\tau_1)
  \hat \psi^{\vphantom{\dagger}}_{nR}(t_1-\tau_1) +
  \mbox{h.c.} \right]
  \nonumber \\ && \mbox{}
  + v_F^2 \int_{\rm c} dt_1 dt_2
  \sum_{mn} (\hat \psi^{{\dagger}}_{mL}(t_1) +
  \hat \psi^{{\dagger}}_{mF}(t_1)) \delta G_{Vmn}(t_1,t_2)
  (\hat \psi^{\vphantom{\dagger}}_{nL}(t_2) +
  \hat \psi^{\vphantom{\dagger}}_{nE}(t_2)),
  \label{eq:SeffVV}
\end{eqnarray}
where, as before, all fermion fields are evaluated at the
position $x = 0^+$.
Although the field $\psi_{Ej}$, $j=1,\ldots,N$ could, in
principle, be
integrated out of the effective action, it is more convenient to
keep them explicit and deal with them at the level of the
perturbation theory in ${\cal S}$. No physical
observables depend on the distribution function $f_L$ of the
field $\psi_E$.

The chemical potential $\mu_V$ of the voltage probe reservoir must
be chosen such as to have no net current flowing into that
reservoir at any time. For the current $I_V$ in the voltage probe
we find, in the adiabatic approximation,
\begin{eqnarray}
  I_V(t) &=&
  \frac{e i v_F}{4 \pi}
  \int d\varepsilon \tr
  \left[G_V^{\rm K}(t;\varepsilon) -
  \left( 1 - \frac{i}{2} D_{t,\varepsilon} \right)
  \left(
  \tilde S(t;\varepsilon) G_{V}^{\rm K}(t;\varepsilon)
  \tilde S^{\dagger}(t;\varepsilon)
  \nonumber \right. \right. \\ && \ \ \left. \left. \mbox{}
  - \tilde s(t;\varepsilon) G_{RR}^{\rm K}(t;\varepsilon)
  \tilde s^{\dagger}(t;\varepsilon)
  - \tilde s(t;\varepsilon)  G_{RV}^{\rm K}(t;\varepsilon)
  \tilde S^{\dagger}(t;\varepsilon)
  - \tilde S(t;\varepsilon)  G_{VR}^{\rm K}(t;\varepsilon)
  \tilde s^{\dagger}(t;\varepsilon)
  \right) \vphantom{\frac{i}{2}}\right].
  \nonumber \\
\end{eqnarray}
where $D_{t,\varepsilon} A B = (\partial A/\partial t) (\partial
B/\partial \varepsilon) - (\partial A/\partial \varepsilon) (\partial
B/\partial t)$, the superscript ``K'' refers to the Keldysh component,
and
\begin{eqnarray}
   G_{RR}(t',s') &=&
  -i \langle T_{\rm c}
  \hat \psi_{R}^{\vphantom{\dagger}}(t')
  \hat \psi_{R}^{\dagger}(s')
  \rangle, \nonumber \\
   G_{RV}(t',s') &=&
  -i \langle T_{\rm c}
  \hat \psi_{R}^{\vphantom{\dagger}}(t')
  \hat \psi_{V}^{\dagger}(s')
  \rangle, \nonumber \\
   G_{VR}(t',s') &=&
  -i \langle T_{\rm c}
  \hat \psi_{V}^{\vphantom{\dagger}}(t')
  \hat \psi_{R}^{\dagger}(s')
  \rangle.
\end{eqnarray}
Calculation of the Green function $G_{RR}(t',s')$ can be done with
help of the effective action (\ref{eq:SeffVV}) derived above. For
the Green functions $G_{RV}(t',s')$ and $G_{VR}(t',s')$, which keep
explicit reference to field $\psi_{V}$, a slightly more complicated
calculation is necessary. After some algebra, we then find
\begin{eqnarray}
  (\tilde G_{RV})_{ij}(t',s') &=&
  - v_F i \left\langle
  \int_{\rm c} dt_1 \int d\tau_1
  \sum_{m} \hat \psi^{\vphantom{\dagger}}_{Ri}(t)
  (\hat \psi^{{\dagger}}_{mL}(t_1)
  + \hat \psi^{{\dagger}}_{mE}(t_1))
  s_{mj}(t_1,t_1-\tau_1) G_{V}(t_1-\tau_1,s)
  \right\rangle,
  \nonumber
  \\
  (\tilde G_{VR})_{ij}(t,s) &=&
  - v_F i \left\langle \int_{\rm c}
  dt_1 \int d\tau_1
  \sum_{m} G_{V}(t,t_1-\tau_1) s^{\dagger}_{im}(t_1-\tau_1,t_1)
  (\hat \psi^{\vphantom{\dagger}}_{mL}(t_1)+
  \hat \psi^{\vphantom{\dagger}}_{mE}(t_1))
  \hat \psi^{\dagger}_{jR}(s)
  \right\rangle.
  \label{eq:GVRaction}
  \label{EQ:GVR}
\end{eqnarray}
\end{widetext}
where the average is taken with respect to the effective action
(\ref{eq:SeffVV}).

This concludes the presentation of the formal framework of the
current calculation in the presence of relaxation. The remainder
of the calculation proceeds along the same lines as in
Sec.\ \ref{sec:current}. The backscattering term in the
effective action (\ref{eq:SeffVV}) is equal to the original effective
action (\ref{eq:SeffA}), but with a sub-unitary scattering matrix $S$
instead of a unitary scattering matrix $S$; the remaining terms in
the effective action do not contain the fields $\psi_R$ and hence do
not contribute to the interaction correction to the current. As a
consequence, the expression for the interaction correction to the
current one obtains in an expansion in the action (\ref{eq:SeffVV})
is formally equal to the expressions obtained in an expansion in
the original action (\ref{eq:SeffA}), but with a sub-unitary
scattering matrix $S(t,t-\tau)$ instead of a unitary scattering
matrix.

There is a crucial difference with the
calculation of Sec.\ \ref{sec:current}, however: The
fact that the action (\ref{eq:SeffVV})
features a sub-unitary scattering matrix means that, for strong
relaxation, the effective action
is small, and an expansion in the action is justified. This
is in contrast to the case of a fully coherent quantum dot, where
we argued that there was no small parameter that justified
an expansion in the action. We may thus conclude that the
interaction correction to the dc conductance and the capacitance
calculated in Refs.\ \onlinecite{kn:brouwer1999c} and
\onlinecite{kn:aleiner2002} describe the case of a quantum
dot with a relaxation time much smaller than the mean
dwell time $\tau_{\rm d}$. In this sense, although
they do not apply to the case of a fully coherent quantum dot,
Refs.\ \onlinecite{kn:brouwer1999c} and
\onlinecite{kn:aleiner2002} do capture the first effect of
coherent scattering inside the dot at temperatures where
scattering inside the dot is still predominantly incoherent.

In light of this, the general trend of the results obtained in
Refs.\ \onlinecite{kn:brouwer1999c} and \onlinecite{kn:aleiner2002}
is that the charging interaction {\em reduces} the effective
dephasing rate: the weight of scattering processes with a short
delay time is increased,\cite{kn:aleiner2002} leading to a
smaller probability to dephase.\cite{kn:clerk2001}
(Since the average delay time
must remain constant, interactions must also increase the weight
of scattering processes with a long delay time. If the
dephasing time is shorter than the average delay time, the net
effect of the interactions is to suppress dephasing.) A smaller
dephasing rate corresponds to a larger weak localization
correction and larger conductance fluctuations, as was found
in Refs. \onlinecite{kn:brouwer1999c,kn:aleiner2002}.
A detailed analysis of the expression for the current in the voltage
probe model for the various limiting cases, as we did for the fully
coherent quantum dot, is of little practical use, since the
non-interacting (Hartree) contribution to the current contains the
dephasing rate, which is an unknown parameter this model.


\bibliography{refs}

\end{document}